\documentclass[a4paper,fleqn,usenatbib]{mnras}
\usepackage[T1]{fontenc}
\usepackage{ae,aecompl}
\usepackage{graphicx}   
\usepackage{amsmath}    
\usepackage{amssymb}    
\usepackage{color,verbatim,url}

\title[Gravitational Lensing Analysis of KiDS]{Gravitational Lensing Analysis of the Kilo Degree Survey}
\author[K. Kuijken et al.]
{Konrad Kuijken$^1$\thanks{Email: kuijken@strw.leidenuniv.nl},
Catherine Heymans$^2$,
Hendrik Hildebrandt$^3$,
Reiko Nakajima$^3$,
\newauthor  
Thomas Erben$^3$,
Jelte T.A. de Jong$^1$,
Massimo Viola$^1$,
Ami Choi$^2$,
Henk Hoekstra$^1$,
\newauthor
Lance Miller$^{4}$,
Edo van Uitert$^{3,5}$,
Alexandra Amon$^2$,
Chris Blake$^6$,
Margot Brouwer$^1$,
\newauthor
Axel Buddendiek$^3$,
Ian Fenech Conti$^{7,8}$,
Martin Eriksen$^1$,
Aniello Grado$^9$,
\newauthor
Joachim Harnois-D\'eraps$^{10}$,
Ewout Helmich$^{1}$,
Ricardo Herbonnet$^1$,
Nancy Irisarri$^1$,
\newauthor 
Thomas Kitching$^{11}$,
Dominik Klaes$^3$,
Francesco Labarbera$^{9}$,
Nicola Napolitano$^{9}$,
\newauthor  
Mario Radovich$^{12}$,
Peter Schneider$^3$,
Crist\'obal Sif\'on$^1$,
Gert Sikkema$^{13}$,
Patrick Simon$^3$,
\newauthor   
Alexandru Tudorica$^3$,
Edwin Valentijn$^{13}$,
Gijs Verdoes Kleijn$^{13}$, 
Ludovic van Waerbeke$^{10}$.
\\
$^{1}$Leiden Observatory, Leiden University, Niels Bohrweg 2, 2333 CA Leiden, The Netherlands.\\
$^{2}$Scottish Universities Physics Alliance, Institute for Astronomy, University of Edinburgh, Royal Observatory, Blackford Hill, Edinburgh, EH9 3HJ, UK.\\ 
$^3$Argelander Institute for Astronomy, University of Bonn, Auf dem H{\"u}gel 71, 53121 Bonn, Germany.\\
$^4$Department of Physics, Oxford University, Keble Road, Oxford OX1 3RH, UK.\\ 
$^5$University College London, Gower Street, London WC1E 6BT, UK.\\
$^6$Centre for Astrophysics \& Supercomputing, Swinburne University of Technology, P.O. Box 218, Hawthorn, VIC 3122, Australia.\\
$^7$Institute of Space Sciences and Astronomy (ISSA), University of Malta, Msida MSD 2080.\\ 
$^8$Department of Physics, University of Malta, Msida, MSD 2080, Malta.\\
$^9$INAF - Osservatorio Astronomico di Capodimonte, Via Moiariello 16, 80131 Napoli, Italy.\\
$^{10}$Department of Physics and Astronomy, University of British Columbia, 6224 Agricultural Road, Vancouver, V6T 1Z1, BC, Canada.\\  
$^{11}$Mullard Space Science Laboratory, University College London, Holmbury St Mary, Dorking, Surrey RH5 6NT, UK.\\
$^{12}$INAF - Osservatorio Astronomico di Padova, via dell'Osservatorio 5, 35122 Padova, Italy.\\
$^{13}$Kapteyn Institute, University of Groningen, PO Box 800, NL 9700 AV Groningen.\\
\vspace{-1cm}
}

\date{Accepted XXX. Received YYY; in original form ZZZ}

\pubyear{2015}

\newcommand{\be}{\begin{equation}}  \newcommand{\ee}{\end{equation}}
  
\newcommand{\ba}{\begin{eqnarray}}\newcommand{\ea}{\end{eqnarray}}

\renewcommand{\d}[0]{{\rm d}}

\renewcommand{\vec}[1]{{\bmath{#1}}}
\newcommand{\mat}[1]{\mathbfss{#1}}

\newcommand{\x}{\vec{x}}
\newcommand{\y}{\vec{y}}
\newcommand{\W}{\mat{W}}

\newcommand{\putfig}[1]{\includegraphics[width=\hsize]{#1}}

\def\gs{\mathrel{\raise1.16pt\hbox{$>$}\kern-7.0pt %
\lower3.06pt\hbox{{$\scriptstyle \sim$}}}}         %
\def\ls{\mathrel{\raise1.16pt\hbox{$<$}\kern-7.0pt %
\lower3.06pt\hbox{{$\scriptstyle \sim$}}}}         %

\begin{document}
\label{firstpage}
\pagerange{\pageref{firstpage}--\pageref{lastpage}}
\maketitle

\begin{abstract}
The Kilo-Degree Survey (KiDS) is a multi-band imaging survey designed for cosmological studies from weak lensing and photometric redshifts. It uses the ESO VLT Survey Telescope with its wide-field camera OmegaCAM. KiDS images are taken in four filters similar to the SDSS $ugri$ bands. The best-seeing time is reserved for deep $r$-band observations. The median 5-$\sigma$ limiting AB magnitude is 24.9 and the median seeing is below 0.7\arcsec. \\
Initial KiDS observations have concentrated on the GAMA regions near the celestial equator, where extensive, highly complete redshift catalogues are available. A total of 109 survey tiles, one square degree each, form the basis of the first set of lensing analyses of halo properties of GAMA galaxies. 9 galaxies per square arcminute enter the lensing analysis, for an effective inverse shear variance of 69 per square arcminute. Accounting for the shape measurement weight, the median redshift of the sources is 0.53. \\
KiDS data processing follows two parallel tracks, one optimized for weak lensing measurement and one for accurate matched-aperture photometry (for photometric redshifts). This technical paper describes the lensing and photometric redshift measurements (including a detailed description of the Gaussian Aperture and Photometry pipeline), summarizes the data quality, and presents extensive tests for systematic errors that might affect the lensing analyses. We also provide first demonstrations of the suitability of the data for cosmological measurements, and describe our blinding procedure for preventing confirmation bias in the scientific analyses.  \\
The KiDS catalogues presented in this paper are released to the community through  \url{http://kids.strw.leidenuniv.nl}.
\end{abstract}

\begin{keywords}
cosmology: observations -- gravitational lensing: weak -- galaxies: photometry -- surveys
\vspace{-0.75cm}
\end{keywords}

\defcitealias{dejong/etal:2015}{deJ+15}
\defcitealias{miller/etal:2013}{M+13}
\defcitealias{heymans/etal:2012}{H+12}
\defcitealias{erben/etal:2013}{E+13}

\section{Introduction}
\label{sec:intro}
Measurements of the gravitational fields around galaxies have for many decades provided firm evidence for `dark matter': galaxies attract their constituent stars, each other and their surroundings more strongly than can reasonably be estimated on the basis of their visible contents (for a historical account of the subject see \citealt{sanders:2014}). Furthermore, observations of the temperature anisotropies of the cosmic background radiation show that most of this dark matter cannot be baryonic \citep{planckXIII:2015}, in agreement with constraints from Big Bang Nucleosynthesis models (e.g., \citealt{fields/olive:2006}). Understanding the distribution of matter in the Universe is therefore a fundamental task of observational cosmology. The cold dark matter model, augmented with increasingly sophisticated galaxy formation recipes, has been very successful in describing, and reproducing the detailed statistical properties of, the large-scale distribution of galaxies. Though important issues remain, the $\Lambda$CDM model is the baseline for interpreting galaxy formation.

A central role in testing galaxy formation and cosmology models is played by observational mass measurements. They provided the first evidence for dark matter as mass discrepancies in galaxies (e.g., \citealt{bosma:1978}; \citealt{rubin/thonnard/ford:1978}; \citealt{faber/gallagher:1979}; \citealt{vanalbada/etal:1985}; \citealt{fabricant/lecar/gorenstein:1980,buote/canizares:1994}) and clusters \citep{zwicky:1937}. Mass measurements also serve to establish the link between the observed galaxies and their dark haloes, whose assembly age and clustering is mass-dependent, as is well described by the halo model \citep{cooray/sheth:2002}. Masses can be obtained from internal and relative kinematics of galaxies and their satellites, from X-ray observations of hydrostatic hot gaseous haloes around galaxies and clusters, and from strong and weak gravitational lensing. 

While kinematics and X-ray mass determinations usually require assumptions of steady-state dynamical equilibrium, gravitational lensing directly probes the projected mass distribution. This model-independent aspect of lensing is very powerful, but comes at a price. Strong lensing measurements are rare and depend on suitable image configurations and mass distributions. These result in complex selection effects which it is essential to understand \citep{blandford/kochanek:1987}. Weak lensing, on the other hand, is intrinsically noisy and thus requires stacking many lenses, except for the most massive galaxy clusters \citep{tyson/etal:1984}. Over the past two decades, telescopes equipped with larger and larger CCD cameras have provided the means to make wide-area weak lensing studies 
possible: most recently from the CFHTLenS analysis of the CFHT Legacy Survey \citep[henceforth \citetalias{heymans/etal:2012}]{heymans/etal:2012}, which targeted galaxies \citep[see for example][]{coupon/etal:2015}, groups and clusters \citep[see for example][]{ford/etal:2015} and the large-scale structure \citep[see for example][]{fu/etal:2014}. 

This paper introduces the first lensing results from a new, large-scale multi-band imaging survey, the Kilo-Degree Survey (KiDS). Like the on-going Dark Energy Survey \citep[for first lensing results from DES science verification data see][]{melchior/etal:2015,vikram/etal:2015} and the HyperSuprimeCam survey \citep[for first lensing results from HSC see][]{Miyazaki/etal:2015}, KiDS aims to exploit the evolution of the density of clustered matter on large scales as a cosmological probe \citep{albrecht/etal:2006,peacock/etal:2006}, as well as to study the distribution of dark matter around galaxies with more accuracy than has been possible thus far from the ground 
\citep[e.g.,][]{mandelbaum/etal:2006b,vanuitert/etal:2011,velander/etal:2014} or space \citep[e.g.,][]{leauthaud/etal:2012}. 
Unlike DES and HSC, which use large allocations of time on 4- and 8-m facility telescopes respectively, KiDS uses a dedicated 2.6-m wide-field imaging telescope, specifically designed for exquisite seeing-limited image quality. It is also unique in that all its survey area overlaps with a deep near-infrared survey, VIKING \citep{edge/etal:2013}, providing extensive information on the spectral energy distribution of galaxies.

In \citet[henceforth \citetalias{dejong/etal:2015}]{dejong/etal:2015} we present the public data release of the first KiDS images and catalogues.  Here we describe the aspects of the survey, data quality and analysis techniques that are particularly relevant for the weak lensing and photometric redshift measurements, and introduce the resulting shape catalogues. Accompanying papers present measurements and analyses of the mass distribution around galaxy groups \citep{viola/etal:2015}, galaxies (van Uitert et al., in preparation), and satellites \citep{sifon/etal:2015}.

This paper is organized as follows. 
\S\ref{sec:data} presents the survey outline and data quality, as well as the data reduction procedures leading up to images and catalogues. \S\ref{sec:shapes} describes how the lensing measurements are made, \S\ref{sec:photom} discusses the photometry pipeline and the derived photometric redshifts, and in \S\ref{sec:sys} a number of tests for systematic errors in the data reduction are presented. Having demonstrated that the KiDS data deliver high-fidelity lensing measurements, in \S\ref{sec:cosmicshear} we calculate the cosmic shear signal from this first instalment of KiDS imaging. Our conclusions are summarized in \S\ref{sec:conclude}. In three appendices we give the mathematical detail of the PSF homogenization and matched-aperture photometry ``\textsc{GAaP}'' pipeline, illustrate some of the quality control plots that are used in the survey production and validation, and provide a guide to the source catalogues which are publicly available to download at \url{http://kids.strw.leidenuniv.nl}.

\begin{figure*}
 \putfig{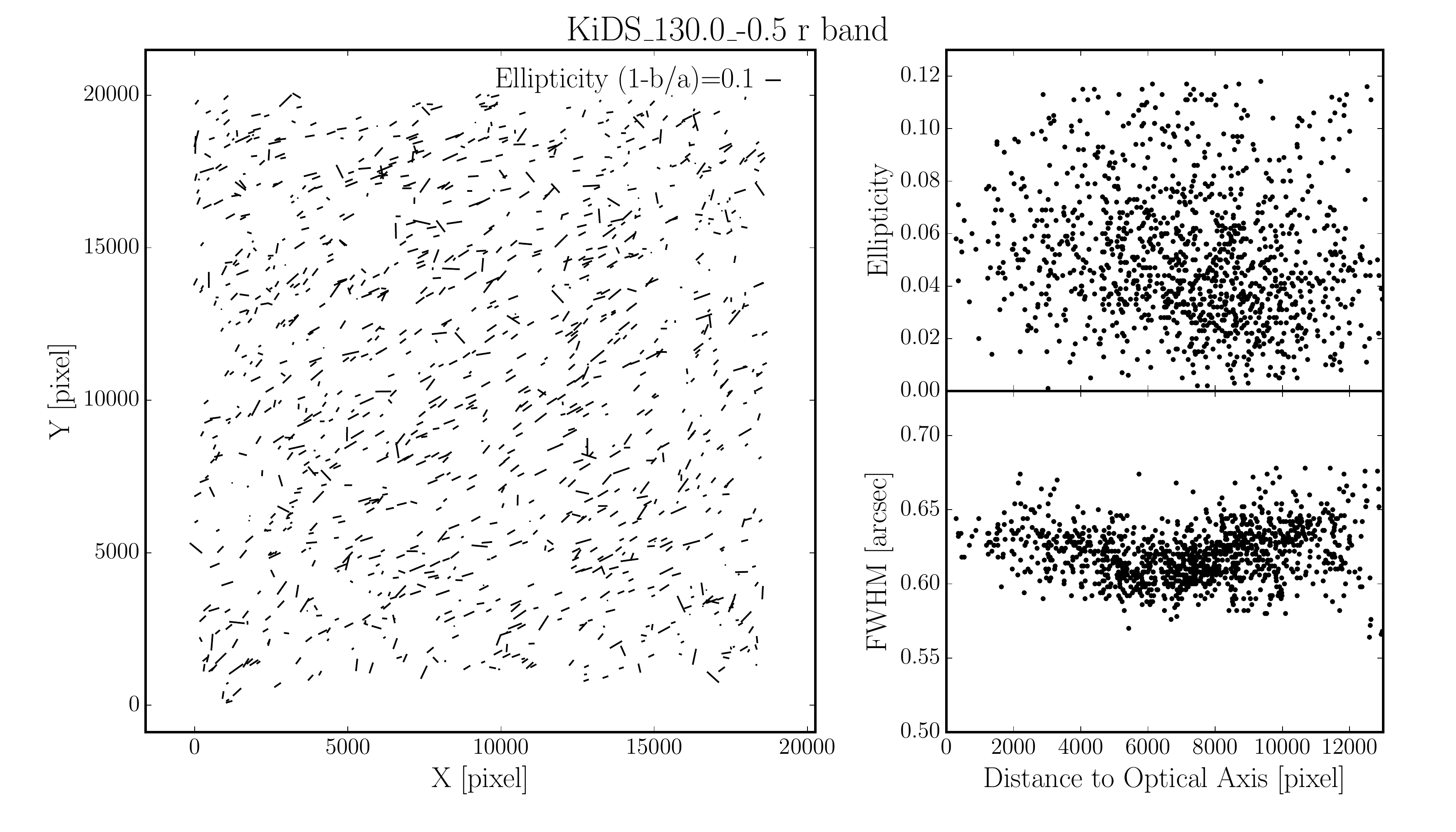}
  \caption{Example of high-quality KiDS data obtained with VST/OmegaCAM. PSF \textsc{SExtractor} parameters shown are for the stacked $r$-band image of tile KIDS\_132.0\_-0.5. {\it Left:} direction and strength of the ellipticities of stars in the field. {\it Right:} PSF ellipticity ({\it top}) and FWHM size ({\it bottom}) vs.\ distance from the centre of the image. 
 \label{fig:example_goodpsf}
}
\end{figure*}

\section{Description of Survey and Data quality}
\label{sec:data}

KiDS \citep{dejong/etal:2013} is a cosmological, multi-band imaging survey designed for weak lensing tomography. It uses the VLT Survey Telescope (VST) on the European Southern Observatory's Paranal observatory. The VST is an active-optics 2.6-m modified Ritchey-Chr\'etien telescope on an alt-az mount, with a 2-lens field corrector and a single instrument at its Cassegrain focus: the 300-megapixel OmegaCAM CCD mosaic imager. The 32 CCDs that make up the `science array' are $4102\times2048$-pixel e2v 44$-$82 devices, which sample the focal plane at a very uniform scale of 0.213\arcsec\ per 15-micron pixel. The chips are 3-edge buttable, and are mounted close together with small gaps of 25-85\arcsec. 
OmegaCAM has thinned CCDs, which avoids some of the problems inherent in deep depletion devices such as the `brighter-fatter' effect that introduces non-linearity into the extraction of PSF shapes from the images \citep[][see also \S\ref{sec:brightfat}]{melchior/etal:2015,niemi/etal:2015}, or the `tree rings' \citep{plazas/etal:2014}.

In order to maintain good image quality over the large field of view OmegaCAM makes use of wavefront sensing. For this purpose two auxiliary CCDs are mounted on the outskirts of the focal plane, vertically displaced $\pm2$mm with respect to the science array. As a result, the star images registered on these chips are significantly out of focus and their shapes and sizes provide the information required to monitor and optimise the optical set-up in real-time. Auto-guiding of both tracking and field rotation is done using two further (in-focus) auxiliary CCDs.
For more details on VST and OmegaCAM see \citet{capaccioli/etal:2012} and \citet{kuijken:2011} and references therein.

The integrated optical design of the telescope and camera makes for uniquely uniform and high-quality images over the full one-square degree field of view, well-matched to the seeing conditions on Paranal. An example `best-case' point spread function (PSF) measured from a co-added stack of five dithered sub-exposures is shown in Fig.~\ref{fig:example_goodpsf}, demonstrating that the system is able to deliver better than 0.6\arcsec\ seeing over the full field even in long exposures with low-level ellipticity distortion. This benign PSF variation can be modelled well and leads to very low residuals in the galaxy ellipticity measurements, (see \S\ref{sec:shapes} below). Furthermore, since there are no instrument changes on the VST the system is mostly stable, and continuously monitored photometrically. For a discussion on the long-term photometric stability of VST/OmegaCAM see \cite{verdoes/etal:2013}.
 
KiDS is part of a suite of three ESO Public Imaging Surveys, which are queue-scheduled together on the VST and observed as conditions and visibility allow \citep{arnaboldi/etal:2013}. The VPHAS+ survey \citep{drew/etal:2014} targets the Southern Galactic plane with short exposures in broad bands and H$\alpha$, and the ATLAS project \citep{shanks/etal:2015} covers some 5000 square degrees of extra-galactic sky in the Southern Galactic Cap to similar depth as the (mostly Northern) Sloan Digital Sky Survey \citep{ahn/etal:2014}.  KiDS, by contrast, aims to survey a 1500 square degree area to considerably greater depth, with the specific goal of measuring weak gravitational lensing masses for galaxies, groups and clusters as well as the power spectrum of the matter distribution on large scales. 

KiDS targets two $\sim10$-degree wide strips on the sky: an equatorial strip between Right Ascension $10^h20^m$ and $15^h50^m$ plus the GAMA G09 field between $08^h30^m$ and $09^h30^m$, and a Southern strip through the South Galactic pole between $22^h00^m$ and $03^h30^m$ (see \citetalias{dejong/etal:2015} for the footprint of the survey).
It makes use of four broad-band interference filters, $ugri$, with bandpasses very similar to the SDSS filters described in \citet{fukugita/etal:1996}.
The observations of a particular KiDS tile in any given filter consist of five dithered sub-exposures (four in the case of the $u$ band), and are taken in immediate succession. This choice means that KiDS is not well suited for the study of variable stars or supernovae, but it does mean that all data for each tile/filter combination are taken in very similar observing conditions, resulting in homogeneous data. The prevailing seeing and sky brightness dictate which observation is scheduled.  The seeing limits for the different filters are matched to the long-term Paranal average, such that the deep, best-seeing $r$-band observations can proceed at the same rate as the shallower $u$ and $g$ exposures. We summarize the observing parameters in Table~\ref{tab:obs}.

The first weak lensing results from KiDS are based on the first two public data releases \citepalias{dejong/etal:2015}, comprising the first 148 square degrees that were observed in all four filters. 109 square degrees from this data set overlap with the unique GAMA spectroscopic galaxy survey \citep{driver/etal:2011,baldry/etal:2014}, and this provides the focus of the early lensing science analyses.

A detailed discussion of the data quality can be found in \citetalias{dejong/etal:2015}; in Table~\ref{tab:obs} we summarize the key quality indicators of PSF sizes and limiting magnitudes. The PSF size distributions reflect that the best dark time is reserved for $r$, with $g$ and $u$ receiving progressively worse seeing time. The seeing distribution of the $i$-band, which is the only filter used in bright time, is very broad. Limiting AB magnitudes (calculated as 5$\sigma$ in a 2\arcsec\ aperture) in $g$ and $r$ are typically $\sim$25, with $u$ significantly shallower. For $i$ band observations, the large variation in seeing and sky brightness results in a wider variation in limiting magnitude than in the other bands. 

\begin{figure}
 \putfig{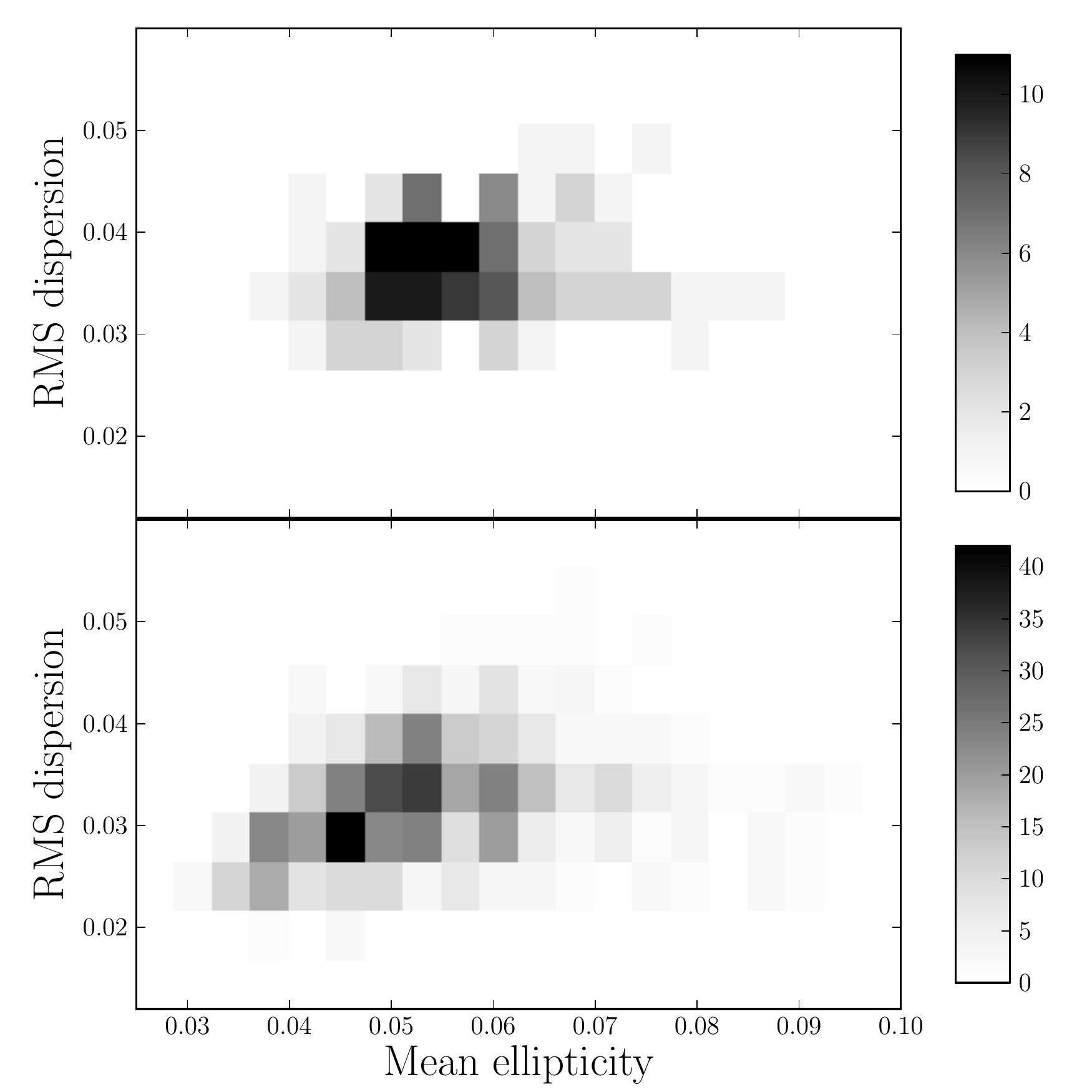}
 \caption{Distribution of mean ellipticities and standard deviations of ellipticities of co-added images in data releases 1 and 2 of KiDS. The values are based on \textsc{SExtractor} ellipticity measurements of the 500 brightest unsaturated stars in each tile. The grey scale indicates the number of survey tiles in each bin. Top: $r$ band only; Bottom: data from all filters.
 \label{fig:psf_ellipticities}
}
\end{figure}

\begin{figure}
\includegraphics[width=\hsize,trim=0.45in 0.95in 6in 0.45in,clip=true]{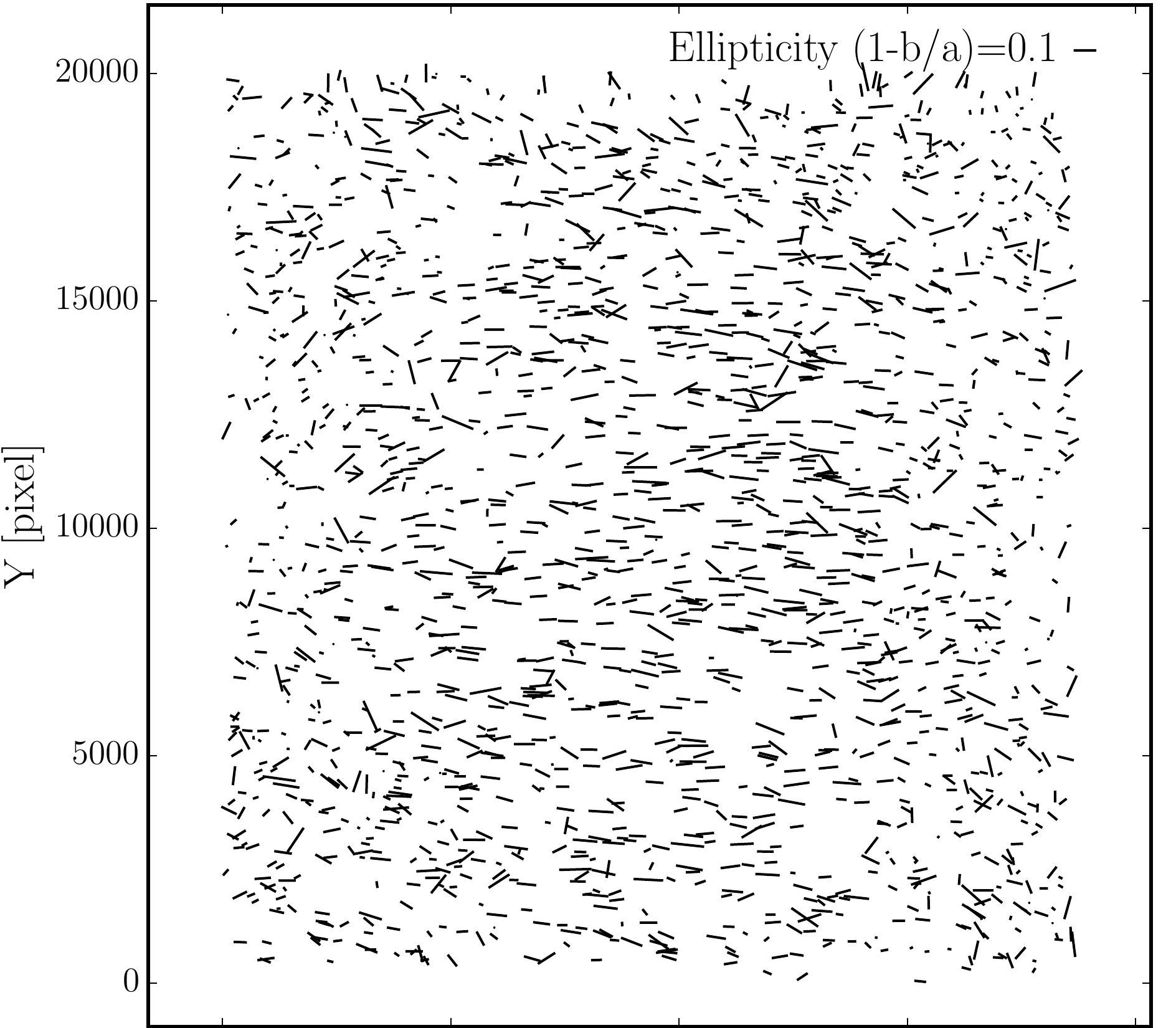}
\includegraphics[width=\hsize,trim=0.45in 0.4in 6in 0.45in,clip=true]{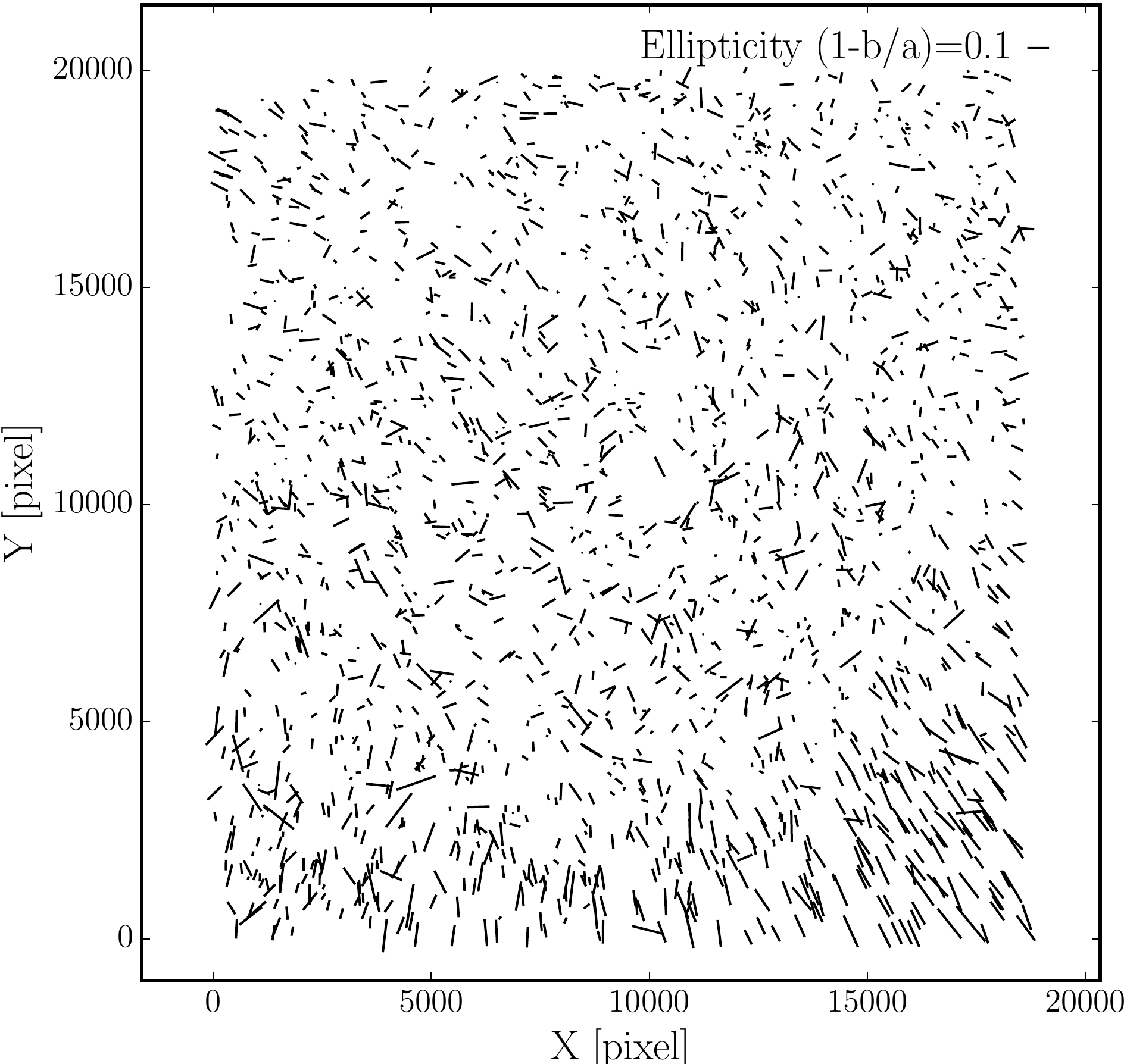}
 \caption{PSF ellipticity patterns caused by a non-optimal optical configuration of the telescope. The curved focal plane  of the VST translates any primary mirror astigmatism into increased ellipticity in the centre of the field (top). A tilt of the secondary mirror results in increased ellipticity near one edge of the field (bottom panel). 
 \label{fig:psf_ellipticity_patterns}
}
\end{figure}

PSF ellipticity is of critical importance for weak lensing studies.  Tile-by-tile statistics of the mean and standard deviation of the PSF ellipticities\footnote{Note that in this section PSF ellipticity is defined as $(1-q)$ where $q=b/a$ is the minor-to-major axis ratio of the star images; this differs from the lensing definition used later on in this paper.} are presented in Fig.~\ref{fig:psf_ellipticities}, and show a typical mean ellipticity of 0.055 and scatter 0.035. Ellipticities do sometimes vary significantly over the field of view, due to focus or alignment errors of the optical system. When such errors arise, the most common ellipticity patterns encountered are an increase in ellipticity either in the centre or towards the corners of the field, and an increase in ellipticity towards one edge. Examples of such PSF ellipticity patterns are illustrated in Fig.~\ref{fig:psf_ellipticity_patterns}. 

\begin{table}
\caption{Observing parameters for the KiDS survey. The longer $r$-band observations are made in the best seeing conditions and are used for galaxy shape measurements, while the remaining bands are used to measure photometric redshifts. Ranges cover $>95$ percent of the data.}
\begin{tabular}{cccclc}
\hline
Filter & Exposure   & Dithers & Seeing   & Limiting  & Moon  \\
       & time (sec) &         & (arcsec) & Magnitude & \\
\hline
$u$ & 960  & 4 & $0.95\pm0.2$ & $24.2\pm0.2$ &dark\\
$g$ & 900  & 5 & $0.8\pm0.2$ & $25.1\pm0.2$ &dark\\
$r$ & 1800 & 5 & $0.7\pm0.2$ & $24.9\pm0.25$&dark\\
$i$ & 1080 & 5 & $0.8\pm0.3$ & $23.7\pm0.7$ &bright\\
\hline
\end{tabular}
\label{tab:obs}
\end{table}


The KiDS data processing pipeline for lensing builds upon the pipeline developed for the CFHTLenS project \citepalias{heymans/etal:2012}.  CFHTLenS reanalysed data from the 154-square degree CFHTLS-Wide survey \citep[see for example][]{fu/etal:2008}, the largest deep cosmological lensing survey completed to date. It is based on new methods for measuring galaxy colours for photometric redshifts, and for obtaining ellipticities, the crucial ingredient for weak lensing. Our KiDS analysis uses further refinements of these techniques.

For historical and practical reasons, KiDS uses different data reduction pipelines for the lensing shape measurements and for the photometry. The latter is based on the 4-band co-added images that are released for general-purpose science through the ESO science archive, while the former uses a lensing-optimised processing pipeline of the $r$-band data only. Integration of both these pipelines and workflows into a single process is underway. Meanwhile, we have taken advantage of the redundancy to perform cross-checks between the different pipelines, for example on star-galaxy separation, masking and photometric calibration, where possible.

Weak lensing measurements are intrinsically noise-dominated; results therefore rely on ensemble averaging so that even small systematic residual shape errors can propagate into the final result and overwhelm the statistical power of the survey. For this reason our dedicated shape measurement pipeline (see \S\ref{sec:shapes}) avoids stacking sub-exposures and resampling of the image pixels. Instead it relies on combining the likelihoods of shape parameters from the different sub-exposures of each source. This part of the reduction was performed only on the $r$-band data, with image calibration and processing using the \textsc{Theli} pipeline \citep[henceforth \citetalias{erben/etal:2013}]{schirmer:2013, erben/etal:2013}, and object detection and classification, PSF modelling and shape measurements using the \emph{lens}fit code \citep[henceforth \citetalias{miller/etal:2013}]{miller/etal:2013}. Before distribution to the team for scientific analysis, the shape measurements were `sabotaged' through a blinding procedure described in \S\ref{sec:blinding}.

The multi-colour photometry was performed tile by tile on stacked images for each of the four bands.  This part of the reduction made use of the \textsc{Astro-WISE} environment \citep{begeman/etal:2013} and optical reduction pipeline \citep{mcfarland/etal:2013}. These multi-band images are released to the ESO archive as part of the second KiDS data release, as described in \citetalias{dejong/etal:2015}.  The lensing-quality reduction of the $r$-band imaging is made available on request.

\section{K\lowercase{i}DS galaxy shapes for lensing}
\label{sec:shapes}
As the lensing data processing of KiDS is built upon the pipeline developed for CFHTLenS, we refer the reader to the CFHTLenS technical papers \citepalias{heymans/etal:2012, miller/etal:2013, erben/etal:2013} for detailed descriptions of the \emph{lens}fit and \textsc{Theli} implementation.  In this section we highlight the differences and improvements implemented for this first KiDS lensing analysis.

\subsection
[Lensing-quality THELI r-band data reduction]
{Lensing-quality T{\sevensize HELI} \textit{r}-band data reduction}
\label{sec:theli}
Our reduction of OmegaCAM data starts from
raw data provided by the ESO archive. Most of the processing algorithms used are
similar to those initially developed for the wide-field imager on the ESO 2.2-m telescope at La Silla, as 
described in \citet{erben/etal:2005}.  A more in-depth description with tests on the \textsc{Theli} data 
products will be published in Erben et al.~(in preparation).

The \textsc{Theli} processing consists of the following steps:
\begin{enumerate}
  \item The basis for all \textsc{Theli} processing is formed by \textit{all} publicly
  	available OmegaCAM data at the time of processing. All data are retrieved
  	from the ESO archive\footnote{ESO data archive: \url{http://archive.eso.org}}.
  \item Science data are corrected for crosstalk effects. We measure significant
    crosstalk between CCDs \#94, \#95 and \#96\footnote{Note that the OmegaCAM CCD's have names ESO\_CCD\_\#65 to \#96, see \citetalias{dejong/etal:2015} for their layout in the focal plane.} \citepalias{dejong/etal:2015}. Each pair of these three CCDs show
    positive or negative crosstalk in both directions. We found that the strength
    of the flux transfer significantly varies on short time-scales and we therefore
    determine new crosstalk coefficients for each KiDS observing block (maximum duration ca.~1800s). 
  \item The characterisation and removal of the instrumental signature (bias, flat field, illumination correction) is performed
  	simultaneously on all data from a two-week period around each new-moon
  	and full-moon phase. Each two-week period of dark or bright time defines an
  	OmegaCAM processing run (see also section 4 of \citealt{erben/etal:2005}),
  	over which we assume that the instrument configuration is stable.
  	The processing run definition by moon phase
  	 also naturally corresponds to the observations with different filters
  	($u$, $g$ and $r$ in dark time and $i$ during bright time).
  \item Photometric zero-points, atmospheric extinction coefficients and colour terms are estimated per complete processing run.
    They are obtained by calibration of \textit{all} science observations in a run that
    overlap with the Data Release 10 of the SDSS \citep{ahn/etal:2014}. Between
    30 and 150 such images, with good airmass coverage, are available per each processing run.
  \item If necessary we correct OmegaCAM data for occasional electronic interference which
    produces coherent horizontal patterns over the whole field of view.
  \item As the last step of the run processing we subtract the sky from all
    individual chips. The resulting single-CCD sub-exposures, 160 per $r$-band tile, form the basis for the later shape analysis with \emph{lens}fit.
  \item All science images belonging to a given KiDS pointing are astrometrically
  	calibrated against the 2MASS catalogue \citep{skrutskie/etal:2006}. At present
  	we only use KiDS data belonging to each individual pointing for its astrometric
  	calibration. A more sophisticated procedure, taking into account overlaps from adjacent
  	pointing as well as data from the overlapping ATLAS survey \citep{shanks/etal:2015}, 
        will be included in the future and should constrain the astrometric solution further near the edges of each tile.
  \item The astrometrically calibrated data are co-added with a weighted mean algorithm.
    The identification of pixels that should not contribute, for example those affected by cosmic rays, and weighting
    of usable pixels is determined as described in \citetalias{erben/etal:2013}.
  \item Finally, \textsc{SExtractor} \citep{bertin/arnouts:1996} is run on the co-added image to generate the source catalogue for the lensing and matched-aperture photometry measurements.
\end{enumerate}

\begin{figure*}
\putfig{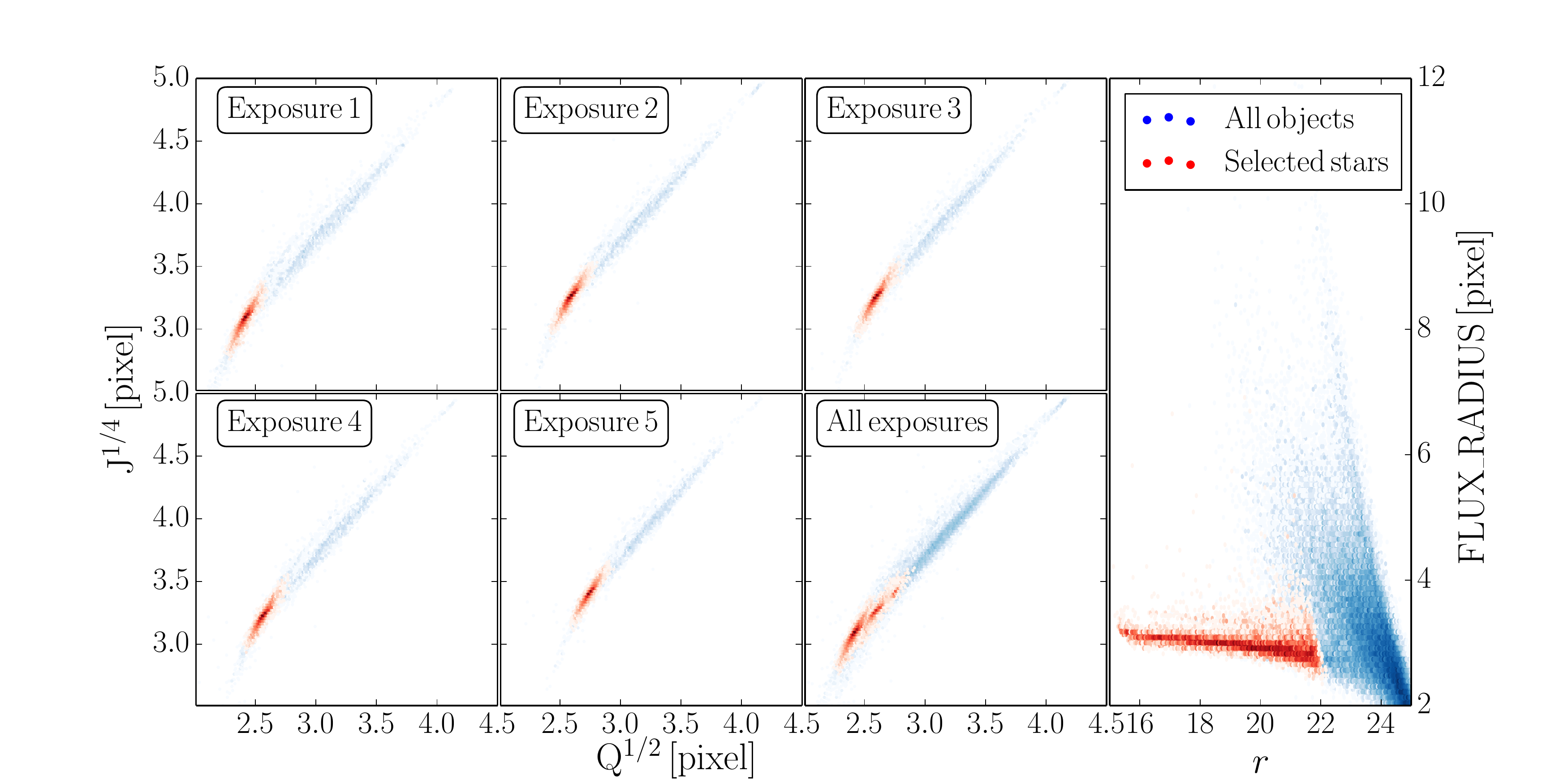}
\caption{Automatic star-galaxy separation based on the second and fourth order moment radii $Q^{1/2}$ and $J^{1/4}$ of individual sources, for a typical KiDS observation. 
Five out of the six square panels show the distributions for the individual sub-exposures, with the objects identified as stars shown in red.  As the seeing differs between the sub-exposures, the combined distribution for the observation, in the sixth square panel, reveals a series of distinct stellar peaks. The right-most panel shows the distribution of these points in the traditional radius-magnitude plane for the co-added image.}
\label{fig:stargal}
\end{figure*}

The final products of the \textsc{Theli} processing are, for each tile, the single-chip $r$-band data, 
the corresponding co-added image with associated weight map and sum 
image, and a source catalogue (see also \citetalias{erben/etal:2013} for a more detailed description of these products).  These images are made publicly available on request.

\subsection{Point spread function}

Knowledge of the point spread function (PSF) is essential for any weak lensing analysis, since the PSF modifies galaxy shapes. The thousands of stars recorded in every KiDS tile provide samples of the PSF across the field. The first steps are to identify these stars among the many galaxies in each image, and to build a PSF model from them.

\subsubsection{Star selection}

High-density, spatially homogeneous  and pure star catalogues are required to construct a good PSF model across the field of view. We outline in this section how we classify stars in order to meet these requirements.   We start by creating a source detection catalogue for each of the 5 sub-exposures in a KiDS field, using \textsc{SExtractor} with a high detection threshold. For each sub-exposure, and every detected object for which FLUX\_AUTO has a signal-to-noise ratio (SNR) larger than 15, we then measure the second-order moments $Q_{ij}$ and the axisymmetric fourth order moment $J$ given by
\be
Q_{ij} = \frac{\int \d^2 \x  \,W(\x)I(\x) \, x_i x_j}
{\int \d^2\x \, W(\x)I(\x)} \, ,
\label{eqn:quadmom}
\ee
\be
J=\frac{\int d^2 \x  \,W(\x)I(\x) \, |\x| ^4 }
{\int \d^2\x \, W(\x)I(\x)} \, .
\ee
In the above equations $I(\x)$ is the surface brightness of the object at position $\x$ measured from the  \textsc{SExtractor} position of the object, and $W(\x)$ is a Gaussian weighting function which we employ to suppress noise at large scales. The width of the weighting function is fixed and we choose it to have a dispersion of 3 pixels, motivated by the typical seeing value of our $r$-band data ($\sim0.7$\arcsec). 

Defining $Q = Q_{11} + Q_{22}$, we note that $Q^{1/2}$ and $J^{1/4}$ are two different measures of the size of an object, and the ratio between these two quantities depends on the concentration of the object's surface density profile.
These two parameters therefore efficiently classify sources according to their sizes and luminosity profiles.

Fig.~\ref{fig:stargal} shows the distribution of detected objects as a function of their second and fourth order moments for the different sub-exposures in an example tile. We see that galaxies are scattered over a wide range of $(Q,J)$ values whereas point sources cluster in a very compact region with low $Q^{1/2}$ and $J^{1/4}$.   The width of this region depends on how strongly the PSF varies across the field of view.

We identify stars in the $Q^{1/2}$--$J^{1/4}$ plane by locating the compact over-density with a `friends of friends' algorithm.  The fixed linking length was empirically determined from a sample of the data.  We require the final star catalogue to contain the largest possible number of objects while minimising contamination by galaxies, as assessed visually by inspecting the stellar-locus in the (half-light radius, magnitude) plane, shown in the right panel of Fig.~\ref{fig:stargal}.  In order to minimise the effect of the PSF variation across the field of view we perform this search in each individual CCD and sub-exposure separately.  This automated method is a significant improvement over the approach taken by CFHTLenS, where the stellar locus was visually identified for each chip using data from the co-added image, for every tile in the full survey.  
 
In a final cleaning stage, we combine the 5 star catalogues for each chip and we count how many times each object has been classified as a star. The final star catalogue requires that an object be classified as a star in at least 3 out of the 5 sub-exposures. In the cases where the object is not observed in all sub-exposures, for example when the object lands in a chip gap or at the edge of the field due to the dithering, we only require the star to be classified as such once.   In Appendix~\ref{app:phot} on quality control,  Fig.~\ref{fig:check_photometry} shows an example distribution of the selected stars across the field of view.  Plots such as these are inspected for each field to ensure that the stellar classification is producing a spatially homogeneous catalogue.   Confirmation of the purity of our star catalogue comes from the PSF modelling where typically less than 1 percent of the objects are rejected as outliers at that stage.

\subsubsection{PSF modelling}
\label{sec:psfres}
For each KiDS sub-exposure, we construct a PSF model that describes the position-dependent shapes of the identified stars.
The PSF model is expressed as a set of amplitudes on a $32 \times 32$ pixel grid, sampled at the CCD detector resolution and normalized so that their sum is unity.
The variation of each pixel value with position in the field takes the form of a two-dimensional polynomial of order $n$, with the added flexibility that the lowest-order coefficients are allowed to differ from CCD to CCD: this allows for a more complex spatial variation of the PSF and also, in principle, allows for discontinuities in the PSF between adjacent detectors.  If the polynomial coefficients up to order $n_\rmn{c}$ are allowed to vary in this way, then the total number of model coefficients per pixel is 
\be
N_\rmn{coeff} = \frac{1}{2} \left[(n+1)(n+2) + (N_\rmn{D}-1)(n_\rmn{c}+1)(n_\rmn{c}+2) \right] 
\ee
with $N_\rmn{D} = 32$, the number of CCD detectors in OmegaCAM.  The coefficients for each PSF pixel are fitted independently and a check is made that the total PSF normalisation is unity at the end of the fitting process. The flux and position of each star are also allowed to be free parameters in the fit, 
with the stars aligned to the pixel grid of the PSF model using a sinc function interpolation.
This approach allows a great deal of flexibility in the PSF model: in particular it does not imprint any additional basis set signature on top of the detector pixel basis. The total number of coefficients is large, but is well constrained by the large number of data measurements (number of pixels times number of stars) in each sub-exposure. Only stars with a high SNR should be used for constructing the PSF model, because otherwise noise on the measurement of the stellar positions will bias the model towards larger sizes.

In order to optimise the functional form of the PSF model, we selected 10 KiDS fields at random and analysed the five $r$-band sub-exposures in each field, varying the polynomial orders $n$ and  $n_\rmn{c}$.  We characterise the PSF ellipticity $\epsilon_{\rm PSF}$ and size $R^2_{\rm PSF}$ of the pixelised model and data as
\be
\epsilon_{\rm PSF} = \frac{Q_{11} - Q_{22} + 2\rmn{i}Q_{12}}{Q_{11} +Q_{22} + 2\sqrt{Q_{11}Q_{22} -Q_{12}^2}} \, ,
\label{eqn:estar}
\ee
\be
R^2_{\rm PSF} = \sqrt{Q_{11}Q_{22} -Q_{12}^2} \, 
\label{eqn:psfsize}
\ee
(cf.\ Eq.~\ref{eqn:quadmom}), with the weight function $W(\x)$ set to a Gaussian of dispersion two pixels.  

For an accurate PSF model the residuals $\delta \epsilon_{\rm PSF} = \epsilon_{\rm PSF}(\hbox{model}) -\epsilon_{\rm PSF}(\hbox{data}) $ and $\delta R^2_{\rm PSF} = R^2_{\rm PSF}(\hbox{model}) - R^2_{\rm PSF}(\hbox{data})$ should be dominated by photon noise, and therefore uncorrelated between neighbouring stars. Following \cite{rowe:2010} we therefore seek to miminise the PSF ellipticity residual auto-correlation, with as few parameters as necessary. This statistic can be estimated from the data as
\be
\langle \delta \epsilon_{\rm PSF}  \delta \epsilon_{\rm PSF}^* \rangle_{\theta} = 
\overline{ \Re\left[ \delta \epsilon_{\rm PSF} (\x_a) \delta \epsilon_{\rm PSF}^* (\x_b) \right]} \, ,
\label{eqn:xi_res}
\ee
where the average is taken over pairs of objects for which $|\x_a - \x_b|$ falls in a bin around angular separation $\theta$, and $\Re$ and $^*$ denote the real part and complex conjugate, respectively.  Analogously, we also measure the correlation function of the residual size $\delta R^2_{\rm PSF}$.

\begin{figure*}
\centering  
\begin{minipage}{.5\textwidth}
  \centering  
  \includegraphics[width=\linewidth]{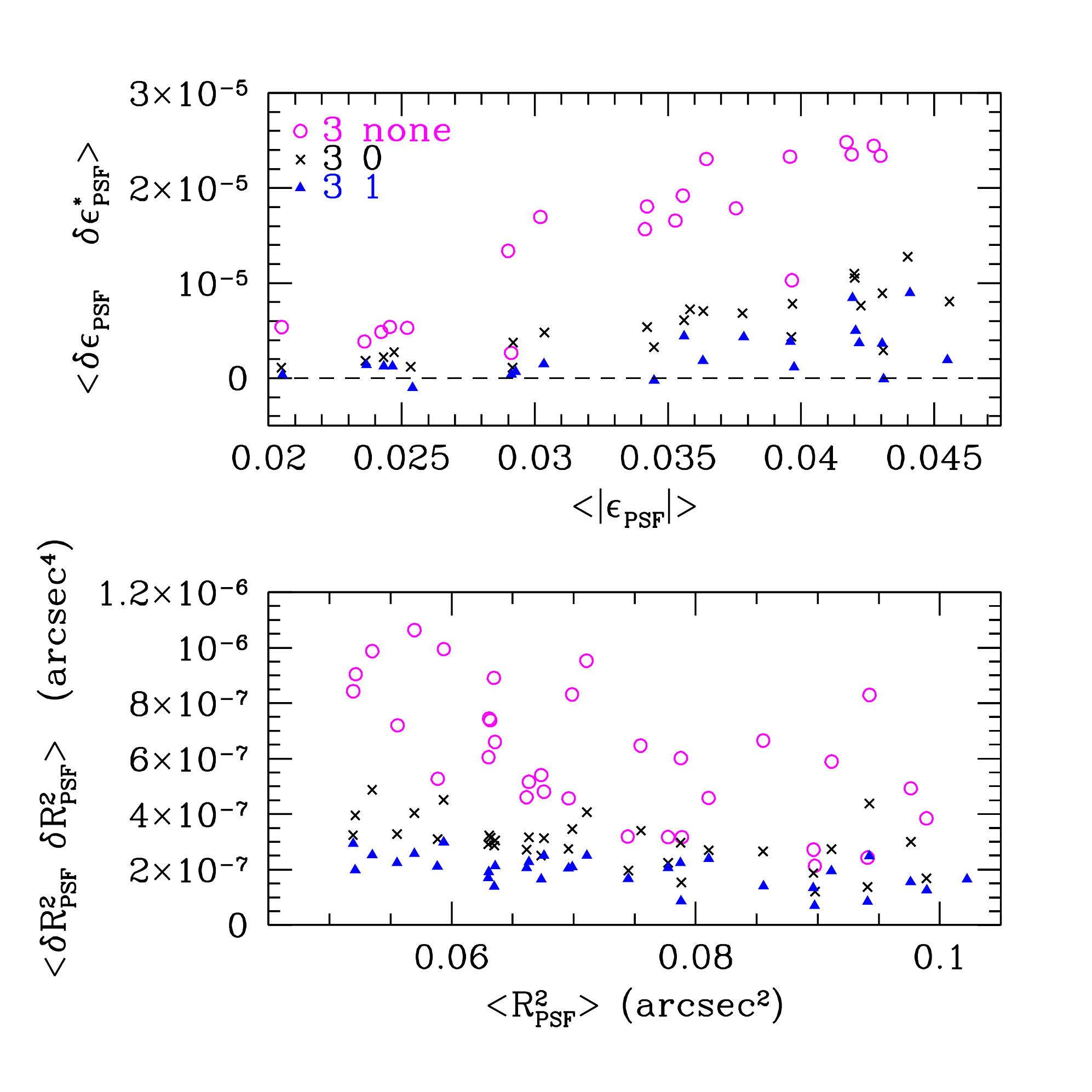}
\end{minipage}%
\begin{minipage}{.5\textwidth}
  \centering  
  \includegraphics[width=\linewidth]{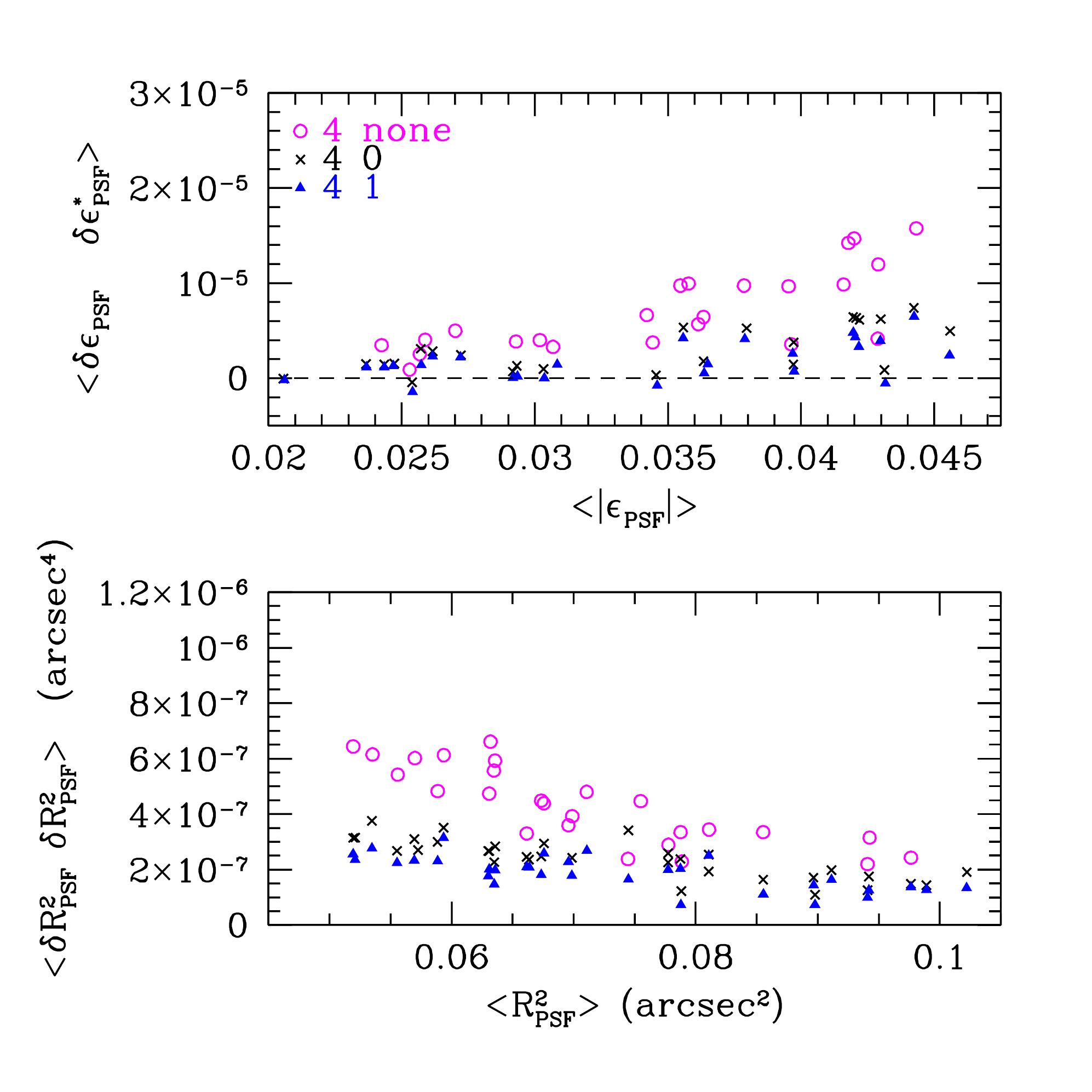}
\end{minipage}
 \caption{Selecting the optimal fitting orders for the PSF model for a sample of representative KiDS observations.  The upper panels show the residual PSF ellipticity correlation, measured at 1 arcmin separation, as a function of the average PSF ellipticity within the sub-exposure.  The lower panels show the two-point residual PSF size correlation measured at 1 arcmin separation as a function of the average PSF size $R^2_{\rm PSF}$.   Each data point represents a different sub-exposure, with the point style indicating the polynomial orders $(n, n_{\rm c})$ of the model. Left: results for $n= 3$; right: results for $n= 4$.}
\label{fig:orderchoice}
\end{figure*}

 Fig.~\ref{fig:orderchoice} shows the residual correlation functions measured at 1 arcmin separation.  We chose this scale as it is the smallest scale that can be reliably measured given the typical star density in the images.  The data come from our sample of KiDS sub-exposures for six different PSF models, with the full field of view polynomial order $n=3$ and 4, and chip-dependent polynomial order $n_\rmn{c} = 0$ and 1.  We also test models without any chip-dependent coefficients, denoted $n_{\rm c} = {\rm none}$ (these models have a total of $\frac12(n+1)(n+2)$ coefficients).  The lower panels of Fig.~\ref{fig:orderchoice} show the residual PSF size correlation as a function of the average PSF size $R^2_{\rm PSF}$.  We see a general trend, that the larger sized PSFs lead to more accurate modelling, which suggests that the impact of undersampling, when imaging the PSF, may be an important effect to model in the future.  The upper panels of Fig.~\ref{fig:orderchoice} show the residual PSF ellipticity correlation as a function of the average PSF ellipticity $| \epsilon_{\rm PSF}|$.  Unsurprisingly, more elliptical PSFs lead to less accurate modelling.  

Comparing the results from the different models, we find a reduction in the residuals with the inclusion of a chip-dependent component to the PSF modelling, favouring $n_\rmn{c} = 1$.  With that choice, we find little difference between the $n=3$ and $n=4$ model, 
selecting the $n= 3$  PSF model, as is has the lowest number of parameters for the two options.  With $n= 3$ and $n_\rmn{c}=1$ we fit $N_\rmn{coeff} = 103$ parameters per model PSF pixel. (With several thousand stars per tile, this large number of parameters can still be determined reliably from the data.) 

Analysing the full KiDS data set with this PSF model, we find residual correlation functions in the range
$\langle \delta R^2_{\rm PSF}  \delta R^2_{\rm PSF} \rangle_{\theta = 1'} = (3.5 \pm 1.3) \times 10^{-7} {\rm arcsec ^4}$, and $\langle \delta \epsilon_{\rm PSF}  \delta \epsilon_{\rm PSF}^* \rangle_{\theta = 1'} = (7.1 \pm 3.5) \times 10^{-6}$.  The size residual correlation remains fairly constant as a function of angular separation, whereas the amplitude of the ellipticity residual correlation decreases with increasing separation, becoming consistent with zero for scales $\theta> 20$\arcmin.   The angular dependence of the PSF ellipticity correlation function and the residuals are shown for an example KiDS field in Appendix~\ref{app:psfmod}.    
Even though we find persistent PSF residual correlations, they are too small to impact our scientific analyses of the data.  For example, \citet{rowe:2010} define a requirement on the systematic PSF ellipticity residual with correlation amplitude $\langle \delta \epsilon_{\rm PSF}  \delta \epsilon_{\rm PSF}^* \rangle_{\theta= 1'} < 5 \times 10^{-5}$, such that it contributes to less than 5 percent of the $\Lambda$CDM cosmic shear lensing signal for source galaxies at $z\sim0.5$.   At larger separations the requirement is more stringent with $\langle \delta \epsilon_{\rm PSF} \delta \epsilon_{\rm PSF}^* \rangle_{\theta= 10'} < 8 \times 10^{-6}$ but, as seen in Fig.~\ref{fig:check_PSF}, the KiDS residual correlation functions are already consistent with zero on these scales.  With the present analysis we therefore easily meet the \citet{rowe:2010} target requirement on PSF ellipticity residuals for the full KiDS data set.     

PSF modelling software development, currently undergoing testing for future data analysis, allows for the central region of the pixel basis PSF model to be oversampled by a factor 3.  Rather than re-centering each star's data to its best fit position, the fitting proceeds by shifting the model to the best-fit data position for each star. These developments improve the sampling of the core of the PSF and avoid the introduction of correlated noise caused by interpolation of the star data in the re-centering process. The disadvantage of this procedure is that the model pixel values become correlated, requiring a joint fit of a large number of parameters, which is computationally expensive.

\begin{figure}
\putfig{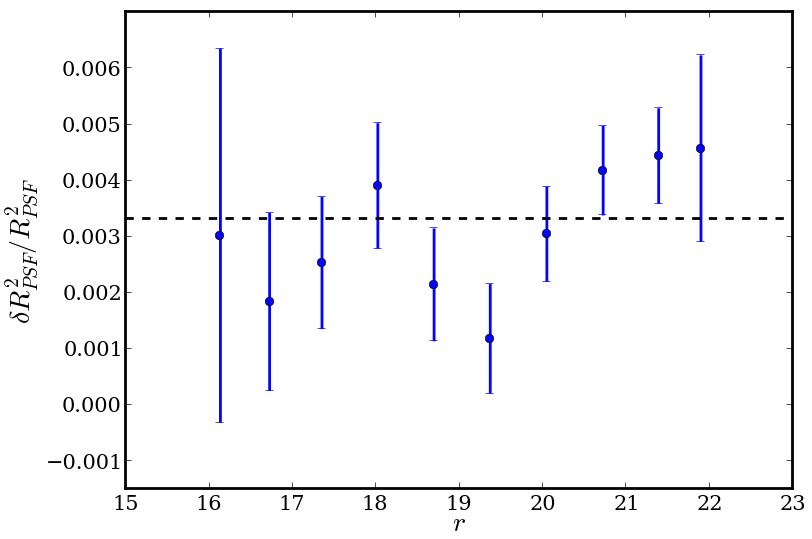}
\caption{The average residual PSF size as a function of the magnitude of the star, showing no significant flux-dependence in the PSF size. The average non-zero residual (shown as a dashed line) is too low to introduce any significant bias in our analysis.}
\label{fig:brightfat}
\end{figure}

\subsubsection{Testing PSF flux dependence}
\label{sec:brightfat}
\citet{melchior/etal:2015} report a significant flux dependence in the PSF size in Dark Energy Survey data. 
The effect is due to the use of modern deep-depletion CCDs in DECam \citep{antilogus/etal:2014}, and is not expected to affect the thinned OmegaCAM detectors used for KiDS. 
This is indeed the case. Fig.~\ref{fig:brightfat} shows the difference between the PSF model size and the star size, averaged over the full KiDS data set, as a function of the star's magnitude.  As the PSF model has no flux-dependence by definition, any detected flux dependence in the size offset between model and data would arise from CCD effects.   Only a very slight trend with star magnitude is seen, more than an order of magnitude smaller than the effect seen by \citet{melchior/etal:2015}.  The origin of the average non-zero residual of $(3.3 \pm 0.3) \times 10^{-3}$ is unclear: most likely it arises from the presence of noise in the size measurement of the data, in comparison to the measurement on the noise-free model, or from not including the effects of undersampling in the PSF modelling.   We conclude that PSF flux-dependence will not be a challenge for the KiDS analysis.

\subsection{Shape measurement with \textbfit{lens}fit}

Weak gravitational lensing induces a coherent distortion in the images of distant galaxies, which we parametrize through the observed complex galaxy ellipticity $\epsilon = \epsilon_1 + \rmn{i} \epsilon_2$.  For a galaxy that is a perfect ellipse, the ellipticity parameters are related to the axial ratio $q$ and orientation $\phi$ as 
\be
\epsilon=
\epsilon_1 + \rmn{i}\epsilon_2 = \left(\frac{1-q}{1+q}\right)\rmn{e}^{2\rmn{i}\phi}\, .
\label{eqn:e1e2}
\ee
Central to any weak lensing study is a data analysis tool that can determine galaxy shapes from imaging data.  We use the \emph{lens}fit code\footnote{See \citetalias{heymans/etal:2012} for a discussion on why \emph{lens}fit is our preferred shape measurement method.} (\citealt{miller/etal:2007,kitching/etal:2008}; \citetalias{miller/etal:2013}) which performs a seven-parameter galaxy model fit ($x,y$ position, flux, scale length $r_\d$, bulge-to-disc ratio and ellipticity $\epsilon_{1,2}$),  simultaneously to all sub-exposures of a given galaxy, taking into account the different PSFs in each sub-exposure and the astrometric solution for each CCD.

\emph{Lens}fit first performs an analytic marginalization over the galaxy model's centroid, flux and bulge fraction, using the priors from \citetalias{miller/etal:2013}. It then numerically marginalizes the resulting joint likelihood distribution $L(\epsilon,r_\d)$ over scale length, incorporating a magnitude-dependent prior derived from high-resolution Hubble Space Telescope (HST) imaging. Finally, for each galaxy a mean likelihood estimate of the ellipticity and an estimated inverse variance weight is derived, as described by \citetalias{miller/etal:2013}. We will refer to this latter quantity as the `lensing weight'.

The KiDS lensing data are obtained in the $r$ band.  We therefore change the \emph{lens}fit scale-length prior with respect to the $i$-band based prior used in the CFHTLenS analysis. For this purpose we repeat the \citetalias{miller/etal:2013} analysis of the \citet{Simard/etal:2002} catalogue of morphological parameters. This catalogue is based on \textsc{GALFIT} galaxy profile fitting \citep{Peng/etal:2010} of HST imaging data, and provides disc and bulge parameters in various wavebands including the F606W filter which is a good match to the KiDS {\it r} band.  Selecting galaxies with $18.5<r_{606}<25.5$, we find the following relation between the median disc scale length and magnitude:
\be 
\ln(r_\d/{\rm arcsec})= -1.320 - 0.278 (r_{606}-23) \, .
\ee
We note that the more extensive HST galaxy morphology analysis by \citet{Griffith/etal:2012} satisfies our requirements in terms of imaging depth and filter choice. However, it is limited to single S\'ersic profile fits which prevents the selection of disc-dominated galaxies with which to determine a scale-length prior for the disc component.  

As discussed in \citetalias{miller/etal:2013}, the measurements do not strongly constrain the shape of the prior of $r_\d$ and we therefore adopt the same functional form \citepalias[appendix B1]{miller/etal:2013}.  For the bulge scale-length prior,
the small numbers of bulge dominated galaxies in the \citet{Simard/etal:2002} catalogue prevent a robust determination.  We continue to fix the half-light radius of the bulge component to be the exponential scale length of the disc component, as motivated in appendix A of  \citetalias{miller/etal:2013}. 

The change of the galaxy size prior is the only significant change in \emph{lens}fit as compared to the CFHTLenS analysis. While appropriate, its effect on the results is small: Hildebrandt et al. (in preparation) present an analysis of the RCSLenS survey where similar changes in the scale-length prior are shown not to impact the measured shear amplitudes by more than a few percent.

\begin{figure}
 \putfig{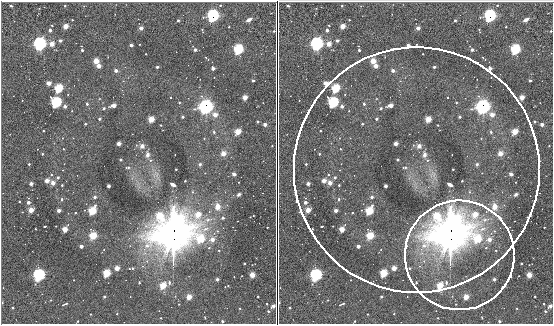}
 \caption{ Left: Small and large stellar haloes, due to reflection from different pairs of surfaces within the VST. 
 Right: Same image, displaying the two types of halo masks demarcated by small and large circles. 
 The reflection halo centroids are offset from the star; relative to the centre of the 
 field-of-view, the small (large) halo centroids lie inwards (outwards). 
 The bright star in these images has an $r$ magnitude of $\sim$10. The large circle has a radius of 210\arcsec. 
 \label{fig:stellar_halo_image}
}
\end{figure}
\subsection{Masking of the KiDS images}
The masking of the $r$-band \textsc{Theli} reduction uses the \textsc{automask}
tool\footnote{\url{http://marvinweb.astro.uni-bonn.de/data_products/THELIWWW/automask.html}}
to generate automated masks, which come in three types.
`Void masks' indicate regions of high spurious object detection and/or a strong density
gradient in the object density distribution \citep[see][]{dietrich/etal:2007}. `Stellar masks'
are generated based on standard stellar catalogs GSC-1
\citep[complete at the bright end,][]{lasker/etal:1996}
and UCAC4 \citep[complete from $r\simeq10$ to $\simeq16$]{zacharias/etal:2012}.
The stellar catalogs are used to mask the brighter stars as well as associated small and large
reflection haloes, using mask radii and centroid offsets that were derived empirically for OmegaCAM
as illustrated in Fig.~\ref{fig:stellar_halo_image}.
Finally, the `asteroid masks' flag asteroids and satellite trails.
The \textsc{automask} algorithms and procedures are described in more detail in \citet{erben/etal:2009}.

\begin{figure}
\includegraphics[width=\hsize,trim=15 0 10 0]{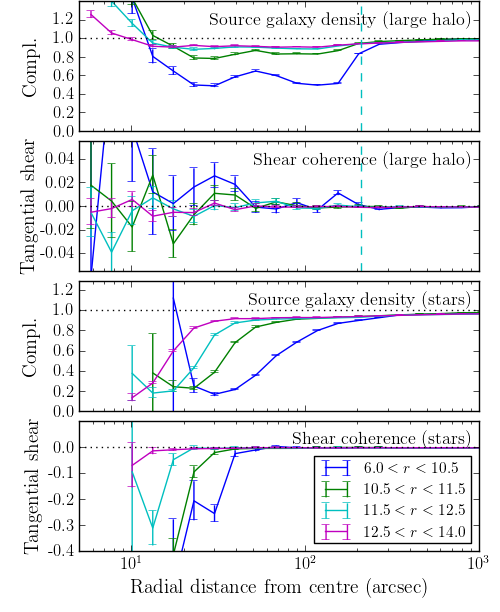}
 \caption{The impact of bright stars on source galaxy counts and galaxy shapes. 
 The upper two panels show the number count completeness and tangential shear measured within the large reflection haloes
 as a function of the radial distance from the centre; the dashed vertical line indicates the 210\arcsec\ radius that is used
 to mask these haloes.  For the very brightest haloes, a coherent tangential alignment of  $\sim$1--2 percent
 can be seen at the edges of the large reflection halo, and on small scales.
 The lower two panels show these same quantities as function of distance from the centre of bright stars.
 We see two effects: a decrease in the source galaxy counts, and a strong, coherent radial
 shape alignment immediately around the star, which can be removed from the sample by applying stellar masks. 
 \label{fig:stellar_halo_signal}
 }
\end{figure}

Fig.~\ref{fig:stellar_halo_signal} shows the effect of bright stars, grouped by magnitude, on the neighbouring
`source galaxies',  defined here as objects with valid shape measurements. 
The upper panel shows the relative source number density within the large reflection haloes as a function of the radial distance from the centre.  The annular halo clearly results in a source count incompleteness out to $\sim 200$\arcsec\ from the halo centre, the severity of which increases with stellar magnitude.
The detection incompleteness is essentially identical whether the source objects are unweighted, or weighted
using the \emph{lens}fit weights, implying that the source density
count deficiency originates in the object detection stage. 

The second panel of Fig.~\ref{fig:stellar_halo_signal}  shows the tangential shear measured by \emph{lens}fit for objects detected within the large reflection haloes, as a function of the distance from its centre.  In general this signal is found to be consistent with zero, on all scales, indicating that the local sky background subtraction performed by \emph{lens}fit removes any bias introduced by the haloes.  The cross shear signal, not shown, is also consistent with zero.  For the brightest stellar sources with $r<10.5$, however, there is a $\sim$2$\sigma$ coherent tangential ellipticity detected at the halo edges at $\sim 170$\arcsec, and on small scales $<50$\arcsec. For this reason we mask and remove the areas with reflection haloes from the scientific analyses.  A similar analysis was also performed within the other, smaller halo seen in Fig.~\ref{fig:stellar_halo_image},
showing identical trends in source count incompleteness and shape coherence.

Based on this analysis, we define two reflection halo masks: a `conservative' mask, with a magnitude limit at $r=11.5$, to indicate the regions of source density incompleteness, and a `nominal' mask that flags regions where there are signs of a coherent shear ($r<10.5$).  The stellar halo masks are based on both the GSC-1 and UCAC4 catalog.

The lower panels of Fig.~\ref{fig:stellar_halo_signal} investigate source incompleteness and radial alignment
of source galaxies around the centre of the bright stars themselves, where no sources are detected within 10\arcsec of the star, as these pixels are typically saturated.   Again we see that the incompleteness and shape
coherence depends strongly on the stellar magnitude and the radial dependence of this effect determines the area masked around each star. All stars in the UCAC4 catalog with $r<14.0$ are masked, with masking radius (in \arcsec) determined from
the stellar magnitude as $R_{\rm mask} = 2.96 r^2 - 81.2r + 569$.  Taking an example $r=11$ magnitude star, $R_{\rm mask} =34$\arcsec, thereby masking the full area within which a significant coherent negative tangential shear is measured.

The automatically generated masks were visually inspected, and additional manual masking was
performed if necessary.  A number of the early observations are affected by 
stray light from bright objects outside the field-of-view as a result of poor baffling of the telescope (see \citetalias{dejong/etal:2015} for some examples). Additionally, a number of missed asteroid and
satellite tracks were masked manually in the co-added image. Manual masking is also used to cover areas
of non-uniformity which the void automask had missed, or for additional stellar halo masks in cases
where the bright stellar catalogues are incomplete.  The manual masking is then inspected by a
single person to check for uniformity.

In total, the automated masks, using the conservative halo reflection scheme, along with the manual masks from the lensing pipeline remove 32 percent of the imaged area. With recent improvements at the VST to reduce scattered light, we anticipate the masked area fraction to reduce in future analyses. For this first analysis of 109 square degrees of KiDS data that overlap with GAMA, the total unmasked area is $A=75.1$ square degrees.

\subsection{Effective number density of lensed galaxies}
\label{sec:shapecat}

In its current implementation, \emph{lens}fit is quite conservative when it comes to rejecting galaxies whose isophotes might be affected by neighbours. The final \emph{lens}fit shape catalogue contains a total of 2.2 million sources with non-zero lensing weight, with an average number density of 8.88 galaxies per square arcmin over the unmasked area $A$ of 75.1 square degrees.  While this raw number density provides information about the number of resolved, relatively isolated galaxies, it does not represent the true statistical power of the survey.  When weights are employed in the analysis to account for the increased uncertainty in the galaxy shape measurements of smaller or fainter objects, the effective number density is reduced.

\citet{chang/etal:2013} propose an effective number density defined as
\be
n_\rmn{eff}=\frac{1}{A} \sum_i \frac{\sigma_\rmn{SN}^2}{\sigma_\rmn{SN}^2 + \sigma_{m,i}^2}\, ,
\ee
where $\sigma_\rmn{SN}$ is the intrinsic ellipticity dispersion (`shape noise') and $\sigma_{m,i}$ is the measurement error for galaxy $i$.  With this definition, $ n_\rmn{eff}$ represents the equivalent number density of high SNR, intrinsic shape-noise dominated sources with ellipticity dispersion $\sigma_\rmn{SN}$, that would yield a shear measurement of the same accuracy. 

As the \emph{lens}fit weights are designed to be an inverse variance weight, $ w_i^{-1} \sim \sigma_\rmn{SN}^2 + \sigma_{m,i}^2$, with the intrinsic ellipticity dispersion fixed to a value $\sigma_\rmn{SN}=0.255$, we can estimate $n_\rmn{eff}$ as
\be
n_\rmn{eff} \approx \sigma_\rmn{SN}^2 \frac{\sum_i w_i}{A} = 4.48\, \hbox{arcmin}^{-2} \, .
\ee
The inverse shear variance per unit area, $\hat{w}$, that the survey provides is thus equal to
\be
\hat{w} = \frac{\sum_i w_i}{A} = 69 \, \hbox{arcmin}^{-2} 
\ee
which corresponds to a 1-$\sigma$ shear uncertainty of $(\hat{w}A)^{-1/2}=0.12/\sqrt{N}$ when averaging $N$ square arc\-minute of survey.
While this definition is useful for forecasting, it makes a number of assumptions; that the shape noise and measurement noise are uncorrelated, that the estimated inverse variance weight is exact, that the intrinsic ellipticity dispersion does not evolve with redshift and that it can be accurately measured from high-SNR imaging of low redshift galaxies. 

\citetalias{heymans/etal:2012} propose an alternative definition of an effective number density defined as
\be
n_\rmn{eff}^*= \frac{1}{A}  \frac{(\sum_i w_i)^2}{\sum_i w_i^2} = 5.98 \hbox{arcmin}^{-2}  \, .
\ee
With this definition, $ n_\rmn{eff}^*$ represents the equivalent number density of sources with unit weight and a total ellipticity dispersion per component of $\sigma_\rmn{\epsilon}$, that would yield a shear measurement of the same accuracy where
\be
\sigma_\epsilon^2 = \frac{1}{2}\frac{\sum_i w_i^2 \epsilon_i \epsilon_i^*}{\sum_i w_i^2} \, .
\ee
For KiDS we measure $\sigma_\epsilon = 0.278$ per ellipticity component, which is very similar to the ellipticity dispersion measured in CFHTLenS.  This definition is useful as it makes no assumptions about how the weight is defined.   As the shot noise component for cosmic shear measurement scales with $\sigma^2/n_\rmn{eff}$, the difference between these two definitions for KiDS would change the expected shot noise error on a cosmic shear survey by $\sim 10$ percent.  

\section{K\lowercase{i}DS Photometry and Photometric Redshifts}
\label{sec:photom}

Without good redshift estimates any weak lensing data set is of limited use, as redshifts are required to determine the critical surface density that sets the physical scale for all lensing-based mass measurements. For the moment, KiDS photometric redshifts are derived from $ugri$ imaging, and are adequate for the first lensing science analyses from the survey (\citealt{viola/etal:2015}; \citealt{sifon/etal:2015}; van Uitert et al., in preparation). Combination with the VIKING near-IR flux measurements will be used to refine the redshifts further in future.

The colours of the galaxies are obtained with `Gaussian Aperture and PSF' (\textsc{GAaP}) photometry, a novel technique that is designed to account for PSF differences between observations in different filter bands while optimizing SNR. The procedure is summarized in \S\ref{sec:gaap} below, and described in detail in Appendix~\ref{app:gaap}.

We base our photometric redshifts on the \textsc{bpz} code of \citet{benitez:2000}. Further details are given in \S\ref{sec:pz} below. Alternative photometric redshift techniques based on machine-learning are also being investigated \citep{cavuoti/etal:2015}, but have not been integrated into the lensing analysis at this point.

\subsection{Data reduction}
\label{sec:AW_data_red}
The KiDS photometric redshifts are based on the co-added images provided in the public data releases. The processing from raw pixel data to these calibrated image stacks is performed with a version of the \textsc{Astro-WISE} pipeline \citep{mcfarland/etal:2013} tuned for KiDS data. We refer the reader to \citetalias{dejong/etal:2015} for a detailed description of all the steps.

There are some small differences between the \textsc{Theli} reduction of the $r$-band data described in \S\ref{sec:theli}, and the four-band \textsc{Astro-WISE} processing. The latter uses a single flat field per filter for the entire data set, since a dome-flat analysis shows that the peak-to-valley variations of the pixel sensitivity were less than 0.5 percent over the period during which the data were taken. Also, the $i$-band data require a de-fringing step, and different recipes are used to create the illumination correction maps (which are applied in pixel space), and the pixel masks that flag cosmic rays and hot/cold pixels. Satellite track removal is automatic (currently implemented on a per-CCD basis). Finally, background structure from shadows cast by scattered light hitting the shields that cover the CCD bond wires are subtracted separately in a line-by-line background removal procedure. All images are visually inspected and masked if necessary before release.


Photometric calibration starts with zero points derived per CCD from nightly standard field observations, tied to SDSS DR8 PSF magnitudes of stars \citep{aihara/etal:2011}. The calibration uses a fixed aperture (6.3\arcsec\ diameter) not corrected for flux losses. Magnitudes are expressed in AB in the instrumental system. For $g$, $r$ and $i$ the photometry is homogenized across all CCDs and dithers for each survey tile individually. In $u$-band the smaller source density often provides insufficient information for this scheme. The resulting photometry is homogeneous within two percent per tile and filter. Due to the rather fragmented distribution of observed tiles in the first two data releases, no global photometric calibration over the whole survey is feasible yet, resulting in random offsets in the absolute zero points of the individual tiles thus obtained. For the GAMA tiles, which overlap with SDSS, we correct these offsets after the fact. Detailed analysis and statistics of the photometric calibration are presented in \citetalias{dejong/etal:2015}.

A global astrometric calibration combining all CCDs and dithers is calculated per filter for each tile using a second order polynomial. 
The de-trended sub-exposures are then re-gridded to a 0.2\arcsec\ pixel scale, photometrically scaled, and co-added to produce the image stacks.

\subsection[Gaussian aperture and PSF photometry (GAaP)]
{Gaussian aperture and PSF photometry (GA{\sevensize A}P)}
\label{sec:gaap}

Photometric redshifts of galaxies require accurate colour measurements. These colours do not need to describe the total light from the galaxy, but they should represent the ratio of the fluxes from the same part of the galaxy in different filter bands. This means that we can optimize SNR by measuring the colours of the brighter, central regions of galaxies without the need to include the noise-dominated low surface brightness outskirts.

Such aperture photometry is complicated by the fact that the PSF is not constant: it varies from sub-exposure to sub-exposure, with position in each image, and with wavelength. 
We correct for PSF variations in two steps. First, we homogenize the PSF within each co-added image to a Gaussian shape without significantly degrading the seeing. The resulting images contain most of the information that is present in the original stacks, with a simpler PSF but correlated noise between neighbouring pixels. Second, we perform aperture photometry using elliptical Gaussian aperture weight functions, and correct analytically for the seeing differences. 


In brief, the PSF Gaussianization of each KiDS tile consists of the following steps:
\begin{enumerate}
\item
We model high-SNR stars in the co-added image with a shapelet expansion \citep{refregier:2003}, using the pixel-fitting method described in \citet{kuijken:2006}. This formalism provides a natural and mathematically convenient framework for PSF modelling and image convolutions. The scale radius (i.e., size of the parent Gaussian in the shapelet expansion) of the shapelets is matched to the worst seeing found in the individual sub-exposures making up the co-added image for each filter. 
\item
We then derive a PSF map by fitting the variation of the shapelet coefficients across the image, using polynomials.
\item
We construct a grid of kernels that yield a Gaussian when convolved with the model PSF, also expressed in the shapelets formalism. The size of the `target' Gaussian is set by the shapelet scale chosen in step (i). We fit the spatial variation of these kernels' coefficients using polynomials, resulting in a kernel map.
\item
Each co-added image is convolved with its kernel map. 
\item
The shapes of the PSF stars on this PSF-Gaussianized image are modelled once again with a shapelet expansion, but now using a larger scale radius in order to measure residual flux at large radii. A map of the residual PSF non-Gaussianities is then made as above, and used to make a perturbative correction to the Gaussianized image to improve the PSF Gaussianity further.
\item
As a result of the convolution (and to a lesser extent, also from the preceding re-gridding before co-addition) the noise in these images is correlated on small scales. We keep track of the noise covariance matrix during the Gaussianization, and account for it in the photometric measurements.
\end{enumerate}

The \textsc{GAaP} photometry is performed from these PSF-Gaussianized, co-added images for all sources in the $r$-band \textsc{Theli}-\emph{lens}fit catalogue. First we pick an elliptical Gaussian aperture for each source, with aperture size, shape and orientation chosen to optimize the SNR of the fluxes, based on the pre-Gaussianization $r$-band image. For major and minor axis lengths $a$ and $b$, and orientation $\alpha$ with respect to the pixel coordinate grid, we construct an `aperture matrix' 
\be
\W=
\left(\!\!
\begin{array}{cc}
a^2\cos^2\alpha+b^2\sin^2\alpha & (a^2-b^2)\sin\alpha\cos\alpha\\
(a^2-b^2)\sin\alpha\cos\alpha&a^2\sin^2\alpha+b^2\cos^2\alpha
\end{array} 
\!\!\right),
\ee
which in turn is used to define the \textsc{GAaP} flux $F_\mat{W}$ as the Gaussian-weighted aperture flux of the \emph{pre-seeing} image of the source, $I_\rmn{pre}(\x)$: 
\be
F_\W 
\equiv
\int\d\x\ I_\rmn{pre}(\x){\rmn e}^{-\frac12\x^\mat{T} \W^{-1}\x} \, .
\label{eq:FW}
\ee
$F_\mat{W}$ is well-defined and manifestly PSF-independent, but since it is defined in terms of the pre-seeing image it is a theoretical construct. However, it is possible to measure this quantity from a Gaussian-smoothed image $I_\rmn{G}= I_\rmn{pre} \otimes G$ (where $G$ is a Gaussian PSF of dispersion $p$ and $\otimes$ denotes convolution) using the identity
\be
F_\mat{W}=
\frac{\det(\W)^\frac12}{\det(\W-p^2\mat{1})^\frac12}
\int\d\x\ I_\rmn{G}(\x){\rmn e}^{-\frac12\x^\mat{T} (\W-p^2\mat{1})^{-1}\x}
\ee
which is valid for any PSF size $p<a,b$ (i.e., as long as the aperture is larger than the PSF). 
$\mat{1}$ denotes the identity matrix.
For a given source, provided the same aperture matrix $\mat{W}$ is used for all bands, Eq.~\ref{eq:FW} shows that this technique returns fluxes that weight different parts of the source consistently. 

A detailed description of the PSF Gaussianization pipeline, propagation of the noise correlation due to the convolution, and a discussion and derivation of the \textsc{GAaP} flux formalism, may be found in Appendix~\ref{app:gaap}. We stress that these aperture magnitudes are not designed to be total magnitudes.


\subsection{Photometric calibration}
\label{sec:photcal}

As described above, the photometric zero points of the co-added images used for the current analysis are calibrated based on nightly standard star field observations, and no global photometric calibration is included. To improve these absolute zero points, a cross-calibration to SDSS is done before the derivation of the photometric redshifts.

We calibrate against the eighth data release of the SDSS \citep{aihara/etal:2011}, which represents the complete SDSS imaging and fully overlaps with the KiDS-GAMA fields. Stars are selected from SDSS and matched to the KiDS multi-colour catalogues. We choose a magnitude range where the OmegaCAM sub-exposures are not saturated and SDSS photometry is sufficiently precise. Over this range we average the differences in the photometry between our \textsc{GAaP} measurements and the SDSS PSF magnitudes in all four bands ($ugri$).  We find no trend with magnitude, confirming that the difference is a pure zero point offset.

The distribution of the differences for all 114 fields is similar to the one shown in \citetalias{dejong/etal:2015}. We find the mean offset to be consistent with zero in the $g$-band and offsets of $\sim$0.02mag, $\sim$0.05mag, and $\sim$0.06mag in the $r$-, $i$, and $u$-bands, respectively. Field-to-field scatter is in the range 2.5--5 percent. The offsets are applied to each field globally relying on the photometric stability of SDSS and the KiDS illumination correction. All subsequent analysis is based on these re-calibrated magnitudes.

\subsection{Photometric redshifts}
\label{sec:pz}

The KiDS photometric redshift estimates are obtained following the methods used for CFHTLenS \citep{hildebrandt/etal:2012}. We use the Bayesian photometric redshift code \textsc{bpz} \citep{benitez:2000}, a spectral template-fitting code, together with the re-calibrated template set by \cite{Capak:2004}.

To assess the accuracy of our photometric redshifts, we also produce stacks from VST data in two fields with deep spectroscopic coverage, the Chandra Deep Field South (CDFS) and the COSMOS field. These data were taken under the VOICE \citep{decicco/etal:2015} project. Total exposure times in these fields are much longer than for typical KiDS observations, but individual sub-exposures are similar to those from KiDS, allowing us to produce stacks with similar depth and seeing as a typical KiDS field. We extract catalogues and photometric redshifts in the same way as for the KiDS tiles, and then match the resulting photometric catalogues with the combined CDFS spectroscopic  catalogue\footnote{\url{http://www.eso.org/sci/activities/garching/projects/goods/MasterSpectroscopy.html}} and a deep zCOSMOS catalogue (zCOSMOS team, private communication). In the following we compare the KiDS photometric redshifts  to the high-confidence spectroscopic redshifts from these catalogues. 

\begin{figure}
\putfig{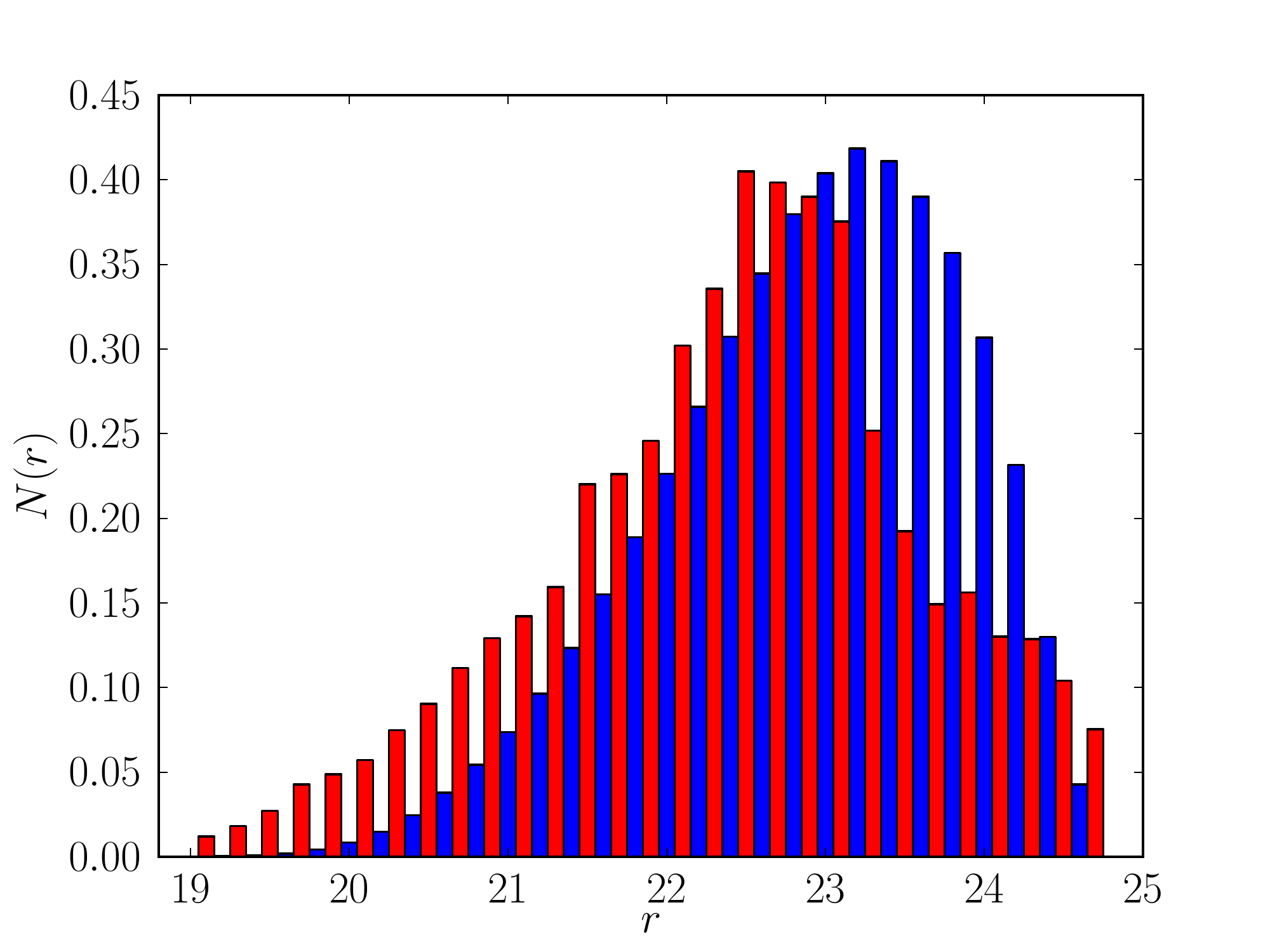}\\
 \caption{Number counts in the $r$ band of the lensing catalogue (blue, weighted by \emph{lens}fit weight) and the spectroscopic catalogue (red, unweighted).}
\label{fig:photo-z_numbercounts}
\end{figure}
 
Fig.~\ref{fig:photo-z_numbercounts} shows the $r$-band magnitude number counts of the lensing catalogue (weighted by the \emph{lens}fit weight, see Sect.~\ref{sec:shapes}) and the spectroscopic matches (unweighted). This deep spectroscopic sample spans the full magnitude range of the lensing sample, with broadly similar distribution, and therefore we do not apply any further weighting. This is also the reason why we concentrate on the zCOSMOS and CDFS fields here. Adding in the numerous bright spectroscopic redshifts from SDSS and GAMA would not add significant information about the performance of the photometric redshifts of the faint KiDS sources.

\begin{figure}
\putfig{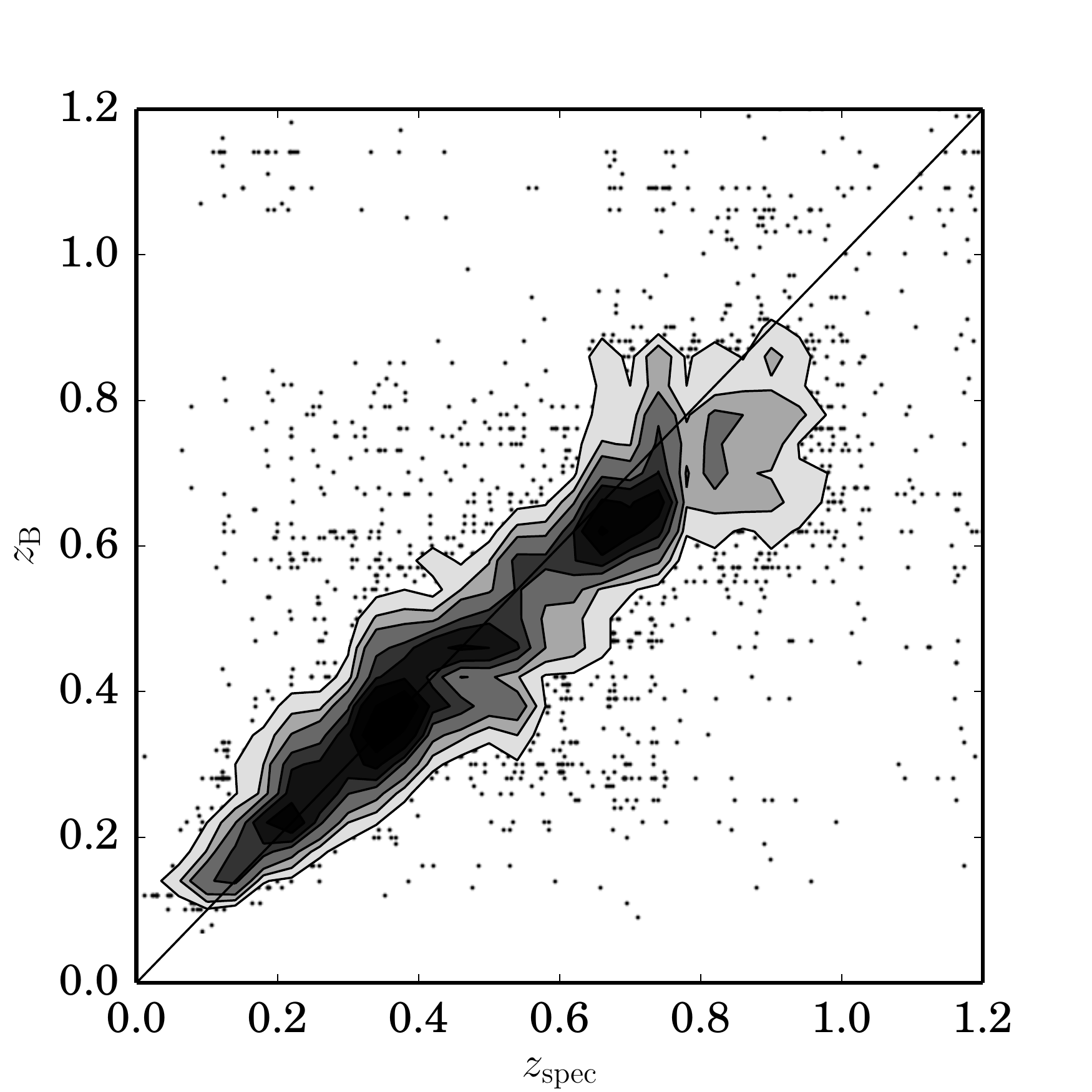}\\
\caption{Photometric redshift vs. spectroscopic redshift in the CDFS and COSMOS fields for objects with $19<r<24$. Contours are spaced in 0.5-$\sigma$ intervals with the outermost contour corresponding to the 2-$\sigma$ level. Photo-$z$ are estimated from four-band $ugri$ data from the VOICE project in the two fields, stacked so as to approximate the KiDS depth and seeing. Spec-$z$ are from the combined ESO CDFS catalogue and a deep zCOSMOS catalogue. For this sample we find a photo-$z$ scatter of 0.054 after rejecting 11 percent of the galaxies as outliers. The photo-$z$ bias for this sample is 0.01.}
\label{fig:zz}
\end{figure}

\begin{figure}
\putfig{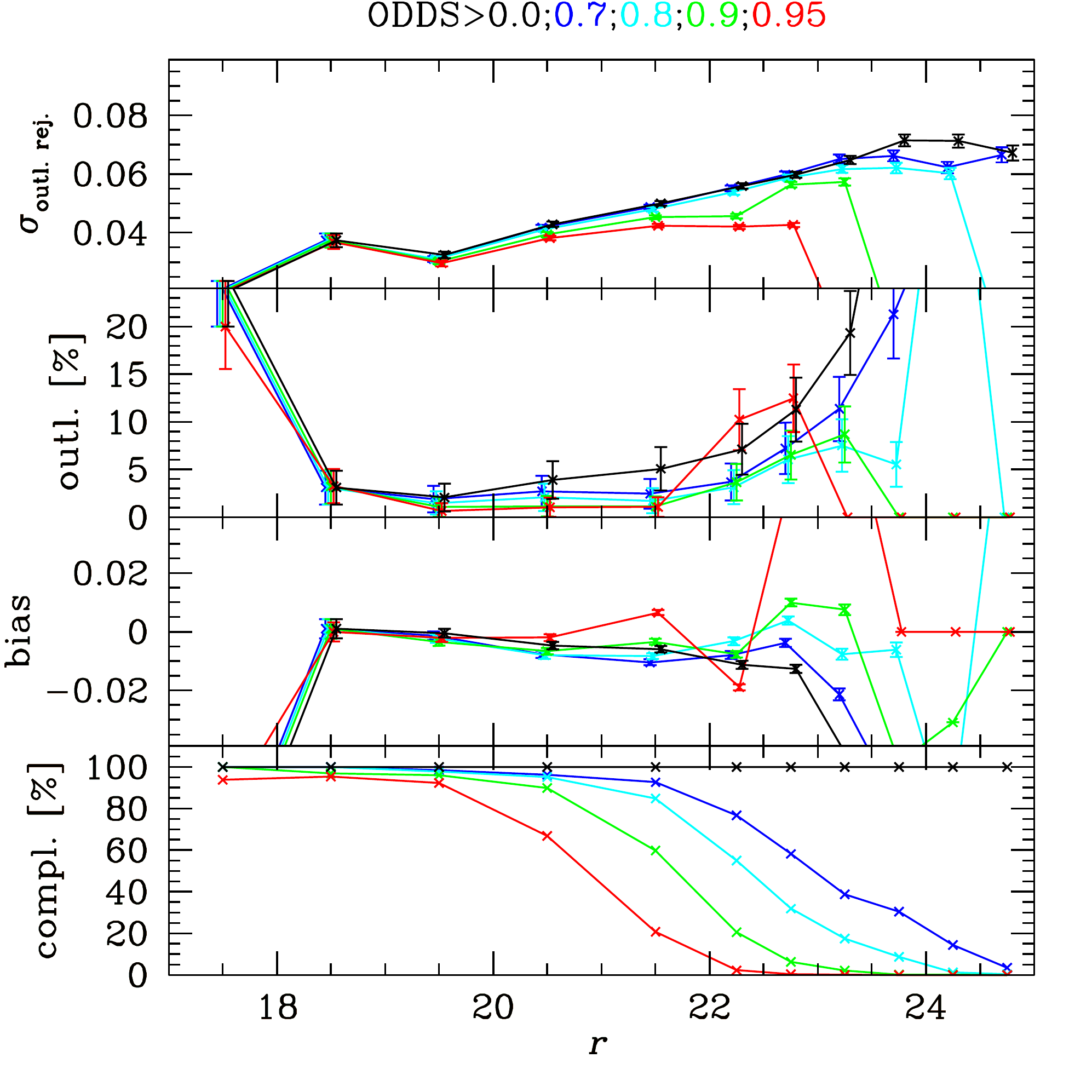}\\
\caption{Statistics of the photometric vs.\ spectroscopic redshift discrepancy $\Delta z$ as a function of $r$-band magnitude in the CDFS and COSMOS fields. From top to bottom: clipped RMS dispersion, outlier fraction, average offset, and fraction of galaxies in each given ODDS cut (normalized to the total).}
\label{fig:photo-z_stats_mag}
\end{figure}

\begin{figure}
\putfig{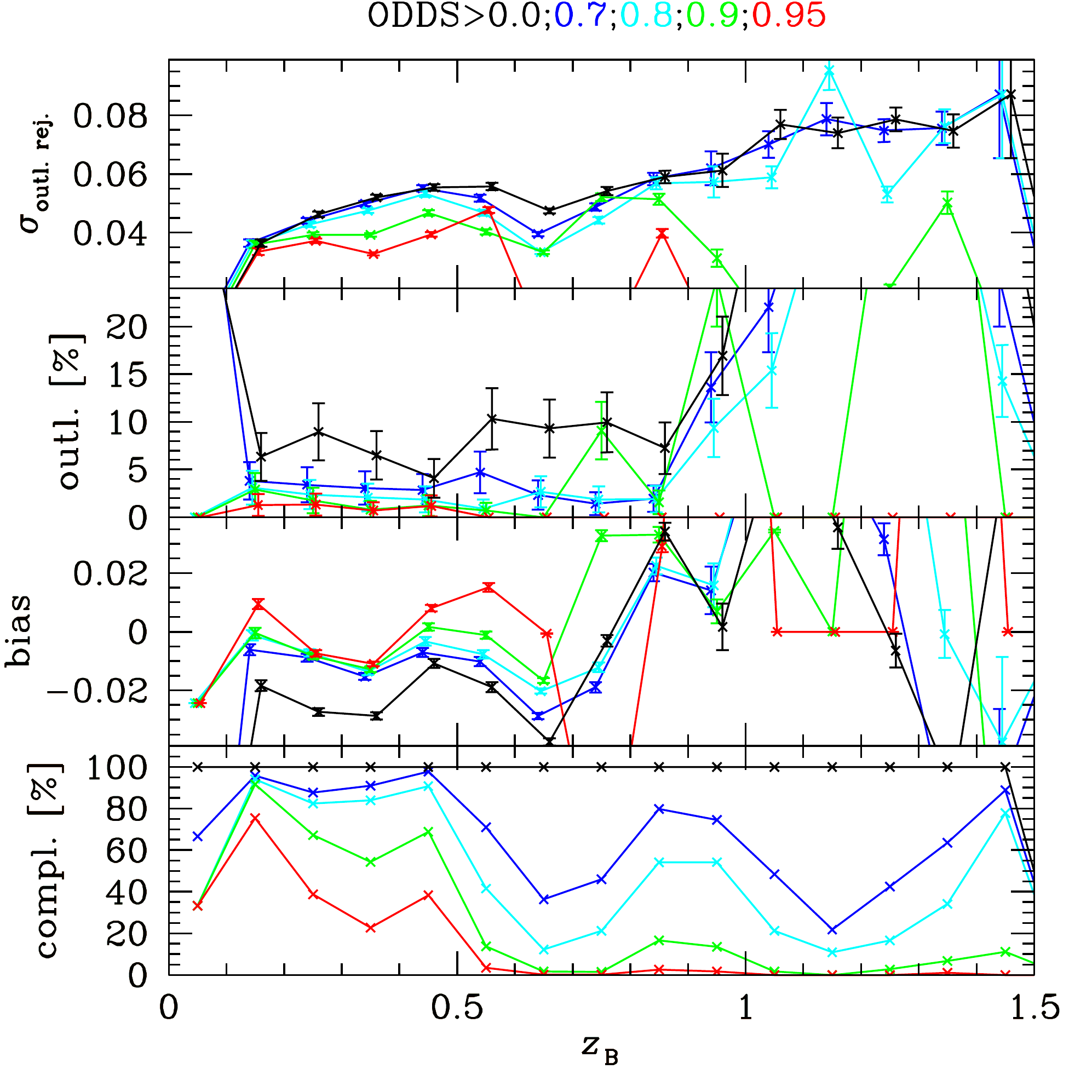}\\
\caption{As Fig.~\ref{fig:photo-z_stats_mag}, but plotted as a function of photometric redshift $z_\rmn{B}$.}
\label{fig:photo-z_stats_z}
\end{figure}

A straight comparison of the Bayesian photometric redshifts, $z_\rmn{B}$, and the spectroscopic redshifts, $z_{\rm spec}$, is shown in Fig.~\ref{fig:zz}. To quantify the level of agreement, we characterize the photometric redshift of each galaxy by the relative error
\begin{equation}
\Delta z=\frac{z_{\rm B}-z_{\rm spec}}{1+z_{\rm spec}}\,.
\end{equation}
and plot its statistics in bins of magnitude and redshift in Figs.~\ref{fig:photo-z_stats_mag} and \ref{fig:photo-z_stats_z}, respectively. 
We use the mean of $\Delta z$ as a measure for the photometric redshift bias, the fraction of objects with $|\Delta z|>0.15$ as the outlier rate, and the RMS scatter after rejection of the outliers as the dispersion. We show the statistics for different cuts on the \textsc{bpz} ODDS parameter \citep[see][]{benitez:2000}, which is a measure of the uni-modality of a galaxy's posterior redshift distribution. Cutting on ODDS usually leads to slightly better photometric redshifts at the expense of losing objects. This is reflected in the completeness fraction, plotted in the bottom panel of Figs.~\ref{fig:photo-z_stats_mag} and \ref{fig:photo-z_stats_z}.

These tests check for the accuracy of the photometric redshift point estimates. Such point estimates can be used to select galaxies in certain redshift regions, to define tomographic redshift bins, and to distinguish between foreground and background galaxies in different lensing applications. The modelling of the lensing measurement, however, makes use of the full photometric redshift posterior probability distributions $p(z)$ that \textsc{bpz} estimates for each galaxy, and in that sense $p(z)$ is the more crucial quantity for the weak lensing science goals. 

We have checked that the summed $p(z)$ posteriors of the galaxies plotted in Fig.~\ref{fig:zz} agree well with their spectroscopic redshift distribution provided we exclude galaxies whose $z_\rmn{B}$ values lie at the extremes of the redshift distribution of the spectroscopic calibration sample.
After some experimentation, based on these results as well as on Fig.~\ref{fig:photo-z_stats_z}, we cut our galaxy catalogue at $0.005<z_{\rm B}<1.2$ in all lensing analyses.

Detailed characterization and testing of the $p(z)$ will be presented in forthcoming papers (Choi et al. in prep., Hildebrandt et al. in prep.). 

\begin{figure*}
\putfig{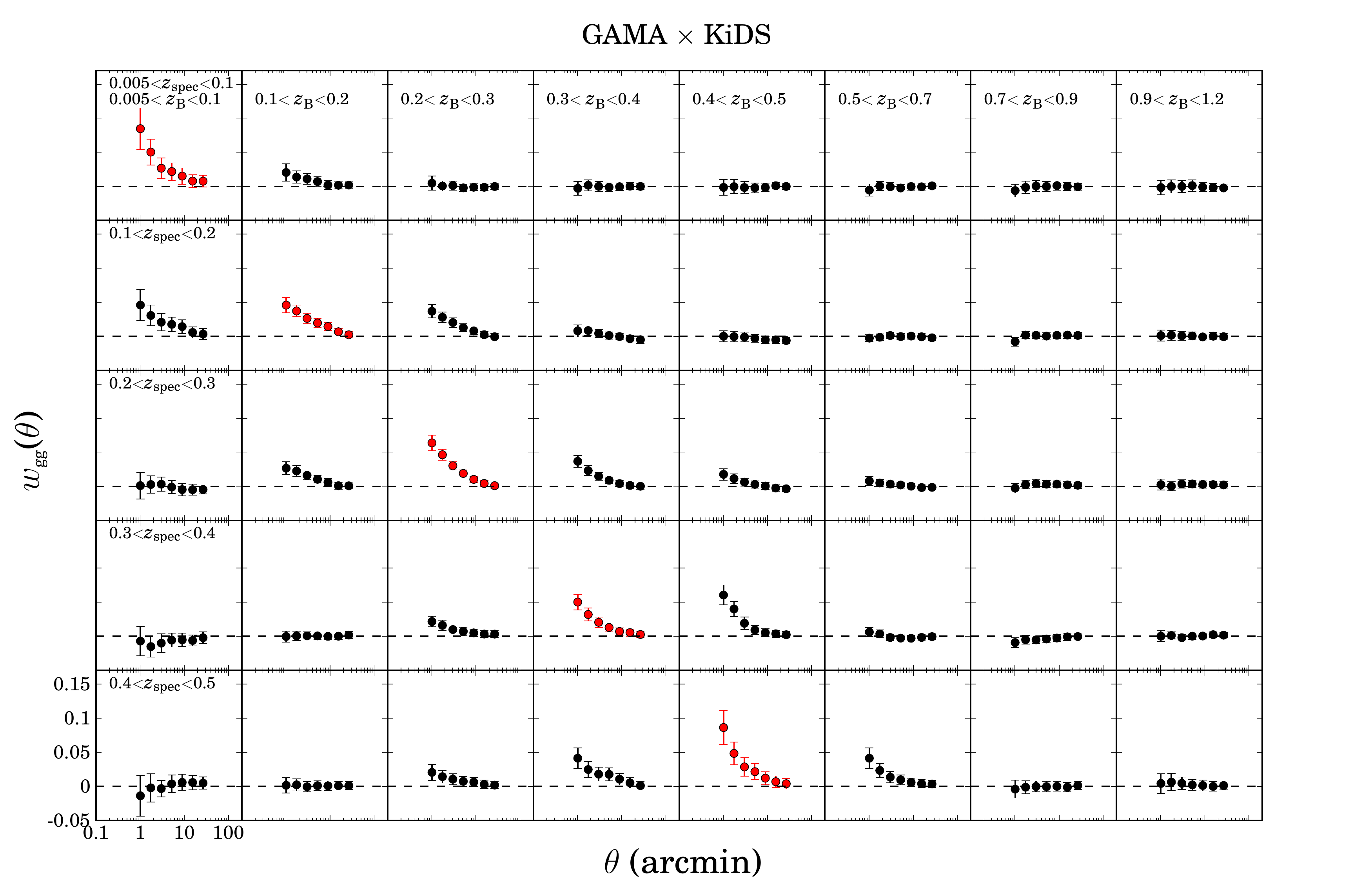}\hfill 
\caption{Angular cross-correlations between KiDS galaxies binned by photometric redshift, and GAMA galaxies binned by spectroscopic redshift. Spectroscopic redshifts increase from top to bottom, and photometric redshifts increase from left to right.}
\label{fig:photo-z_spec-z_cross}
\end{figure*}

\subsection{Galaxy Clustering analysis}
As a further test of our photometric redshifts, following \citep{newman/:2008} we calculate the angular cross-correlation of the positions of GAMA and KiDS galaxies on the sky, grouped by spectroscopic (GAMA) and photometric (KiDS) redshifts.  Galaxies that are physically close will produce a strong clustering signal, and hence this measurement can validate photometric redshift estimates.

GAMA is a highly complete spectroscopic survey down to a limiting magnitude of $r<19.8$, measuring redshifts out to $z_{\rm spec}=0.5$.  We group the GAMA galaxies into five redshift bins $i$ of width $\Delta z_{\rm spec} \simeq 0.1$.  We limit the KiDS galaxies to $r<24$, and group them into eight photometric redshift bins $j$, listed in Fig.~\ref{fig:photo-z_spec-z_cross}.  The photometric redshifts extend beyond the GAMA redshift range to $z_{\rm B}=1.2$.  The projected angular clustering statistic $w_\rmn{gg}^{ij}(\theta)$, between spectroscopic bins $i$ and photometric bins $j$, is then estimated using the \citet{landy/szalay:1993} estimator by means of the \textsc{athena} code \citep{kilbinger/bonnett/coupon:2014}. Errors are calculated using a jackknife analysis.  We focus on angular scales $1\arcmin< \theta < 30\arcmin$, where the upper angular scale is set by signal-to-noise constraints, and the lower angular scale is chosen to reduce the impact of scale dependent galaxy bias on the measurements \citep{schulz/:2010}.  The results are shown in Fig.~\ref{fig:photo-z_spec-z_cross} with the spectroscopic redshift bin $i$ increasing from top to bottom, and the photometric redshift bin $j$ increasing from left to right. 

The strongest angular clustering is found when the spectroscopic and photometric redshift sample span the same redshift range, which can be seen along the `diagonal' of Fig.~\ref{fig:photo-z_spec-z_cross} where $i=j$.  This is anticipated if there is no significant bias in the photometric redshift measurement $z_{\rm B}$.  As the photometric redshifts have an associated scatter, we also see clustering between adjacent spectroscopic and photometric redshift bins.   With the exception of the $0.2<z_{\rm B}<0.3$ bin, we find non-zero clustering only in matching or adjacent bins, implying that the photometric redshift scatter is less than the spectroscopic bin width $\Delta z = 0.1$.  This is consistent with the analysis presented in \S\ref{sec:pz} which found the scatter $\sigma_z < 0.08$ out to $z_{\rm B}=1.2$.  

A correlation between the positions of galaxies in widely separated redshift bins would indicate the presence of catastrophic errors in the KiDS photometric redshifts.  We see this to some extent in the non-zero clustering measured between the $0.2<z_{\rm B}<0.3$ and $0.4<z_{\rm spec}<0.5$ galaxy samples, indicating that a small fraction of the photometric redshifts in this bin are actually at a higher redshift.  This measurement could be used to infer the true redshift distribution of this galaxy sample \citep[see for example][which is beyond the scope of this paper]{mcquinn/white:2013}.  For all other photometric redshift bins, we find the clustering signal to be consistent with zero for all bin combinations separated by $\Delta z=0.1$ or more.  We can therefore conclude that the fraction of `catastrophic outliers' is low, in agreement with the direct spectroscopic-photometric redshift comparison presented in \S\ref{sec:pz}.  

We consider this analysis as a validation of our redshift estimates. A similar conclusion is drawn from the analysis of the cross-correlation between different photometric redshifts bins of KiDS galaxies, presented in \citetalias{dejong/etal:2015}, which extends the cross-correlation between bins beyond redshift $z=0.5$ which cannot be probed with the GAMA catalogues.

\subsection{The combined shear-photometric redshift catalogue}
\label{sec:photzsample}

\begin{figure}
\includegraphics[width=\hsize,trim=2mm 0 11mm 10mm]{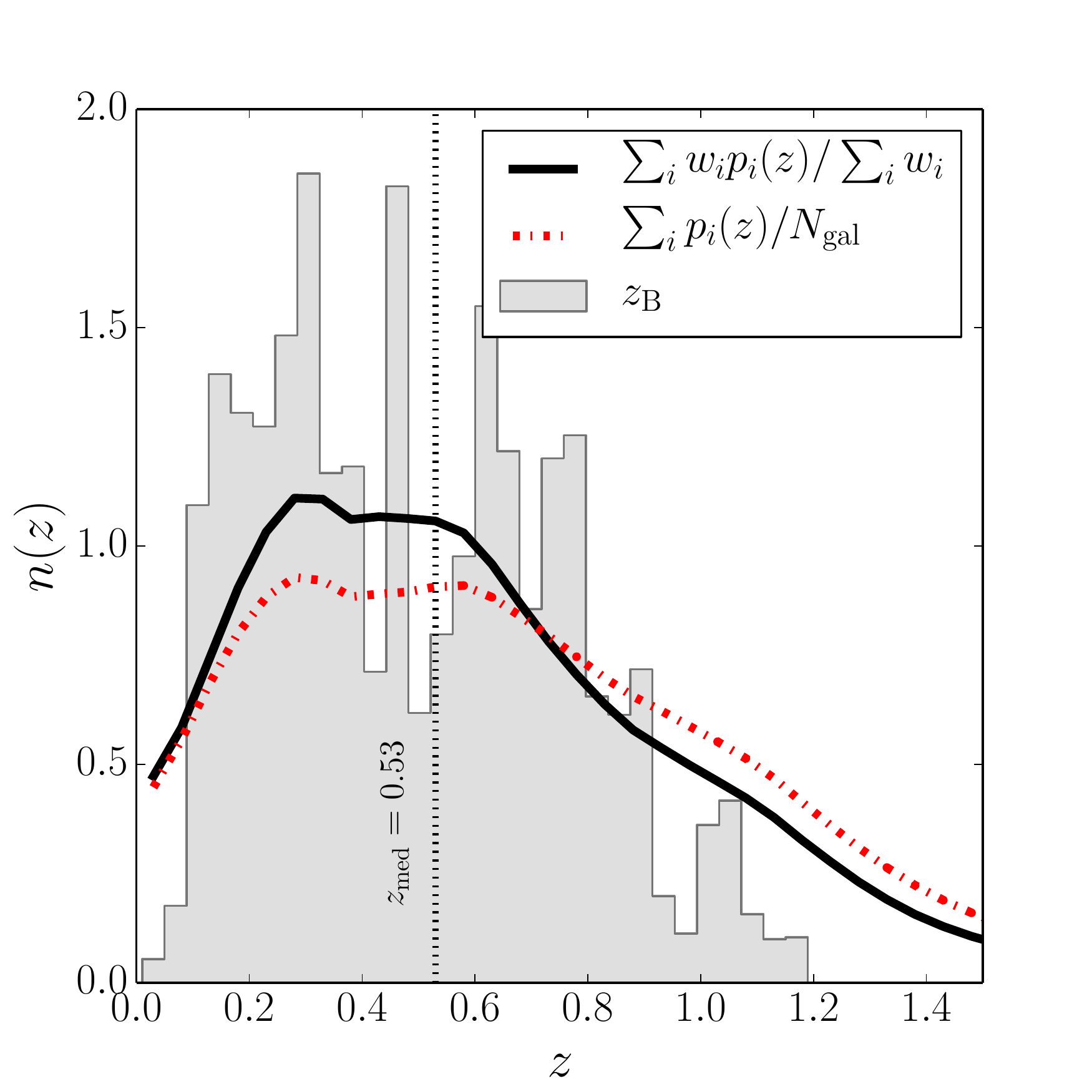}
\includegraphics[width=\hsize,trim=0 0 0 0]{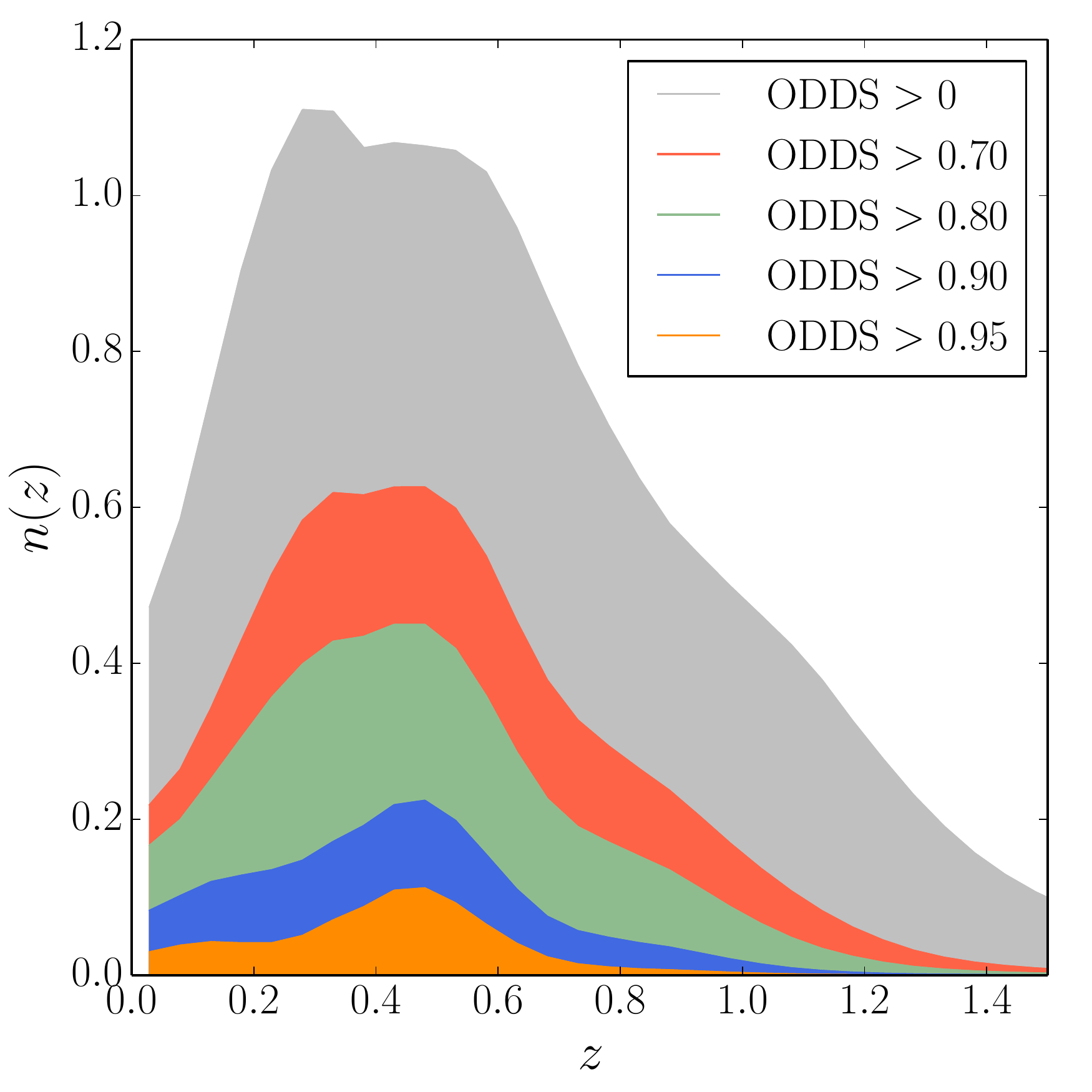}
\caption{The galaxy photometric redshift distribution. Upper panel: summed posterior redshift distributions $n(z)$, with (solid line) and without (dashed line) weighting by the \emph{lens}fit weight. The effective median redshift of the lensing survey is $z_\rmn{m}=0.53$. The histogram shown in this panel shows the distribution of the $z_\rmn{B}$ point estimates of the photometric redshift.
Lower panel: the lensing weighted posterior $n(z)$ distributions of galaxies in progressively lower ODDS categories (see text).}
\label{fig:nofz}
\end{figure}

In \S\ref{sec:pz} we defined a photometric redshift selection criterion $0.005<z_{\rm B}<1.2$ to ensure a good level of accuracy in the photometric redshifts.  We now combine that redshift selection with the shape measurement analysis by also selecting galaxies with a \emph{lens}fit weight $w>0$ \citepalias[this cut excludes all galaxies for which no shape measurement was obtained, see][]{miller/etal:2013}.    The upper panel of Fig.~\ref{fig:nofz} compares three redshift distributions for this sample of galaxies, showing the distribution of the $z_{\rm B}$ point estimates of the photometric redshift, and the weighted and unweighted sums of the associated posterior distributions $p(z)$.  The weighted distribution, plotted as the thick solid line, is the one most relevant for our analysis: it is the effective redshift distribution of the lensing information, and has a median redshift of $z_m=0.53$. 

The weights used in the lensing analysis favour higher SNR galaxies which are typically at lower redshift in this flux-limited survey, and hence the weighted median redshift is lower than that of the unweighted sample (which has $z_m=0.63$).  Indeed if the shape measurement criterion $w>0$ had not been applied, the unweighted median redshift would be even higher with $z_m=0.66$.   This is illustrated in the lower panel of Fig.~\ref{fig:nofz}, which shows the effective redshift distribution for galaxies with different \textsc{bpz} ODDS parameters: the more precise photometric redshifts, with high ODDS, also tend to be at lower redshifts (e.g., the weighted median redshift for galaxies with ODDS$>0.9$ is 0.43).
As the ODDS value decreases, so does the accuracy of each individual photometric redshift, owing to multiple peaks in each galaxy's posterior distribution that result from degeneracies in the redshift solution. In the stacked posterior shown in Fig.~\ref{fig:nofz}, these degeneracies are responsible for the shape of the distribution at the peak. 

\begin{figure}
\putfig{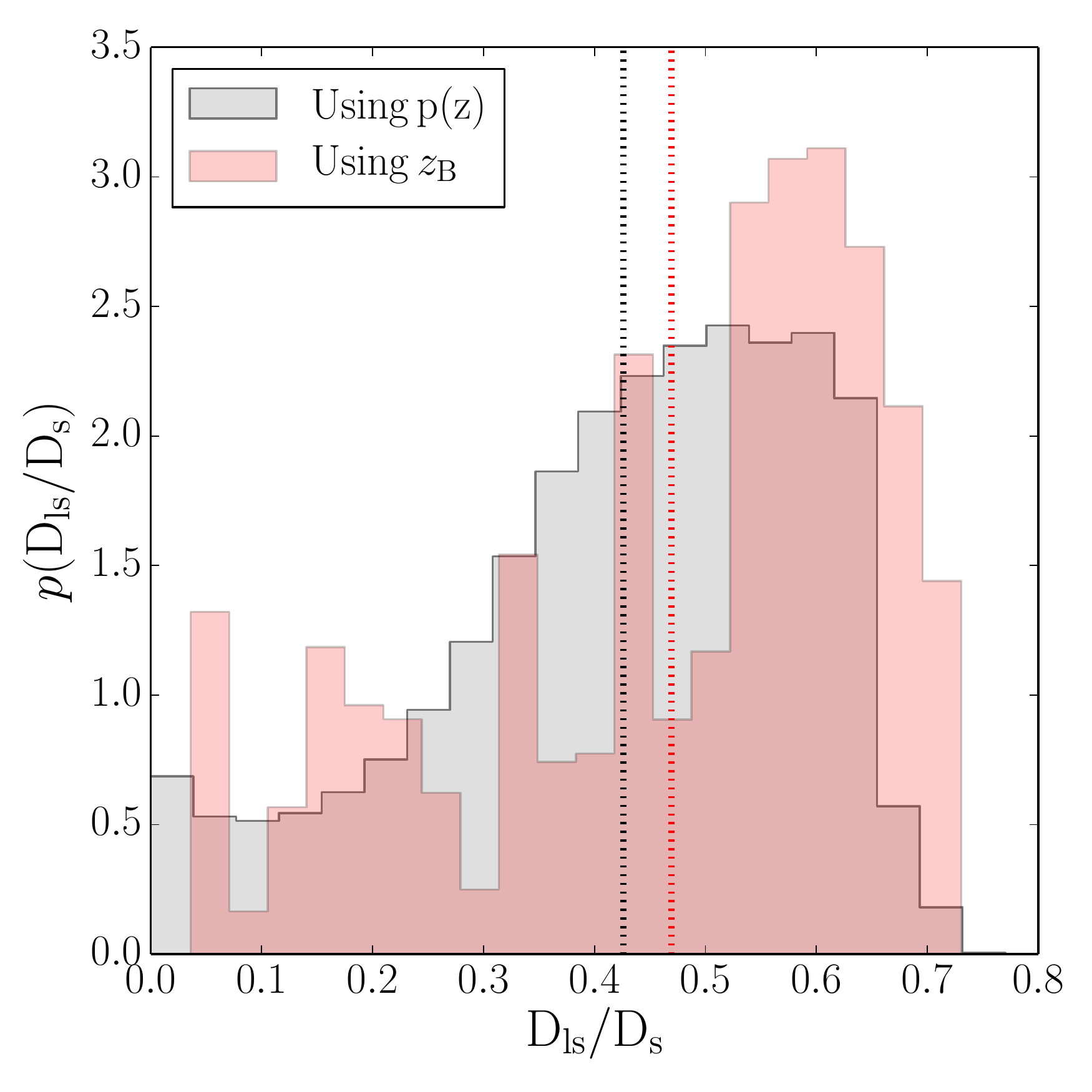}
\caption{Effect of using the full photometric redshift posterior $p(z)$, or the point estimate $z_\rmn{B}$ to determine the angular diameter distance ratio $D_\rmn{ls}/D_\rmn{s}$ for a lens galaxy at redshifts $z_\rmn{l} = 0.25$.   The average distance ratio $D_\rmn{ls}/D_\rmn{s}$ sets the lensing efficiency and differs by 10 percent depending which redshift measure is used (dashed lines).}
\label{fig:beta}
\end{figure}

Fig.~\ref{fig:nofz} illustrates the importance of using the full posteriors $p(z)$ instead of the best-fit photometric redshifts $z_\rmn{B}$ to define the survey redshift distribution.  The point estimates are more prone to artefacts associated with the particular filter set used.  They also do not reflect the full information content of the photometry.  As an illustration of how using $z_\rmn{B}$ could bias a lensing analysis, Fig.~\ref{fig:beta} shows the measured angular diameter distance ratio $D_\rmn{ls}/D_\rmn{s}$ for a lens at redshift $z_\rmn{l}=0.25$.  Using $z_\rmn{B}$ or $p(z)$ to determine the redshift distribution of the background lensed sources, changes the average distance ratio by $\sim 10$ percent.  As the distance ratio defines the lensing efficiency of sources at different redshifts, using $z_\rmn{B}$ instead of $p(z)$ would result in an underestimate of the lensing surface mass density by $\sim 10$ percent. 

\section{Tests for systematic errors in the K\lowercase{i}DS lensing catalogue}
\label{sec:sys}

Different science cases require different levels of accuracy in the shear and photometric redshift catalogues.  It is common to model calibration corrections to shear measurement in terms of a multiplicative term $m$ and additive terms $c_k$ such that
\be
\epsilon_k^{\rm obs} = (1+m) \epsilon_k^{\rm true} + c_k \, ,   \qquad (k=1,2)\, ,
\ee
where $\epsilon_k^{\rm obs}$ are the observed ellipticity parameters, and $\epsilon_k^{\rm true}$ the true galaxy ellipticity parameters \citep{heymans/etal:2006}. 
\citet{massey/etal:2013} present a compilation of possible sources of such correction terms, and calculate requirements on their amplitudes for different kinds of analysis.
In an ideal shape measurement method, both $m$ and $c_k$ would be zero. In reality however, these corrections need to be determined so the data can be calibrated, and then systematics tests performed to ensure the calibration is robust.  

Our first series of lensing science papers measure shear-position correlation statistics, also known as galaxy-galaxy lensing, where the tangential shear of background galaxies is determined relative to the position of foreground structures. As this measurement is taken as an azimuthal average, it is very insensitive to additive correction terms $c_k$ except on scales comparable to the survey boundaries. It is, however, very sensitive to the accuracy of the measured multiplicative calibration $m$, an error which leads directly to a bias in the mass determined from the lensing measurement.   Furthermore, these measurements rely on a good knowledge of the photometric redshift distribution to determine the level of foreground contamination in the background source sample and hence the level of dilution expected in the measured lensing signal.  In this section we therefore first describe the analysis done to validate the multiplicative calibration $m$ used, and then verify that the redshift scaling of the galaxy-galaxy lensing signal is consistent with the expectation based on the photometric redshift error distributions.

In this technical paper we also present the first demonstration of the suitability of the data for cosmological measurements through two-point shear statistics. Such an analysis places more stringent requirements on the accuracy of the shear catalogue, in particular the additive corrections $c_k$.  We therefore perform an additional set of tests, following \citetalias{heymans/etal:2012}, first selecting fields where the cross-correlation between the measured shear signal and the PSF pattern is consistent with zero systematics.  We then empirically determine the $c_k$ terms from the remaining data.  

\begin{figure}
\includegraphics[width=\hsize]{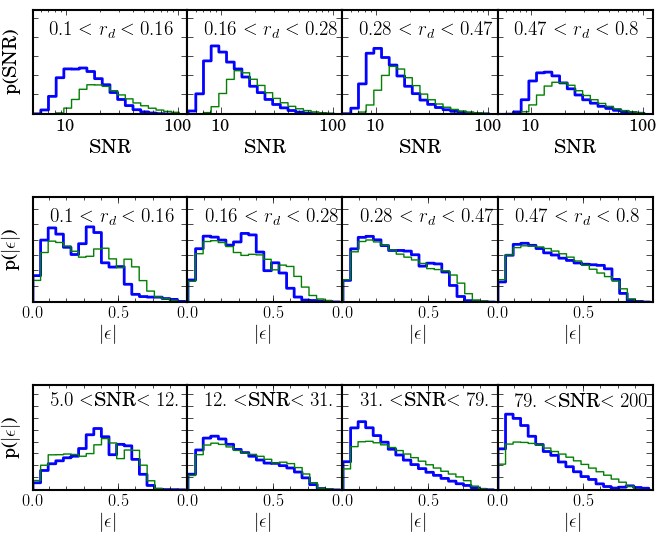}
\caption{Comparison of the observed properties of galaxies in the image simulations from \citet{miller/etal:2013} (thin lines) to the observed properties of galaxies in KiDS (thick lines).  The upper panels compare the signal-to-noise ratio (SNR) distributions in bins of increasing galaxy size (in arcseconds). The ellipticity distributions can be compared as a function of galaxy size (middle panels) and SNR (lower panels).}
\label{fig:sim_data_comp} 
\end{figure}

\subsection{Multiplicative calibration}
The multiplicative calibration term $m$ can only be determined through the analysis of image simulations where the true galaxy shapes are known.  \citetalias{miller/etal:2013} describe the CFHT MegaCam image simulations against which \emph{lens}fit  was calibrated extensively in the CFHTLenS analysis. The primary aim of these simulations was to correct for noise bias \citep{hirata/etal:2004,refregier/etal:2012,melchior/viola:2012}. 
On average the noise bias resulted in a $\sim5$ percent correction to the measured shear, with more significant corrections for smaller, fainter galaxies.  This analysis provided a calibration correction that depends on the \emph{lens}fit parameters SNR and size $r_\d$ as
\be
m(\hbox{SNR}, r_\d) = \frac{\beta}{\log_{10} \hbox{SNR}} \exp(-\alpha \, r_\d \, \hbox{SNR})\, ,
\label{eqn:mcalmod}
\ee
with $\alpha=0.306\,\rmn{arcsec}^{-1}$ and $\beta=-0.37$.

The  {\it r}-band KiDS VST-OmegaCAM imaging differs from the simulated {\it i}-band CFHT MegaCam imaging in a few key respects.  The pixel scales differ: $\theta_{\rm pix}=0.213\arcsec$ for OmegaCAM and $\theta_{\rm pix}=0.186\arcsec$ for MegaCam.  The KiDS data are shallower than CFHTLenS, and while the mean PSF FWHM values for the two sets of lensing data are the same (0.64\arcsec), the average KiDS PSF ellipticity is $\sim 15$ percent smaller than the average CFHTLenS PSF.  We verify in two different ways that this CFHTLenS correction is suitable to use for KiDS: (i) using a re-sampling technique such that the simulated catalogues better match the KiDS data, and (ii) by comparing the galaxy-galaxy lensing signal around bright galaxies in CFHTLenS and KiDS for progressively fainter source samples.

\begin{figure*}
\putfig{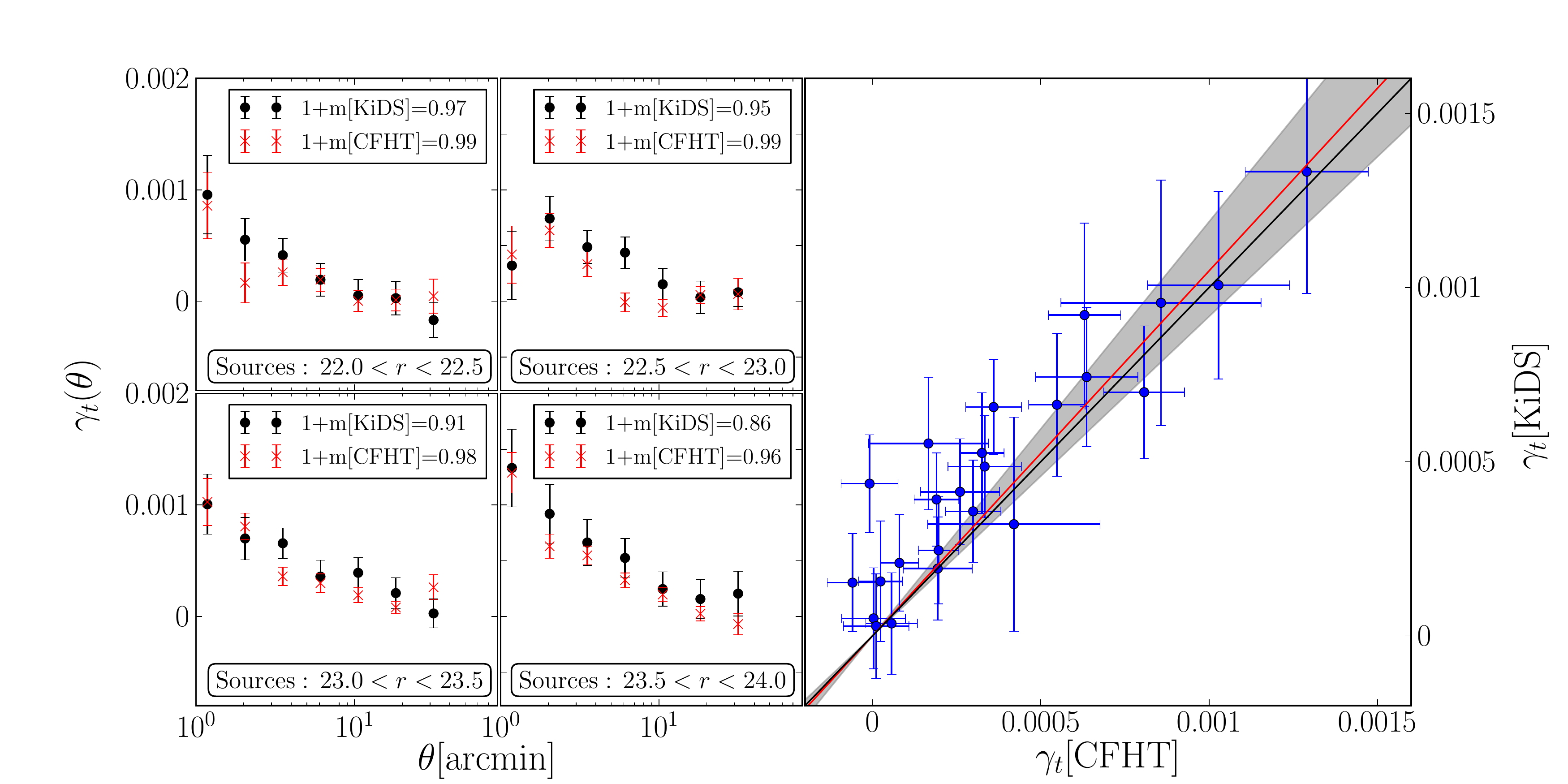}
\caption{Four panels on the left: KiDS vs.\ CFHTLenS comparison of the average tangential shear around galaxies with $20<r<21$ measured from progressively fainter source populations. The insets show the average multiplicative correction factor, as derived from Eq.~\ref{eqn:mcalmod} and applied to the plotted measurements. Right-hand panel: bin by bin comparison of the shear values for $\theta$ between 1 and 20 arcmin. The red line shows the best-fit linear regression, and the grey zone the corresponding 1-$\sigma$ uncertainty (errors on both axes are taken into account).}
\label{fig:shearcomp}
\end{figure*}

\subsubsection{Re-sampled image simulations}
\label{sec:resamp}
Fig.~\ref{fig:sim_data_comp} compares the measured properties of galaxies in the image simulations from \citetalias{miller/etal:2013} (thin lines) to the properties of galaxies in KiDS (thick lines). The upper panels compare the SNR distributions in bins of increasing galaxy size\footnote{In principle this comparison should be made in terms of the relative galaxy-to-PSF size, but as the KiDS and CFHTLenS imaging have similar seeing distributions we work with galaxy size in arcseconds.} showing that the image simulations have a deficit of small galaxies.  \citetalias{miller/etal:2013} concluded this arose from an overestimate of the true PSF size when creating the image simulations. Compared to the image simulations, which are a good match to the SNR distribution of the CFHTLenS data, we also see a higher proportion of low SNR galaxies in KiDS.  This arises because CFHTLenS imposed a magnitude limit $i<24.7$ on their galaxy sample, based on the depth to which photometric redshifts were considered reliable. For KiDS, we do not include a similar imposed fixed magnitude limit, see Fig.~\ref{fig:photo-z_numbercounts}, as the depth of the survey is within the limits covered by deep spectroscopic surveys.  

Comparing the ellipticity distributions as a function of galaxy size (middle panels) and SNR (lower panels) in Fig.~\ref{fig:sim_data_comp}, we see an excess of simulated galaxies of large ellipticity in the high-SNR regime.  As shown in \citet{viola/etal:2014} and \citet{hoekstra/etal:2015}, calibration corrections can be sensitive to the ellipticity distribution. For the purposes of the analysis of our first 100 square degrees, we re-sample the simulated galaxy catalogues from \citetalias{miller/etal:2013} such that the simulated ensemble galaxy properties match the KiDS data in terms of size, SNR and ellipticity. This is possible as the image simulations from \citetalias{miller/etal:2013} simulated two complete CFHTLenS surveys.  Hence while there is a deficit of small, low SNR galaxies in the simulations, relative to the global populations, there are sufficient numbers with which to validate the calibration scheme $m$ from Eq.~\ref{eqn:mcalmod}, for KiDS, in this under-represented regime.   

We sample galaxies from the image simulations, such that the correlations that exist between observed size, observed SNR and observed ellipticity in the data are retained. As \emph{lens}fit performs a joint parameter fit of galaxy ellipticity and size, selecting galaxies based on their observed size will introduce a selection bias on galaxy ellipticity. It is therefore critically important not to subject \emph{lens}fit catalogues to any `cleaning criterion', for example rejecting small galaxies based on the \emph{lens}fit size estimate.  Instead we use the \emph{lens}fit weights to optimally combine the shape measurements.  Following \citetalias{miller/etal:2013} we determine the accuracy of the CFHTLenS calibration correction for KiDS by calculating
\be
\delta m = \frac{\sum_{ik} \left[1+m(\hbox{SNR}, r_\d)\right] w_{i} (\epsilon_{ik}^{\rm obs} - \epsilon_{ik}^{\rm true}) }{2\sum_i w_i} = -0.04 \pm 0.02 \, ,
\ee
where the sum is taken over the simulated galaxies $i$ in the re-sampled image simulation catalogues, weighted by the observed \emph{lens}fit weights $w_i$, and calculated for both components $k$ of the ellipticity.  We find that the CFHTLenS calibration correction underestimates the calibration required for KiDS by a few percent\footnote{We note an error in the calculation of Eq.~\ref{eqn:mcalmod} used in the first KiDS lensing analyses (\citealt{viola/etal:2015}; \citealt{sifon/etal:2015}; van Uitert et al., in preparation) that did not correctly account for the different MegaCam and OmegaCAM pixel scales.  By luck this error erroneously increased the average value of $m$, such that the KiDS-correction $\delta m$ was reduced to $\delta m = -0.03 \pm 0.02$.}, which is within the current statistical error budget for the early science presented in \citet{viola/etal:2015}, \citet{sifon/etal:2015} and van Uitert et al. (in preparation).   We also verified that this underestimate did not vary significantly as a function of galaxy SNR, as it arises from the increased fraction of small galaxies in the sample.
A new suite of KiDS image simulations are in production using the \textsc{GALSIM} software \citep{rowe/etal:2015}, in preparation for future analyses in which the larger area surveyed will demand a more accurate calibration scheme.

\subsubsection{Galaxy-galaxy lensing at different signal to noise ratio: KiDS vs. CFHTLenS}
\label{sec:mtest}
In this section we apply an additional consistency check to confirm the findings of the image simulation re-sampling analysis, using real data.  We verify that the SNR dependence of the multiplicative calibration is robust by comparing galaxy-galaxy shear measurements from observations of different depths.   To divorce this test from any uncertainties in photometric redshift, we define lens and source samples purely by $r$-band magnitude.  We then compare the dimensionless, $m$-calibrated tangential shear profile $\gamma_{\rm t}(\theta)$ measured with KiDS and with the deeper CFHTLenS data \citepalias{erben/etal:2013}. The lens samples are selected with $20<r<21$, and four source samples are selected in half-magnitude bins from $r=22$ to $r=24$.  For the brightest sources the average calibration corrections from Eq.~\ref{eqn:mcalmod} are only a few percent for both surveys, but the faintest bin includes a 14 percent calibration correction for KiDS compared to a 4 percent correction for CFHTLenS.  Figure.~\ref{fig:shearcomp} shows the good agreement between the calibrated KiDS and CFHTLenS tangential shear profiles, measured between 1 and 20 arcmin, for the four different source samples. To quantify the consistency we perform a direct bin by bin comparison of the measured shears in the right-hand panel of Fig.~\ref{fig:shearcomp}. Fitting a simple proportionality relation to the points, using uncorrelated bootstrap errors, as motivated by the results of the analytical prescription described in \citet{viola/etal:2015}, we find a best-fit ratio of (KiDS/CFHTLenS)=$1.05 \pm 0.13$.

\subsection{Testing redshift scaling with galaxy-galaxy lensing}
As objects get fainter, our ability to measure shape, photometry and photometric redshifts degrades.   On the other hand the fainter galaxies tend to be at higher redshifts, and therefore they experience a stronger lensing distortion. Measuring the dependence of the lensing signal with source redshift can in principle provide tight constraints on the growth of structure and geometry of the Universe.  It is therefore imperative to perform a cosmology-insensitive joint test of the shear-redshift catalogue and determine whether any redshift-dependent shear bias exists.   In \citetalias{heymans/etal:2012} a galaxy-galaxy lensing test of shear-redshift-scaling was designed that was found to be only very weakly sensitive to the fiducial cosmology assumed in the analysis. The mean tangential shear $\gamma_\rmn{t}$ is measured around a sample of lens galaxies for a series of source galaxies split by increasing photometric redshift, $z_{\rm B}$. We approximate the mass distribution of the galaxies in the lens sample as simple isothermal spheres with a fixed velocity dispersion $\sigma_\rmn{v}$.  The predicted tangential shear around the lens sample $i$, measured from source sample $j$, is then given by
\be
\gamma_{\rm t}^{ij} (\theta) = \frac{2\pi}{\theta} \left( \frac{\sigma_\rmn{v}}{c} \right)^2 \Big\langle \frac{D_{\rm ls}}{D_{\rm s}} \Big\rangle_{ij} \, .
\label{eq:gamasis}
\ee
Here $c$ is the speed of light, and $D_\rmn{ls}/D_\rmn{s}$ is the ratio between the angular diameter distances from the lens to the source, and from the observer to the source. The average of this ratio depends on the effective redshift distribution of the lens and source sample \citep[see for example][]{bartelmann/schneider:2001}.  For a fixed lens sample, we should recover consistent measurements of $\sigma_\rmn{v}$, independent of which source sample is used. Any discrepancy indicates either a poor knowledge of the photometric redshift distribution for that source sample, a redshift-dependent shear measurement bias, or a strong redshift dependence in the velocity dispersion $\sigma_\rmn{v}$ of the lenses within their foreground redshift bin.

\begin{figure}
\includegraphics[width=\hsize]{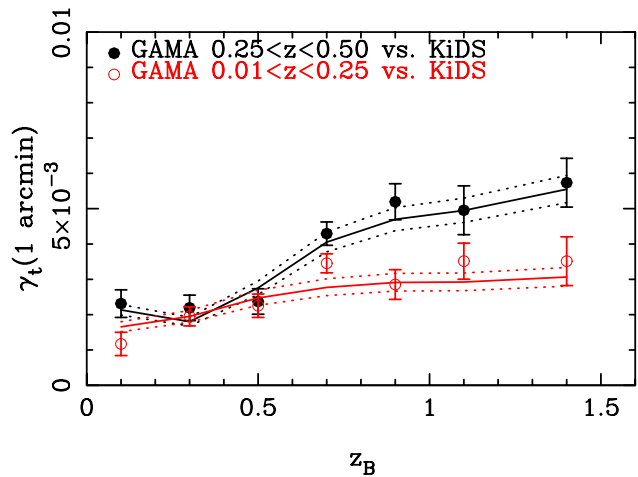}
\caption{The tangential shear measured at one arcminute as a function of the average redshift of the source sample, for two samples of GAMA lenses with spectroscopic redshifts between $0.25<z_s<0.5$ (filled) and $z_s<0.25$ (open).  The solid line shows the predicted signal from the best-fit SIS model, with the dashed lines showing the 68 percent confidence interval.}
\label{fig:zscaling}
\end{figure}

Fig.~\ref{fig:zscaling} shows the tangential shear determined at one arcminute, for source galaxies in seven bins of $z_\rmn{B}$ spanning $0.005<z_\rmn{B}<1.5$. Two samples of lens galaxies from GAMA were used, with spectroscopic redshifts between $0.25<z_\rmn{s}<0.5$ (filled) and $z_\rmn{s}<0.25$ (open).  

The solid line connects the predicted signals from the best-fit SIS model, assuming a Planck cosmology, taking into account the full redshift posterior $p(z)$ for the sources in each bin. The amplitude of the model is set by fitting to all sources with photometric redshifts $0.2<z_{\rm B}<1.0$, which is considered to be the safest photometric redshift range based on the results presented in Fig.~\ref{fig:photo-z_stats_z}. The dashed lines show the 68 percent confidence intervals on the model amplitudes.

As expected, the signal increases as the average redshift of the source sample increases.  We also see that the signal and model do not tend to zero for low $z_\rmn{B}$, even though the mean source photometric redshift is in front of the lens.  This is a result of a non-zero fraction of catastrophic outliers in the photometric redshift sample that are actually at high redshift, causing a significant tangential shear signal.  By taking account of the full photometric redshift posterior probability distributions of the sources, the knowledge of catastrophic outliers enters the model, generating an upturn at low source redshift (note that such low-$z_\rmn{B}$ galaxies which are actually at high redshift do not show up in the cross-correlations in Fig.~\ref{fig:photo-z_spec-z_cross} as they fall outside the GAMA redshift range).  This analysis shows that, within the current SNR of the measurement, our shear-redshift catalogue is not subject to significant redshift-dependent shear biases.

\subsection{Field Selection for cosmic shear test}
\label{sec:passfail}
\citetalias{heymans/etal:2012} describe a method to identify observations with significant residual contamination of the galaxy shapes by the PSF.  It involves comparing the correlation between galaxy and PSF shape, measured in the data and with mock catalogues.  As a result 25 percent of the CFHTLenS tiles were flagged as unsuitable for cosmic shear science; nonetheless these data could be retained for the galaxy-galaxy lensing analyses as the azimuthal averaging renders the measurement essentially insensitive to additive PSF errors.  We follow CFHTLenS in not applying field selection for our first series of galaxy-galaxy lensing science papers, but repeat the \citetalias{heymans/etal:2012} analysis on KiDS in order to assess its future competitiveness for cosmic shear science.  We summarize the key steps of the analysis, and refer the reader to \citetalias{heymans/etal:2012} for a detailed description.

The ellipticity estimate for each source can be written as
\be
\epsilon^\rmn{obs}=\epsilon^\rmn{int}+\gamma+\eta+A_{\rmn{sys},i}\epsilon^i_{\rm PSF} \, ,
\ee
where $\epsilon^\rmn{int}$ is the intrinsic galaxy ellipticity, $\gamma$ is the true cosmological shear that we wish to detect, and $\eta$  is the random noise on the shear measurement whose amplitude depends on the size and shape of the galaxy in addition to the SNR of the observations. The final term reflects residual amounts of PSF contamination from the various sub-exposures $i$ that `print through' to the final galaxy ellipticities. Even though the coefficients $A_{\rmn{sys},i}$ should be very small for good shape measurement pipelines, this term can generate significant coherent correlations when the shapes of many galaxies on the same tile are averaged.

From a set of $N$ sub-exposures of a part of the sky ($N=5$ in the case of KiDS $r$-band data) \citetalias{heymans/etal:2012} define a vector of star-galaxy cross-correlation coefficients $\bxi_\rmn{sg}$, with one element per sub-exposure: 
\be
\bxi_\rmn{sg} = 
\langle \epsilon^\rmn{obs} \bepsilon^*_\rmn{PSF} \rangle  =  
\langle \epsilon^\rmn{int} \bepsilon^*_\rmn{PSF} \rangle +  
\langle \gamma \, \bepsilon^*_\rmn{PSF} \rangle +  
\langle \eta \, \bepsilon^*_\rmn{PSF} \rangle + 
\mat{C} \vec{A}_{\rm sys} \, ,
\label{eqn:sg}
\ee
where the average is taken over all galaxies in the pointing.
Here $\bepsilon_\rmn{PSF}$ is a vector of PSF ellipticity patterns, one per sub-exposure, determined from the PSF model at the locations of the source galaxies in each sub-exposure.  $\mat{C}$ is a matrix whose elements $C_{ij} = \langle \epsilon_{\rm PSF}^i {\epsilon_{\rm PSF}^j}^* \rangle$ give the average covariance of PSF ellipticities between the sub-exposures. The complex conjugate of the ellipticity is denoted with a $*$, and only the real part of the averages in Eq.~\ref{eqn:sg} is kept (as in Eq.~\ref{eqn:xi_res}). We have assumed that $\vec{A}_\rmn{sys}$ does not vary across the field of view. 

\begin{figure}
\putfig{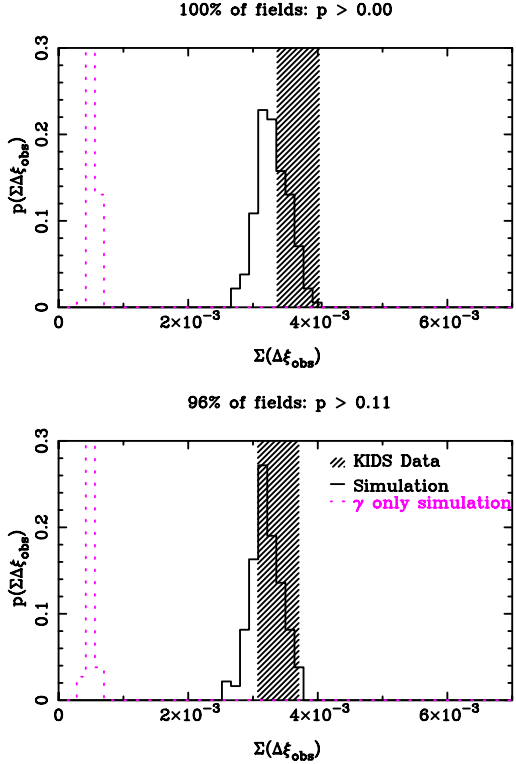}
\caption{Field selection based on the degree of correlation between the PSF ellipticity pattern and the galaxy ellipticities as compared to simulated data. See the text for the definition of $\sum(\Delta\xi_\rmn{obs})$, which quantifies the degree of residual PSF contamination in measurements of the two-point shear correlation function. The histogram shows the expected range of this statistic in simulations, and the hashed region indicates the measured value $\pm$ the 1-$\sigma$ bootstrap error. For comparison the dashed histogram shows the expected range for shape-noise free simulations. Top: all 109 KiDS fields. Bottom: result of field selection (see text for details).}
\label{fig:pu_test}
\end{figure}

For a sufficiently wide area, the first three terms of Eq.~\ref{eqn:sg} average to zero, in which case $\vec{A_\rmn{sys}}=\mat{C}^{-1}\bxi_\rmn{sg}$. The contribution of this systematic ellipticity error to the two-point shear correlation function, $\langle \epsilon^\rmn{obs} {\epsilon^\rmn{obs}}^* \rangle$ is then given by 
\be
\Delta \xi_\rmn{obs}  = \bxi_\rmn{sg}^\mat{T} \mat{C}^{-1} \bxi_\rmn{sg} \, .
\label{eqn:deltaeobs}
\ee
We wish to use $\bxi_\rmn{sg}$ as a diagnostic with which to identify those tiles where, for whatever reason, the PSF modelling has left significant residuals that would contaminate the shear-shear correlation function. The KiDS data are taken in square-degree tiles, and on these scales the measurement of $\bxi_\rmn{sg}$ will have contributions from the first three noise terms in Eq.~\ref{eqn:sg} through chance alignments between the different noise, PSF and cosmic shear fields.  We therefore estimate the expected amplitude of $\Delta \xi_\rmn{obs}$, a positive quantity, from a series of 184 simulated KiDS data sets each containing 109 systematics-free one-square degree mock catalogues. These mock catalogues are populated to match the intrinsic ellipticity and measurement noise in the data.  A correlated cosmic shear signal is also added, drawn from the $N$-body simulations of \citet{Harnois-Deraps/etal:2012}, following the effective galaxy redshift distribution $n(z)$ of KiDS shown in Fig.~\ref{fig:nofz}.  Fig.~\ref{fig:pu_test} shows the distribution of $\sum (\Delta \xi_{\rm obs})$, where the sum is taken over all 109 mock fields, for the 184 different mock realizations of KiDS.  The dashed line shows the result we would have obtained if the mock catalogues had contained a cosmic shear signal only, to emphasize that the two-point star-galaxy cross correlation function will be non-zero even in the absence of ellipticity noise.  We then measure the average star-galaxy cross correlation coefficient for each field observed, with the result summed over all fields shown as the hashed rectangle in the upper panel.  The difference between the expected result from the mock simulations and the data shows that some fields do indeed contain strong PSF residuals.  To isolate these fields we determine a probability $p$ for each field that $\Delta \xi_{\rm obs}$ is consistent with zero systematics (see \citetalias{heymans/etal:2012} for details).  We then set a threshold on this probability such that the data (shown hashed) match the expected distribution from the simulations, a requirement met when $p>0.11$. We find that this procedure rejects only 4 of our fields (3.7 percent, cf.~25 percent for CFHTLenS), suggesting that the PSF modelling in KiDS is of a significantly higher quality than in CFHTLenS, as could have been expected owing to the clean OmegaCAM PSF.

\subsection{Additive calibration correction}
\label{sec:addcalcor}
\begin{figure}
\putfig{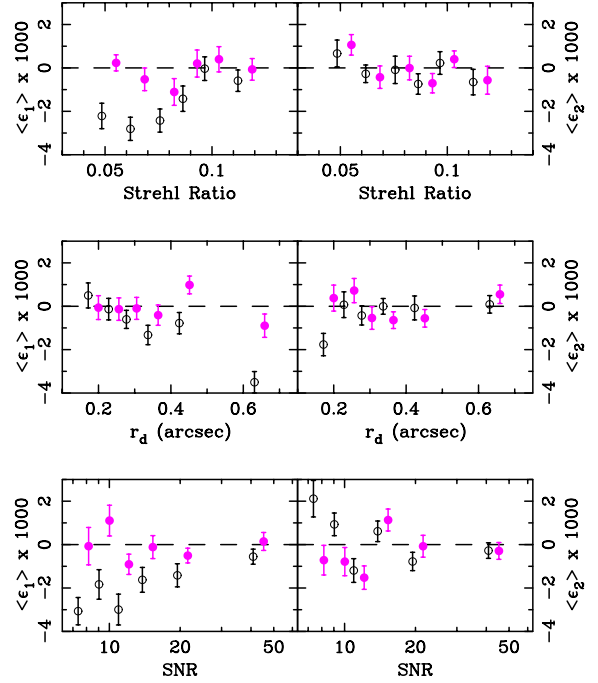}
\caption{The weighted mean ellipticity components $\langle \epsilon_1 \rangle$ (left) and $\langle \epsilon_2 \rangle$ (right), as a function of PSF Strehl ratio (upper), galaxy size (middle) and galaxy SNR (lower).  The points are shown before (open symbols) and after (closed symbols) the empirical calibration has been applied, with the latter offset horizontally for clarity.}
\label{fig:c_corr_before_after}
\end{figure}

For the data that passed our field selection (\S\ref{sec:passfail}) we measure the average weighted ellipticity components $\langle \epsilon_{1,2} \rangle$.  For a KiDS-size survey, in the absence of systematic error, these should be consistent with zero.  As with the analysis of CFHTLenS \citepalias{heymans/etal:2012}, we find a small residual shear signal in KiDS at the level of $\sim 10^{-3}$ (shown in Fig.~\ref{fig:c_corr_before_after}). The dependence on galaxy size and SNR is different though. In CFHTLenS small, high-SNR galaxies were found to be the dominant source of the residual signal in $\langle \epsilon_2 \rangle$ whereas $\langle \epsilon_1 \rangle$ was consistent with zero: instead, for KiDS we find that the lowest SNR galaxies dominate the residual, which is stronger in $\langle \epsilon_1 \rangle$. In addition we see a strong dependence of $\langle \epsilon_1 \rangle$ on the Strehl ratio (defined here as the fraction of light in the PSF model that falls into the central pixel), which could be a sign of error due to undersampling of the PSF. Indeed, with typical pixel-to-seeing ratio of 0.25 for CFHTLenS and 0.3 for KiDS, we expect KiDS to be more prone to such errors. Future analyses of KiDS will therefore include a PSF modelling method that correctly accounts for the under-sampling (Miller et al. in prep).  For this first release, however, we follow the CFHTLenS strategy of calibrating and removing this small systematic effect empirically. Note that the first lensing analyses are based on tangential shear averages and are therefore not affected by such additive errors as long as the analysis is not affected by the survey boundaries: for the current data set we see no sign of additive effects out to projected radii of $2h^{-1}\rmn{Mpc}$ \citep{viola/etal:2015}. 

\begin{figure}
\putfig{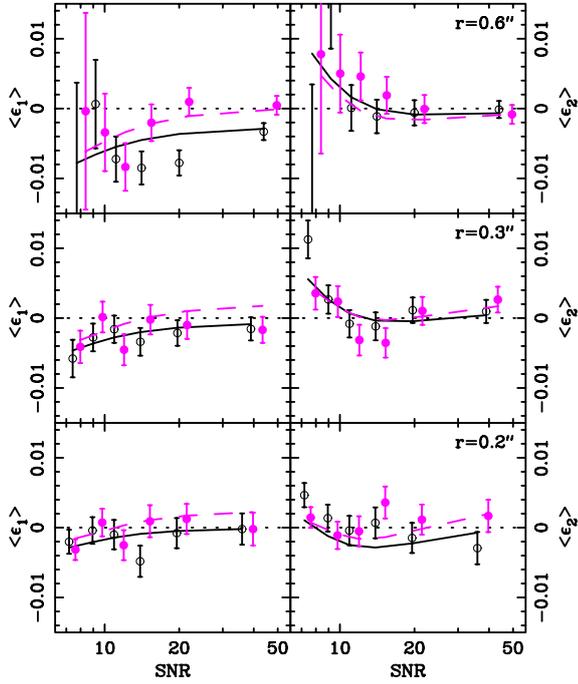}
\caption{The measured dependence of $\langle \epsilon_1 \rangle$ (left) and $\langle \epsilon_2 \rangle$ (right) as a function of SNR, for three different size bins (panels upper to lower $r = 0.6\arcsec, 0.3\arcsec, 0.2\arcsec$), and two different Strehl ratio bins with Strehl $=0.05$ (open symbols)) and  Strehl $=0.1$ (closed symbols). The corresponding best-fitting models are shown as solid (Strehl = $0.05$) and dashed (Strehl $=0.1$) lines.  }
\label{fig:c_corr}
\end{figure}

Using all the data that passed the field selection in \S\ref{sec:passfail}, we bin the data in three dimensions with six bins in size and SNR, and three bins in Strehl ratio, and fit a 3D second-order polynomial model to the bins\footnote{For our first set of galaxy-galaxy lensing papers, an earlier version of the additive correction was applied that used a third-order polynomial fit to a 3D binning with ten bins on each axis. On further inspection this sub-optimal set-up was discovered to introduce a low level of spurious noise into the shape measurement. As the shear-position correlations were found to be insensitive to the additive correction we only updated the additive calibration for the cosmological analysis demonstration in this paper.}.  Fig.~\ref{fig:c_corr} presents example slices from the data cube and the model fit.
Applying the $c$-correction to the shear catalogue changes the one-point statistics $\left\langle (\epsilon_1,\epsilon_2)\right\rangle$ from $(-0.0015,-0.0002)$ to $(0.0004,0.0004)$, with a 1-$\sigma$ uncertainty of 0.0003.  This is sufficiently small that it will not impact the measurement of the two-point shear correlation function presented in \S~\ref{sec:cosmicshear}.  This level of residual shear will however impact future degree-scale cosmological shear measurements, requiring improvements in the calibration scheme for future data releases.

%

\section{Cosmic shear measurement}
\label{sec:cosmicshear}
The measurement of weak gravitational lensing by large-scale structure, often referred to as `cosmic shear', has the ability to set tight constraints on both standard cosmological parameters \citep[see for example][and references therein]{heymans/etal:2013}, and a range of modified gravity scenarios \citep{simpson/etal:2013,planckXIV:2015}. While the amount of data analysed in this paper represents less than 10 percent of the final KiDS area, in this section we argue that the data quality is at the level that the full survey will indeed provide high-fidelity cosmic shear measurements. It also provides a practical demonstration of our blinding scheme, which has been designed to counter user confirmation bias in future KiDS cosmic shear analyses.

\subsection{Blinding the KiDS weak lensing catalogues}
\label{sec:blinding}
In the post-Planck precision cosmology era, one the challenges facing new cosmological observations is confirmation bias \citep[e.g.,][]{croft/dailey:2011}.  Many new surveys are therefore following the approach, particularly favoured by the particle physics community, of performing a `blind' analysis.    The first stage of such an analysis is the verification and validation of software packages through the analysis of mock simulated data.  The KiDS $N$-body simulations span 30,000 square degrees with a WMAP9 cosmology \citep{hinshaw/etal:2013}, and are an extension of the suite of lensing simulations described in \citet{Harnois-Deraps/VanWaerbeke:2015}.  
With these simulations we can verify the analysis methods for galaxy-galaxy lensing, galaxy-cluster lensing and tomographic cosmology, and also determine covariance matrices for the analysis of the data. 

This mock data strategy does not prevent confirmation bias in the analysis of the real data, where potentially unknown sources of systematic error increase the complexity of the analysis. For example, choices are currently made about which sub-exposures or pointings to excise from the analysis based on the outcome of a range of systematic tests on the shear measured in these regions. Choices are also made as to which length scales to include in the analysis of correlation functions or power spectra, which binning to use, and which photometric redshift ranges to trust. It is therefore important to build blinding into our data analysis such that these choices are informed purely through scientific rationale, and not influenced by the results of independent experiments.

An example of an early blind cosmological data analysis is \citet{davis/etal:2007} where the analysis team was given supernova data in which the redshifts had been stretched. This strategy of manipulating the data with a small multiplicative perturbation has also been used by other groups, but has the drawback that when the data are finally unblinded, the analysis has to be re-run. This potentially allows for low-level adjustments in the re-analysis, for example choosing which scales to include. We have therefore designed an alternative blinding scheme that prevents this, by ensuring that the true data are analysed along with the perturbed versions.

All KiDS weak lensing catalogues analysed contain four sets of ellipticity data: the true data, and three versions that have been manipulated by an unknown amount. Specifically, the magnitudes of the ellipticities in column $A=1,2,3,4$ of the catalogues are `curved' with a function 
\be
\epsilon_{\rmn A} = \epsilon \left(\rmn{e}^{k_\rmn{A}[1-(\epsilon/\epsilon_\rmn{max})^2]^2}\right)
\label{eq:blind}
\ee
parametrized by a single value $k_\rmn{A}$ such that $\epsilon_\rmn{max}$, the maximum ellipticity in the catalogues, is left invariant under this remapping. The values $\lbrace k_\rmn{A}\rbrace$ are unknown, except that for one of them, the true data, $k_\rmn{A}$ is equal to zero. The differences between the $k_\rmn{A}$ can easily be reconstructed by dividing the shear columns, but this provides no information as to which column contains the true ellipticities. The values of $k_\rmn{A}$ were limited to $|k_\rmn{A}|<0.2$, in order to satisfy two conditions. On the one hand, the effect of the transformation should be sufficiently large that it effectively blinds KiDS to confirmation bias with CMB measurements from Planck, by changing the results up to $\sim10\sigma$ in terms of the Planck error on the amplitude of the matter density power spectrum \citep{planckXIII:2015}. At the same time Eq.~\ref{eq:blind} must not distort the lensing values to such an extent that it creates suspicious effects in galaxy-galaxy lensing, ellipticity distributions, SNR or redshift scaling. 
We asked a trusted colleague, external to the team, to set the values of $k_\rmn{A}$ through a \textsc{Python} executable that takes the original lensing catalogues output from \emph{lens}fit (see \S\ref{sec:shapes}), manipulates the ellipticity columns, according to Eq.~\ref{eq:blind}, and outputs a new catalogue with the additional blind columns inserted in an order unknown to any member of the KiDS team.  

The team members agreed that they would not wilfully unblind themselves by attempting to back-track the data manipulation to discover which column contains the original data.  All analyses are carried out on all four sets of columns, including systematics tests, empirical corrections and covariance matrix estimation.  Different fields may pass or fail the systematics tests in different blinded columns and this has been taken into account in the final analysis. Even though this setup incurred a factor of four increase in the computational analysis time, we felt this was a necessary step to make, whilst also encouraging the good practice of creating, verifying and validating `press-of-the-button' end-to-end analysis scripts. In order to allow for phased unblinding, team members add an additional individual layer of blinding by not labelling their results with the blinded column number used.  Pre-publication, our results were sent to our external who provided the blinding key, and can verify that the results presented in both this paper, and our first scientific analyses \citep{viola/etal:2015, sifon/etal:2015} were not changed after it was revealed to the authors which column contained the true shear.  We show an example of the blinding scheme in action in the next section, where we present the cosmic shear measurement from the four blinded shear measurements. 

Thus far our blinding is limited to the shape measurements only and future blinding will also include manipulation of galaxy weights and potentially photometric redshifts, stellar masses and galaxy luminosities.  As our first analysis covers less than 10 per cent of the final KiDS area, the blinding described here had only a small effect on the early science results presented in the accompanying \citet{viola/etal:2015} and \citet{sifon/etal:2015} papers. We agreed however, that it was important to implement this blinding scheme from the beginning, in order to learn from this `dry run' in preparation for the future larger-area KiDS cosmological analyses.  

\subsection{Second order weak lensing statistics}
To detect weak lensing by large-scale structures and extract cosmological parameter constraints and information about systematics from the data, a wide
range of different two-point statistics have been proposed \citep[see][for a comprehensive discussion of the relationship between these statistics]{SchvWKM02,COSEBIS}.  These real-space statistics all derive from the observed angular two-point correlation function $\hat{\xi}_{\pm}$ which can be estimated from the data as follows:
\be
\hat{\xi}_{\pm}(\theta) = \frac{\sum_\theta w_a w_b \left[ \epsilon_\rmn{t} (\x_a) \epsilon_\rmn{t} (\x_b) \, \pm \, \epsilon_\times (\x_a) \epsilon_\times (\x_b)
\right]}{
\sum_\theta w_a w_b } \, .
\label{eqn:xipm_est}
\ee
Using inverse variance weights $w$, the sum is taken over pairs of galaxies with angular separation $|\x_a - \x_b|=\theta \pm \Delta \theta /2 $, where $\Delta \theta$ is the width of the bin\footnote{Note that the final reported angular scale of the bin should not be the mid-point of angular range selected, but the weighted average separation of the galaxy pairs in that bin.}.   The tangential and cross components of the ellipticities $\epsilon_{\rmn{t},\times}$ are measured with respect to the vector joining each pair of correlated objects \citep{bartelmann/schneider:2001}.     

Weak gravitational lensing produces curl-free gradient distortions (E-mode), and contributes only to the curl distortions (B-mode) at small angular scales, $\theta < 1$ arcmin, mainly due to source redshift clustering \citep{schneider/etal:2002}.  Decomposing the weak lensing signal into E and B modes therefore provides a method with which to gauge the contribution to the overall shear correlation signal from non-lensing sources.  These could arise from residual systematics in the shape measurement method, or from the intrinsic alignment of nearby galaxies \citep[see][and references therein]{troxel/ishak:2015}.  

\citet{Crittenden/etal:2002} show that the shear correlation functions, estimated in Eq.~\ref{eqn:xipm_est}, can be decomposed
into the E- and B-type correlators 
\be
\xi_\rmn{E}(\theta)=\frac{\xi_+(\theta)+\xi'(\theta)}{2}
\qquad\hbox{and}\qquad
\xi_\rmn{B}(\theta)=\frac{\xi_+(\theta)-\xi'(\theta)}{2}
\, ,
\label{eqn:xieb}
\ee
where
\be
\xi'(\theta)=\xi_-(\theta)+4\int_\theta^\infty \frac{\d\vartheta}{\vartheta} \xi_-(\vartheta)
        -12\theta^2 \int_\theta^\infty \frac{\d\vartheta}{\vartheta^3}\xi_-(\vartheta)\, .
\label{eqn:xipr}
\ee
The measured E-mode $\xi_\rmn{E}(\theta)$ is related to the underlying non-linear matter power spectrum $P_\delta$ that we wish to probe, via
\be
\xi_\pm(\theta) = \frac{1}{2\pi}\int \d\ell \,\ell \,P_\kappa(\ell) \, J_{0,4}(\ell \theta) \, , 
\label{eqn:xiGG}
\ee
where $J_{0,4} (\ell \theta)$ is the zeroth (for $\xi_+$) or fourth (for $\xi_- $) order Bessel function of the first kind. $P_\kappa(\ell)$ is the convergence power spectrum at angular wave number $\ell$ 
\be 
P_\kappa(\ell) = \int_0^{w_{\rm H}} \d w \, 
\frac{q(w)^2}{a(w)^2} \, P_\delta \left( \frac{\ell}{f_K(w)},w \right),
\label{eqn:Pkappa} 
\ee
where $a(w)$ is the dimensionless scale factor corresponding to the comoving radial distance $w$, and $w_H$ is the horizon distance.  The lensing efficiency function $q(w)$ is given by
\be
q(w) = \frac{3 H_0^2 \Omega_{\rm m}}{2c^2} \int_w^{w_{\rm H}}\, \d w'\ n(w') 
\frac{f_K(w'-w)}{f_K(w')}, 
\label{eqn:qk} 
\ee
where $n(w)\d w$ is the effective number of galaxies in $\d w$, normalized so that $\int n(w)\d w = 1$. $f_K(w)$ is the angular diameter distance out to comoving radial distance $w$, 
$H_0$ is the Hubble parameter and $\Omega_\rmn{m}$ the matter density parameter at $z=0$.  For more details see \citet{bartelmann/schneider:2001} and references therein.

\begin{figure}
\putfig{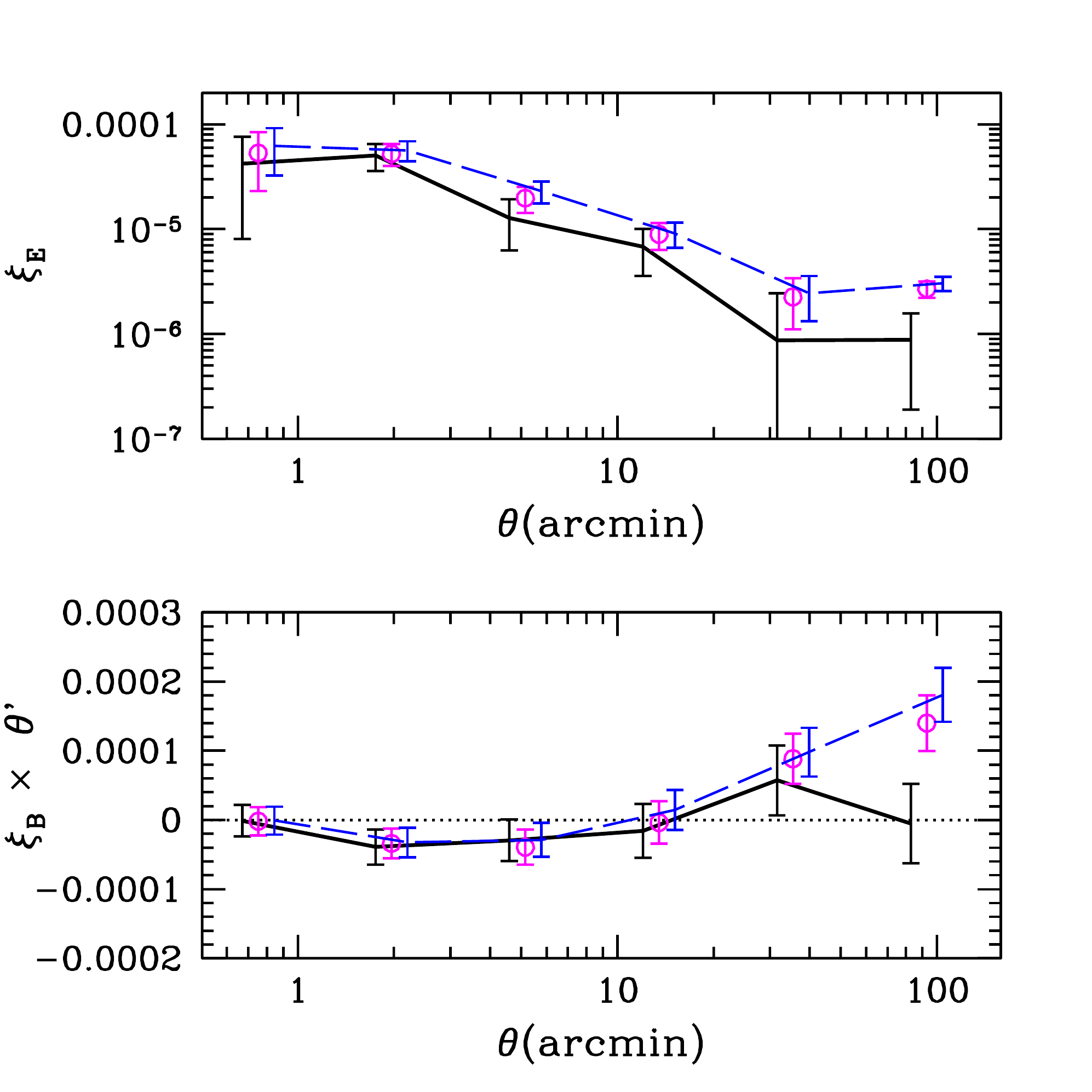}
\caption{Comparison of the E-type (upper) and B-type (lower) shear correlation functions measured using all the data (dashed); after the application of the field selection (open points); and after the application of both field selection and the additive calibration correction (solid).  Without these two corrections the B-mode, which is an indicator of non-lensing systematic errors, becomes significantly non-zero on large scales. Note that the B-mode vertical axis has been multiplied by $\theta$ (in arcminutes) in order to emphasize the differences from a zero signal.}
\label{fig:xi_EB}
\end{figure}

\begin{figure}
\putfig{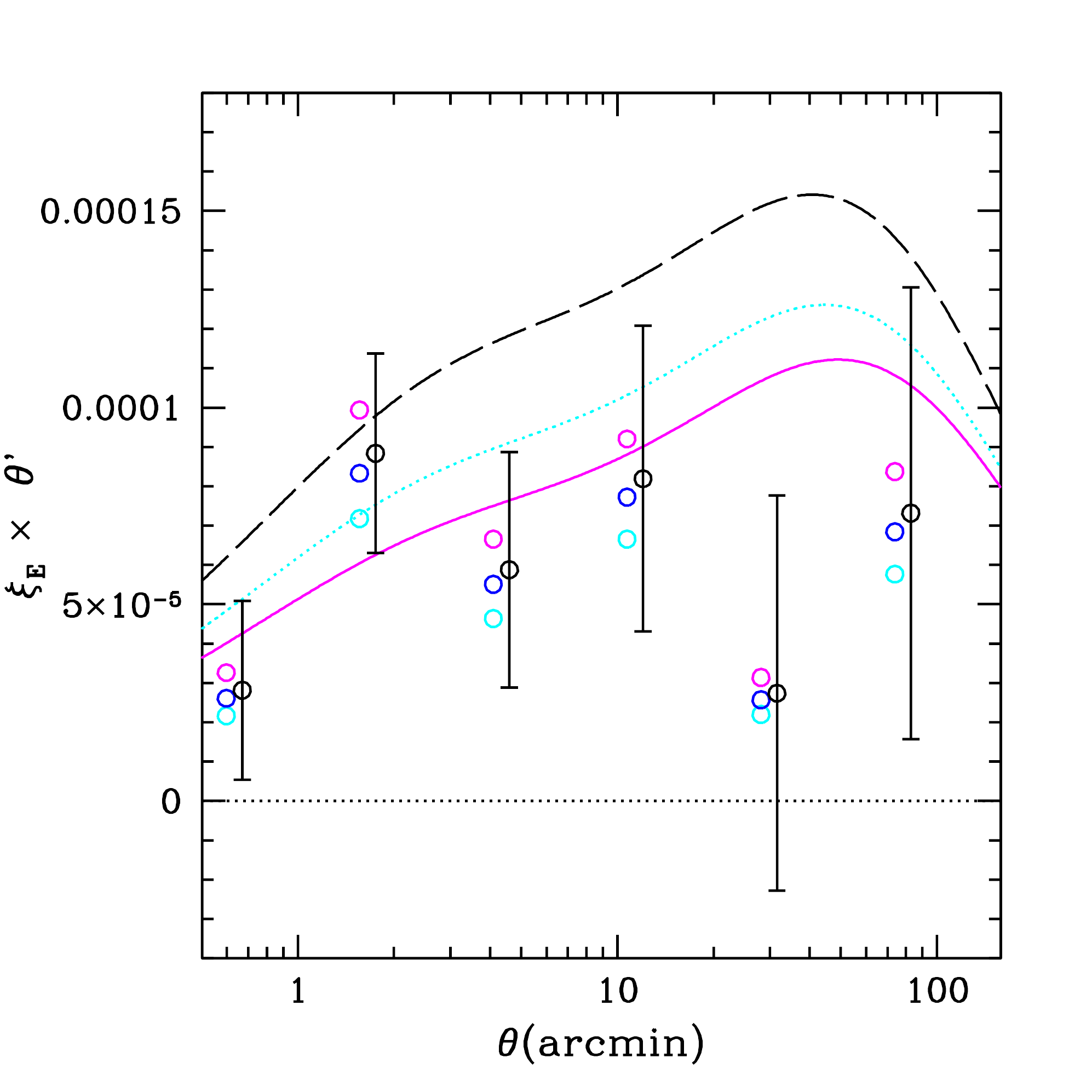}
\caption{The E-type shear correlation functions from the 105 tiles of KiDS data that pass the PSF systematics tests in \S\ref{sec:passfail}.  Measurements from all four blindings are shown, with the true shear measurement indicated with an error bar.  For comparison the E-mode signal expected from three different $\Lambda$CDM cosmological models are shown; Planck cosmology using the TT spectra (dashed), and EE spectra (dotted) along with the best-fit CFHTLenS result (solid).  Note that the vertical axis has been multiplied by $\theta$ (in arcminutes) in order to improve the visualisation by enhancing the differences.}
\label{fig:xi_E_theory}
\end{figure}

\subsection{KiDS shear correlation data and survey parameters}
\label{sec:shear_corr}
Fig.~\ref{fig:xi_EB} presents the derived E- and B-type shear correlation functions, from Eq.~\ref{eqn:xieb}.  These were calculated following the method in \citet{Pen/etal:2002}, using 4000 finely binned measurements of the shear correlation function $\xi_{\pm}(\theta)$ spanning $9\arcsec< \theta<4^\circ$ in equal bins of $\log \theta$.  As our data extend over many degrees, but not to infinity, we use a fiducial cosmological model to determine the integrand in Eq.~\ref{eqn:xipr}, splitting the
integrals into two.  The first is calculated from the observations directly, extending from $\theta$ to $\theta_{\rm max}$ where  $\theta_{\rm max}= 4^\circ$.  The second extends from $\theta_{\rm max}$ to $\infty$ and is calculated by inserting $\xi_-(\theta)$ calculated from Eq.~\ref{eqn:xiGG} assuming the KiDS redshift distribution and the best-fit Planck cosmology \citep{planckXIII:2015}. 
This model dependent part of the integrand sums to $\sim 10^{-7}$ for the three cosmological models that are compared in Fig.~\ref{fig:xi_E_theory}.  This model dependence prevents cosmological parameter estimation directly from the E-mode signal.  The analysis is still a valid diagnostic test for residual systematics, however, as the model-dependent addition to Eq.~\ref{eqn:xipr} is less than 10 percent of the total signal on the largest angular scales probed. The errors are estimated following \citet{Pen/etal:2002}, treating each noisy finely binned raw shear correlation measurement as uncorrelated with the others.  We then propagate these uncorrelated errors through to a final correlated error on the coarsely binned E- and B-type shear correlation functions.  This approximation is sufficient for this diagnostic test as the current KiDS area is relatively small such that for the majority of scales the data are shot-noise dominated.

Focussing first on the measured E mode presented in the upper panel of Fig.~\ref{fig:xi_EB}, the small effect of removing the 4 percent of fields that failed the selection stage (\S\ref{sec:passfail}) can be seen, as well as the result of subsequent application of the additive calibration correction (\S\ref{sec:addcalcor}). The impact of this two-step calibration can also be seen in the B-mode signal (lower panel), which is consistent with zero on all scales, demonstrating excellent control of systematic errors in shape measurement with KiDS.  Without the field selection or additive ellipticity corrections, however, we find a significant B-mode signal on scales $\theta>10'$. In preparation for future releases we are currently implementing a number of improvements in both the data reduction pipeline, PSF modelling and shape measurement analysis which are designed to reduce the significance of the calibration corrections on our analysis.

To illustrate how the implemented blinding scheme modified the results, Fig.~\ref{fig:xi_E_theory} compares the E-mode measured from all four blindings, with the true shear measurement indicated as the data point with Poisson error bars.  For comparison the E-mode signal expected from a range of $\Lambda$CDM cosmological models are also shown, using the effective weighted redshift distribution shown in Fig.~\ref{fig:nofz}, as estimated from the weighted sum of the photometric redshift probability distributions $p(z)$.   The three cosmological models use the Planck results from table 3 of \citet{planckXIII:2015} showing the difference between the cosmology fit to the TT spectra (dashed) and EE spectra (dotted) along with the best-fit CFHTLenS result (solid) from \citet{kilbinger/etal:2013}. 

\section{Conclusions}
\label{sec:conclude}

In this paper we present the first lensing analysis of the Kilo-Degree Survey (KiDS) data obtained at the VLT Survey Telescope (VST) at ESO's Paranal Observatory. KiDS is a multi-band survey specifically designed for weak lensing tomography, that takes advantage of the very good image quality at the VST. A particular advantage of the VST, where the camera operates at an f/5 Cassegrain focus, compared to much faster wide-field prime-focus cameras, is the simplicity and generally low amplitude of the ellipticity pattern, as well as the uniformity of the size of the point spread function (PSF) over the full field of view.

The KiDS lensing analysis draws heavily on heritage from the CFHTLenS project \citep{heymans/etal:2012}, in particular in the use of \textsc{Theli} \citep{erben/etal:2013} and \emph{lens}fit \citep{miller/etal:2013} for measuring galaxy shapes (\S\ref{sec:shapes}), and \textsc{bpz} \citep{benitez:2000} for photometric redshifts \citep{hildebrandt/etal:2012}. As input for the photometric redshifts, aperture-matched colours are derived from PSF Gaussianization of the public data release of the \textsc{Astro-WISE} reduction of the KiDS images \citep{dejong/etal:2015}, and subsequent Gaussian Aperture and PSF (\textsc{GAaP}) photometry. This procedure, which was developed specifically for KiDS, is described in detail in \S\ref{sec:photom} and Appendix~\ref{app:gaap}. The resulting shear/photometric redshift catalogues are available to the community (Appendix~\ref{app:data}), and form the basis of three companion scientific analyses \citealt{sifon/etal:2015}; \citealt{viola/etal:2015}; van Uitert et al., in preparation) that exploit the overlap of these data with the GAMA spectroscopic survey \citep{driver/etal:2011}. The KiDS lensing catalogues contain 8.88 galaxies per square arcminute with non-zero lensing weight, cover an unmasked area of 75 square degrees, and provide an inverse shear variance of 69 per square arcminute. The median redshift of the summed posterior photometric redshift distributions of the galaxies, accounting for the \emph{lens}fit weight, is 0.53.

Considerable attention was paid to quantifying and correcting the lensing estimates for additive and multiplicative bias. In order to validate the galaxy ellipticities, we carried out extensive tests (\S\ref{sec:sys}). All indications are that the data are indeed `lensing-quality.' For example, the degree of star-galaxy shape correlation in the KiDS data is essentially consistent with the expectations from realistic simulated cosmic shear fields, with just 4 percent of the tiles falling outside expected parameter ranges, and the amplitude of galaxy-galaxy lensing around magnitude-limited foreground lenses scales in the same way as it did in CFHTLenS even though the depths of the surveys differ. Taking advantage of the GAMA overlap, we also tested the way the tangential shear around galaxies at known (spectroscopic) redshift scales with the (photometric) redshift of the sources. Also here we recover the expected dependence, which gives us confidence in both the photometric redshifts and the shears we measure.

Finally, in \S\ref{sec:cosmicshear} we present a first measurement of the cosmic shear correlation function from these data. Though admittedly still noisy, the results are consistent with previous measurements, and show negligible B-mode signal, demonstrating the high fidelity of the KiDS lensing data.

KiDS observations continue at the VST, and as the area of the survey grows more refined cosmological lensing measurements will follow.

\section*{Acknowledgments}

We are grateful to Matthias Bartelmann for being our external blinder, revealing which of the four catalogues analysed was the true unblinded catalogue at the end of this study, to Giovanni Covone and Mattia Vaccari for providing the VOICE data, to all the members of the KiDS weak lensing team who supported this work, and to the GAMA team for their spectroscopic catalogues. 
We also thank Mike Jarvis and Martin Kilbinger for \textsc{corr2} and \textsc{athena}, the correlation function measurement software used in this analysis.  We acknowledge support from the European Research Council under FP7 grant number 279396 (MV,MC,CS,RH,ME,H.Ho) and 240185 (AC and CH). EvU acknowledges support from an STFC Ernest Rutherford Research Grant, grant reference ST/L00285X/1. RN and EvU acknowledge support from the German Federal Ministry for Economic Affairs and Energy (BMWi) provided via DLR under project no.50QE1103. HHi is supported by the DFG Emmy Noether grant Hi 1495/2-1. 
JHD  and LvW are funded by the NSERC of Canada, and LvW by CIfAR. TDK is supported by a Royal Society URF.
CB acknowledges the support of the Australian Research Council through the award of a Future Fellowship.  This work is supported by the Netherlands Organisation for Scientific Research (NWO) through grants 614.001.103 and 614.061.610, by the Dutch Research School for Astronomy (NOVA), and  by the Deutsche Forschungsgemeinschaft in the framework of the TR33 'The Dark Universe'.
Based on data products from observations made with ESO Telescopes at the La Silla Paranal Observatory
under programme IDs 177.A-3016, 177.A-3017 and 177.A-3018, and on data products produced by Target/OmegaCEN, INAF-OACN, INAF-OAPD and the KiDS production team, on behalf of the KiDS consortium.

{\small \textit{Author Contributions:} All authors contributed to the development and writing of this paper. The authorship list is given in three groups: the lead authors (KK, CH, HHi, RN, TE, JdJ, MV), followed by two alphabetical groups. The first alphabetical group includes those who are key contributors to both the scientific analysis and the data products. The second group covers those who have either made a significant contribution to the data products, or to the scientific analysis.}

\bibliographystyle{mnras}
\bibliography{KiDS_technical}

\appendix

\section[PSF Gaussianization and GAaP photometry]
{PSF Gaussianization and GA{\sevensize A}P photometry}
\label{app:gaap}

In this Appendix we describe the processing steps involved in the PSF Gaussianization, and the subsequent Gaussian-aperture photometry. The aim is to obtain multi-colour photometry that is insensitive to difference in PSF between the different bands, yet optimizes SNR by down-weighting the outer parts of the images. It is important to stress that the resulting fluxes \emph{do not} provide total magnitudes of the sources; they are mainly intended for consistent estimation of the spectral energy distribution of the high-SNR parts of a source.

The procedures presented here build on the ideas presented in \citet{kuijken:2008}, but differ in that here (i) an explicit pixel-space PSF Gaussianization convolution is performed, and (ii) the aperture shape and size can be specified independently of the PSF size.

\subsection{PSF Gaussianization}
We start from a co-added image of a KiDS tile, observed with a particular filter, and construct a convolution kernel that modifies the PSF everywhere so as to make it a circularly symmetric Gaussian. The first step is therefore to model the PSF from the many star images that are found on every KiDS tile.

To describe the PSF we model the stars with shapelet expansions following \citet{refregier:2003}, and use these to construct and apply a suitable spatially varying convolution kernel. Shapelets are Gaussians multiplied by polynomials and form elementary, compact, orthonormal 2D functions which can be used to fit an image to arbitrary precision. The shapelet with scale radius $\beta$ and Cartesian orders $(a,b)$ is (for $a,b=0,1,2,\ldots$)
\be
S_{ab}^\beta(x,y)=\frac{H_a(x/\beta)H_b(y/\beta) } {\beta\sqrt{2^{a+b}\pi a! b!}}  {\rmn e}^{-(x^2+y^2)/2\beta^2} \, ,
\ee
where $H_a(x)$ is a Hermite polynomial, familiar from the eigenstates of the quantum harmonic oscillator. Many useful properties of shapelets, such as their behaviour under infinitesimal translation, rotation, magnification, shear, and convolution can be derived \citep{refregier:2003}, and are used below.  We use the implementation of shapelets of \cite{kuijken:2006}.

Among many applications, shapelet decompositions have been used to characterize the galaxy populations (e.g., \citealt{kelly/mckay:2004}; \citealt{melchior/meneghetti/bartelmann:2007}), to measure weak lensing shear (e.g., \citealt{refregier/bacon:2003}; \citealt{kuijken:2006}) and flexion (e.g., \citealt{bacon/etal:2006,massey/etal:2007,velander/etal:2011}) and for PSF-corrected photometry \citep{kuijken:2008}.  
Even though the shapelets are formulated in Cartesian coordinates, truncating the expansion at maximum combined order $N=a+b$ results in an orientation-invariant subspace of possible shapes that can be described. Such truncated shapelet expansions are most effective at modelling structure at radii between $\beta/\sqrt{N+1}$ and $\beta\times\sqrt{N+1}$. At large radii they asymptotically approach a Gaussian.

Because of the orthonormality of the elementary shapelets, any source image $I(x,y)$ can be described as a sum $\sum_{ab} s_{ab} S_{ab}^\beta(x,y)$ with coefficients
\be 
s_{ab}=\int\,\d x\,\d y\, I(x,y) S_{ab}^\beta(x,y).
\ee 
Since our data are pixellated we do not use this integral relation, but rather make a least-squares fit of a truncated shapelet model to the image of our source. We fit all pixels within a radius $(4+\sqrt{N})\beta$ of the centre of the source. Typically we truncate the series at $N=10$.
 
Using this formalism, our `PSF Gaussianization' procedure, which we apply to each survey tile and filter separately, is as follows.

\begin{figure*}
\putfig{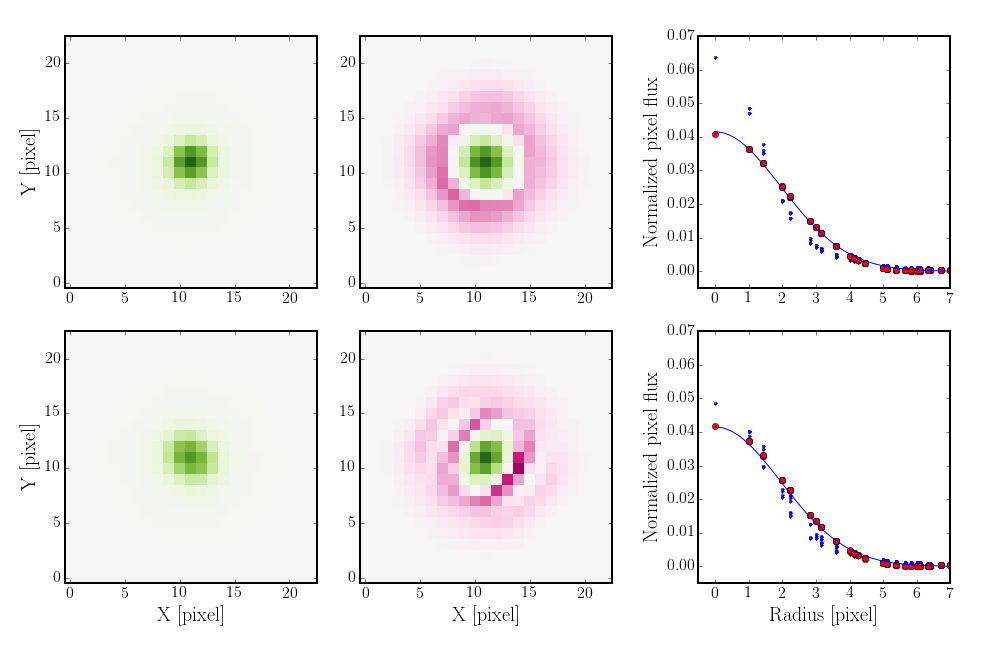}
\caption{Illustration of the PSF Gaussianization and homogenization, for a KiDS field with significant PSF variation. The PSF model is shown on the left, the Gaussianization kernel in the middle, and a comparison of the radial profiles of the resulting convolved PSF (red circles), the target Gaussian (line) and the original PSF (blue dots) on the right. The top row of plots show the PSF in the centre of the field, where it is rather round; the second row shows a much more elliptical, and wider, PSF in one of the corners. The colour scale is the same for each PSF, running from zero (white) to the peak value in the top panel (dark green). Each kernel has been normalized separately to its peak value. Green pixels are positive; the red scale runs from 0 to $-5$ percent of the peak, to highlight the mildly negative regions of the kernel.}
\label{fig:gpsf}
\end{figure*}

\subsubsection{Fit shapelet models to all stars}
First we identify high-SNR unsaturated stars in the images, using the traditional flux vs. radius plot \citep{kaiser/squires/broadhurst:1995} obtained from a \textsc{SExtractor} run on the individual sub-exposures. Typically several thousand such stars can be found per tile. For each CCD the PSF size (FWHM) is averaged over the four or five sub-exposures through a given filter, and these 32 values are fitted with a second-order 2-D polynomial to give a rough map of the average seeing for the tile. The scale radius $\beta_{\rm p}$ for the shapelet model for the PSF is then fixed at (1.15/2.35)$\times$ the largest FWHM value in this map (this choice makes the FWHM of the $S_{00}$ shapelet 15 percent wider than the PSF, enabling the higher-order terms to fit the inner structure of the PSF as well as the wings. It will also be the size of the target Gaussian PSF, see below).

Once $\beta_{\rm p}$ is chosen for a particular tile and filter, we determine shapelet parameters up to order $a+b=10$ for each star using least-squares fitting (66 coefficients per star). This truncation order is set by the pixel size, which is typically about one-third of the scale radius.
All shapelet models use Cartesian pixel coordinates $\x=(x,y)$ with respect to a centre position $\bxi=(\xi,\eta)$ of the star: we define this centre as the position for which $s_{10}=s_{01}=0$ and determine it iteratively.

\subsubsection{Interpolate the PSF model}
The PSF models are then interpolated across the image by means of 4th-order polynomial fits of each shapelet coefficient versus $\xi$ and $\eta$ position (15 spatial variation coefficients per shapelet term). In this step outliers are rejected iteratively, resulting in a smooth model of the PSF variation across the image. The PSF model for a particular co-added image is thus described as a linear combination of $15\times66=990$ terms $s_{ab;kl}$:
\be
P(x,y,\xi,\eta)=
\sum_{a=0;b=0}^{a+b\le10} s_{ab} (\xi,\eta) S_{ab}^{\beta_{\rm p}}(x,y)\, ,
\ee
where
\be
s_{ab}(\xi,\eta)=\sum_{k=0;l=0}^{k+l\le4} s_{ab;kl} (\xi-\xi_0)^k (\eta-\eta_0)^l 
\label{eq:shmap}
\ee
and $(\xi_0,\eta_0)$ is the centre of the image.

\subsubsection{Construct the Gaussianization kernel}
\label{sec:kernel}
Once we have a map of the PSF, the next step is to construct a convolution kernel that renders the PSF Gaussian. Also here the shapelet formalism is convenient, since shapelets behave nicely under convolution. \citet{refregier:2003} provides recurrence relations for calculating the matrix elements $C_{nlm}^{\beta_{\rm o}\beta_1\beta_2}$ that express the convolution of two basis functions as a new shapelet series of arbitrary scale radius $\beta_{\rm o}$:
\be
S_{ab}^{\beta_1}(x,y) \otimes S_{cd}^{\beta_2}(x,y) = 
\sum_{e,f=0}^\infty C_{eac}^{\beta_{\rm o}\beta_1\beta_2} C_{fbd}^{\beta_{\rm o}\beta_1\beta_2} S_{ef}^{\beta_{\rm o}}(x,y) .
\ee
The convolution of an image with shapelet coefficients $s_{ab}$ and scale radius $\beta$ with a PSF that has shapelet coefficients $p_{ab}$ and scale radius $\beta_{\rm p}$ can then be written as a new shapelet, with scale radius $\beta_{\rm o}$ and coefficients
\be
(p\otimes s)_{mn}=\sum_{ab} \!\left(\sum_{cd} 
C_{mac}^{\beta_{\rm o}\beta\beta_{\rm p}} 
C_{nbd}^{\beta_{\rm o}\beta\beta_{\rm p}} 
       p_{cd} \right) \!
s_{ab}
\equiv
\sum_{ab} P_{mnab} s_{ab},\ ,
\label{eq:psfmat}
\ee
where the expression in brackets, the `PSF matrix' $P$, gives the linear transformation from pre- to post-convolved source coefficients. Note that the sum in Eq.~\ref{eq:psfmat} runs over an infinite number of terms, so that the series needs to be truncated in practice. 

Given a PSF, the corresponding Gaussianization kernel is a shapelet which, when convolved with the PSF, gives the target Gaussian (for which $s_{00}=(2\sqrt\pi\beta_{\rm p})^{-1}$ and all other coefficients are zero): i.e., we seek those shapelet coefficients $k_{ab}$ for which
\be
\sum_{ab}^{a+b\le N_k}P_{mnab} k_{ab} -\frac{\delta_{mn}}{2\sqrt\pi\beta_{\rm p}}=0 \qquad\hbox{for}\quad m+n\le N.
\ee
In order to regularise the calculation of the kernel we truncate $k_{ab}$ at shapelet order $N_k= 8$, whereas the PSF is modelled to order $N=10$. This over-constrained set of equations for the $k_{ab}$ is then solved by minimizing the residuals in the least-squares sense. Kernels are constructed in this way for 121 $(\xi,\eta)$ positions on an 11$\times$11 grid covering the image, explicitly normalized to unit integral, and their coefficients' variation across the image fitted with a 5th order polynomial $K(x,y;\xi,\eta)$ (see Eq.~\ref{eq:shmap}). This spatially varying kernel map is now ready for convolution with the co-added image. 

We set the dispersion of the target Gaussian PSF to be equal to the scale radius $\beta_{\rm p}$ of the shapelet expansion of the PSF. We have found that this choice preserves the seeing of the image, while avoiding significant deconvolution (and the associated noise amplification) anywhere on the image.  As a final check, we make sure that the peak of this target PSF is everywhere lower than 90 percent of the central height of the original PSF, and if it is not, we increase the value of $\beta_{\rm p}$ until it is.
An example of the PSF model and the effect of the corresponding kernel is shown in Fig.~\ref{fig:gpsf}.

Note that this smooth model for the PSF (and hence also the kernel) variation across the co-added image does not allow for discrete jumps in the PSF, such as might be caused by seeing variations between the dithered sub-exposures, or misalignments of adjacent chips. While we have no indications that such effects adversely impact the quality of the photometric redshifts, in future releases we plan to construct separate Gaussianization kernels for each sub-exposure, and to apply these before co-addition.

\subsubsection{Image convolution}

Convolving the co-added images by the kernel is mathematically most convenient in Fourier space, especially because of the shapelet formulation: shapelets have simple Fourier transforms which can straightforwardly be multiplied into the Fourier transform of the image. Since the kernel is compact we
can speed up the convolution by splitting the image into smaller segments and processing these separately. In practice we split the (over 18000$\times$18000 pixel) co-added images into segments of 512$\times$512 pixels, which appears to be a reasonable optimum on most machines. This step can also be parallelised. Each segment $I(x,y)$ is background-subtracted, edge-tapered to zero and padded around the edges, with the width $W$ of the edge set to four times $\beta_{\rm p}$ (see \S\ref{sec:kernel}). This preprocessed segment is then convolved with the kernels $K_{\rm bl}, K_{\rm tr}$, etc., at the bottom-left, top-right, etc., corners of the segments, and the four convolutions are averaged as
\begin{eqnarray}
G(x,y) &=&(1-t)(1-u)(I\otimes K_{\rm bl}) + (1-t)u(I\otimes K_{\rm tl})
\nonumber\\
&& + t(1-u)(I\otimes K_{\rm br}) +tu(I\otimes K_{\rm tr}) \, ,
\end{eqnarray}
effectively convolving $I$ with a bilinear interpolation of the kernel map. (Here $t$ and $u$ run from 0 to 1 along the $x$ and $y$ axes of the segment, respectively). The edge of the resulting convolved segment is then trimmed by a width $2W$, the background is added back in, and the pixels are pasted into the final output convolved image.

To prevent numerical issues, pixels that deviate greatly from their neighbours are clipped before the image segment is Fourier transformed and convolved; copies of the kernel of appropriate amplitude are instead added into the convolved image at the locations of the clipped pixels.

\subsubsection{Noise propagation: noise auto-correlation function}
\label{sec:noisecov}

Convolution of an image correlates pixel noise, and this needs to be taken into account in measurements based on the Gaussianized co-adds. If $G(\bxi)$ is the result of convolving image $I(\bxi)$ by a kernel $K(\x)$, the covariance $C^G$ between the errors on two pixel values $G(\bxi)$ and $G(\bxi+\bdelta)$ is
\begin{eqnarray}
\hbox{Cov}\left( \int \d\x I(\x)K(\bxi-\x), \int \d\y I(\y) K(\bxi+\bdelta-\y)\right) \nonumber\\
=\int \!\! \int \d\x \d\y \ \hbox{Cov}\left(I(\x),I(\y)\right) 
K(\bxi-\x) K(\bxi+\bdelta-\y) .
\end{eqnarray}
Our kernels are compact, and for all but the brightest sources the pixel noise in the original images is background-limited and slowly-varying. We can therefore assume that the covariance between neighbouring pixels depends mostly on the vector distance between those pixels and write
\be
\hbox{Cov}\left(I(\x),I(\y)\right)\simeq C^I(\x-\y) =C^I(\y-\x).
\label{eq:csymm}
\ee
(Note that strictly speaking the pixel values in the image are $I(\x)$ times the area of a pixel; for clarity we assume without loss of generality that this area is equal to one.) 

Writing $\x=\y-\x'$, $\y=\y'-\bxi$, and using the symmetry of $C^I$ (Eq.~\ref{eq:csymm}) we obtain
\begin{eqnarray}
C^G(\bdelta)&=&\int \!\! \int \d\x'\d\y' C^I(\x') K(\x'-\y') K(\bdelta-\y')\nonumber\\
&=&\int \d\x' C^I(\x') (K\!*\!K)(\bdelta-\x')\, ,
\label{eq:noisecov}
\end{eqnarray}
where $K\!*\!K$ denotes the autocorrelation function (ACF) of the kernel $K$. The noise covariance matrix $C^G$ of the convolved image is therefore the convolution of the original covariance matrix $C^I$ with $K\!*\!K$.
In our application the kernel ACF is calculated analytically using the shapelets formalism. 
As the shape of the kernel varies across the image, so does its ACF: we model the kernel ACF as a shapelet map (see Eq.~\ref{eq:shmap}) of the same order as the kernel, with scale radius $\beta_\rmn{p}\sqrt2$.

Even before Gaussianization, neighbouring pixels in our incoming co-added images have non-zero covariance as a result of re-gridding after astrometric calibration, particularly since our images are resampled on to slightly finer pixels (0.2\arcsec) than the native 0.213\arcsec\ scale of OmegaCAM. The \textsc{Swarp} code \citep{bertin/etal:2002} that does this resampling and co-addition produces a weight image that gives the inverse variance on each pixel value, but does not report  covariances. Rather than calculating $C^I$ analytically using the form of the \textsc{Swarp} interpolation kernel, we estimate the covariances numerically from the images themselves. An exact measurement is not critical here because in any case the dominant covariance contribution in Eq.~\ref{eq:noisecov} comes from the Gaussianization kernel.

Assuming that our incoming co-added frame $I$ consists of roughly Gaussian background noise plus positive sources, we estimate the shape of its covariance matrix, $C^I(\bdelta)/C^I({\bmath 0})$, from the statistics of pixel values which are likely to be source-free. The amplitude will vary across the image as a result of variations in background, dither patterns, CCD sensitivity variations etc., but we assume that the shape of $C^I$ is constant. The amplitude $C^I({\bmath 0})$ is set by scaling $C^I$ so that its peak value agrees with the \textsc{Swarp} weight image value at the centre of the source under consideration.

The noise variance $C^I({\bf 0})$ of the incoming co-added frame is estimated empirically from the statistics of `source-free' pixel values, which we select as those below the median (background) level $B$. We estimate the width of the distribution from their median difference from $B$, dividing by 0.675 to obtain a standard deviation estimate $\tilde\sigma$ assuming these statistics to be approximately Gaussian:
\be
C^I({\bf 0})\simeq\tilde{\sigma}^2 \,,\qquad\hbox{where}\quad
\tilde{\sigma}=m/0.675
\ee
and $m$ is the median of $\vert I(\bxi)-B\vert$ for those pixels with $I<B$.

The covariance $C^I(\bdelta)$ between pixel values separated by $\bdelta$ is similarly derived from the distribution of values of $I_\pm\equiv\frac12\left(I(\bxi)\pm I(\bxi+\bdelta)\right)$. We obtain
\be
C^I(\bdelta)\simeq \tilde\sigma^2_+ - \tilde\sigma^2_- \,,
\ee
where $\tilde\sigma_\pm=m_\pm/0.675$, and the $m_\pm$ are the medians of $\vert I_\pm-B\vert$  for those pixel pairs with $I_+<B$.

These estimated covariances are then modeled as a low-order shapelet, and convolved with the ACF of the Gaussianization kernel (Eq.~\ref{eq:noisecov}) to give the total noise covariance matrix. 

\subsubsection{Residual correction}
\label{sec:tweak}
\begin{figure}
\includegraphics[width=0.4\hsize]{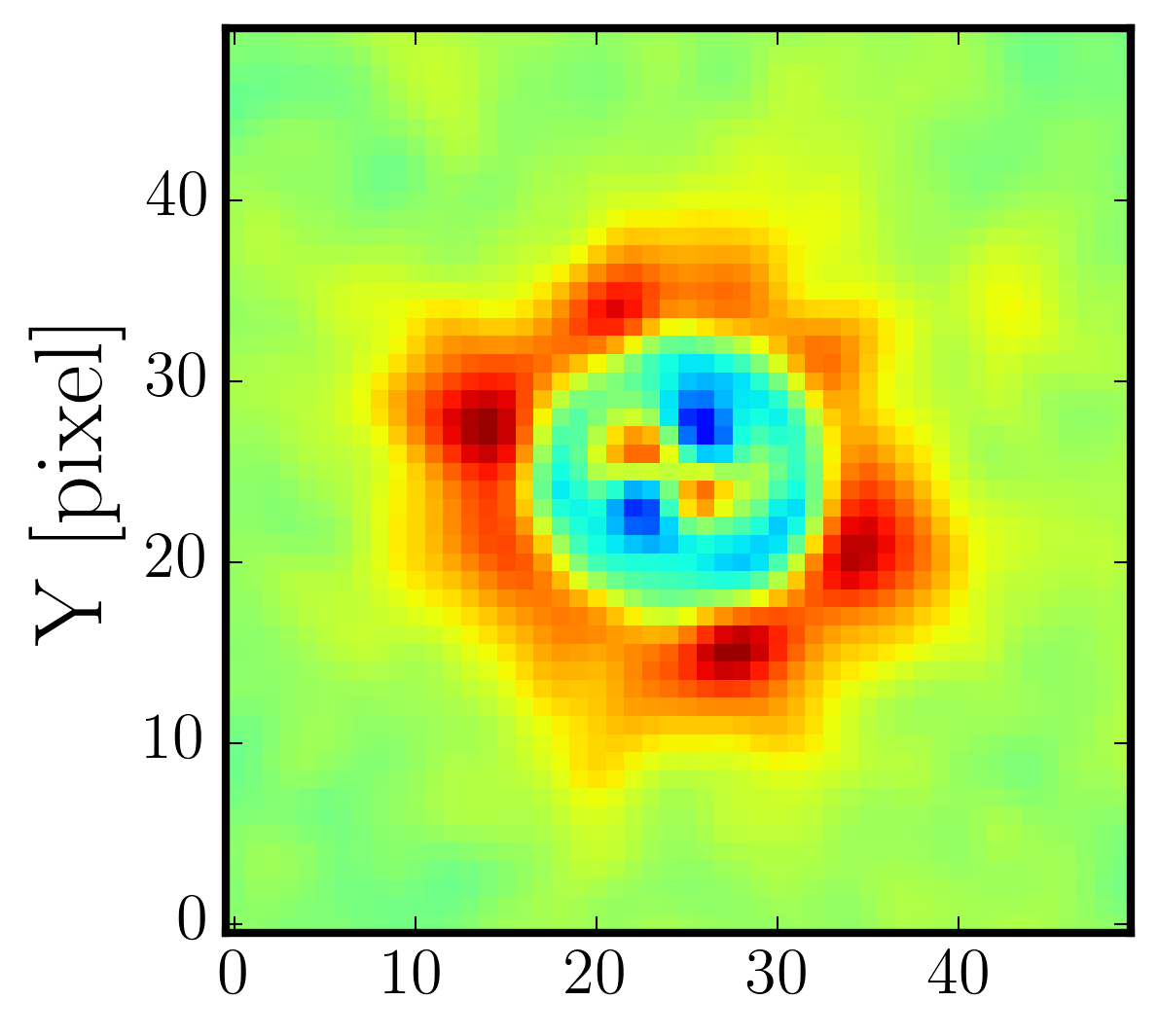}\hfill
\includegraphics[width=0.6\hsize]{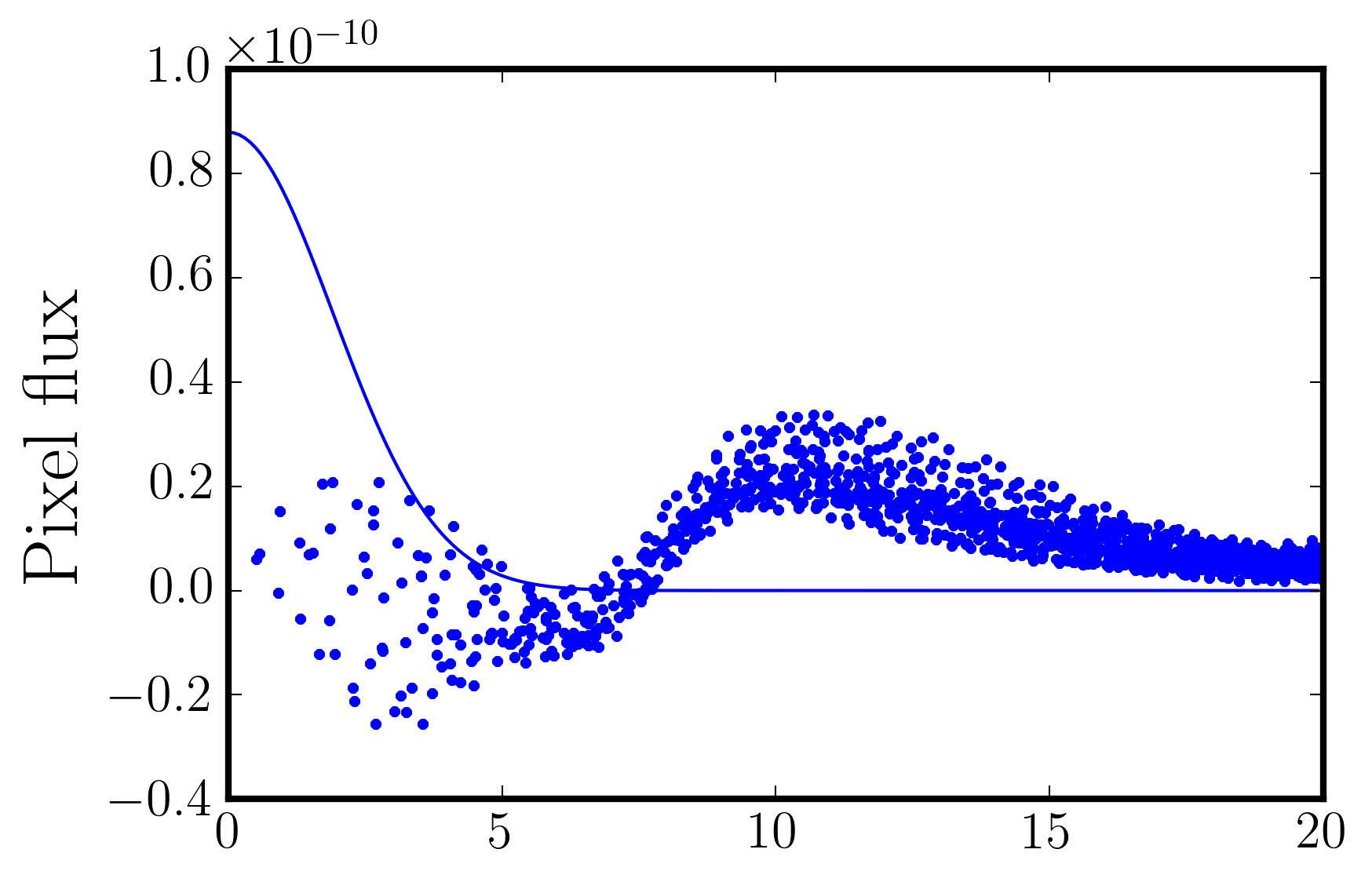}\\
\includegraphics[width=0.4\hsize]{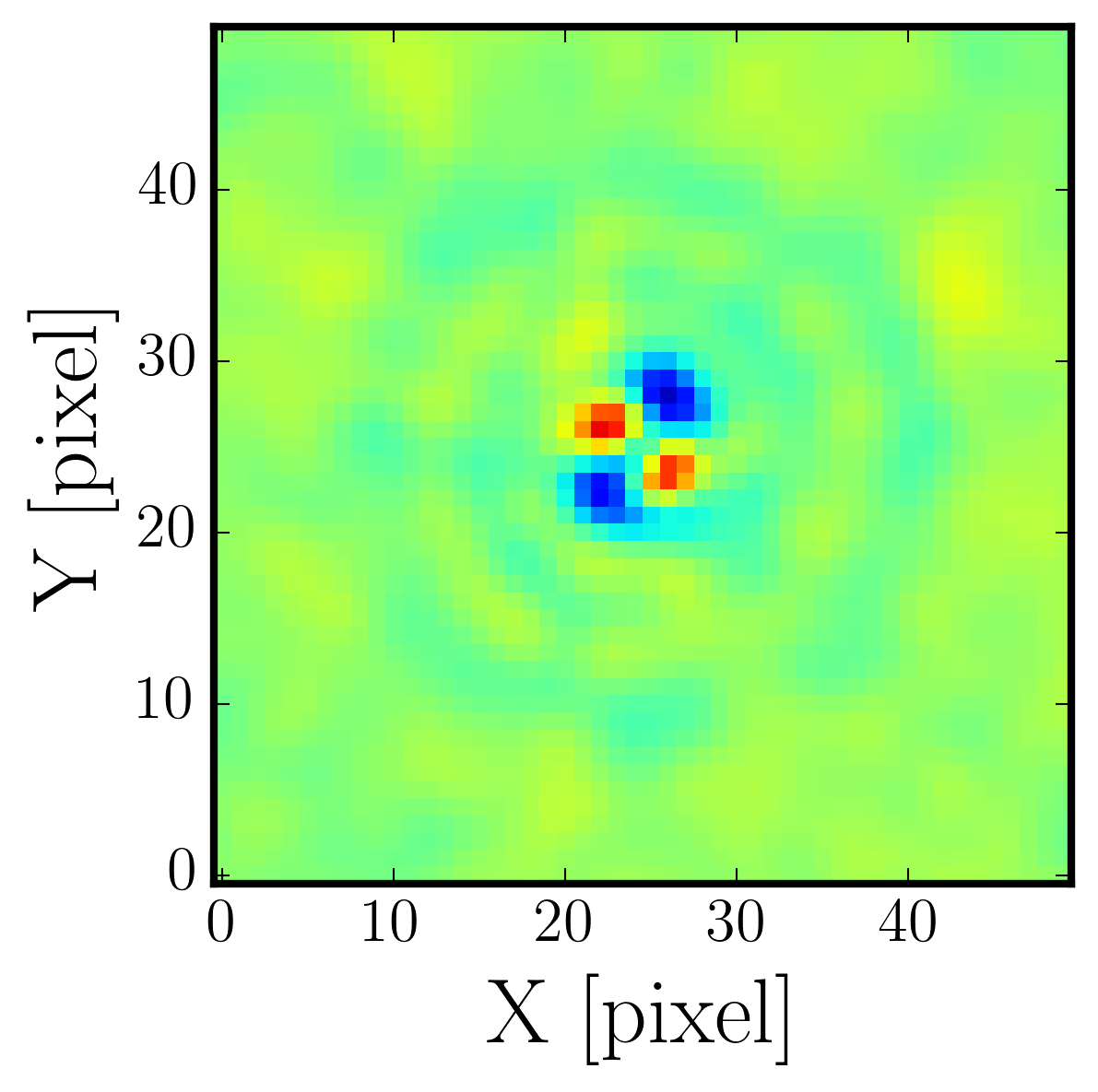}\hfill
\includegraphics[width=0.6\hsize]{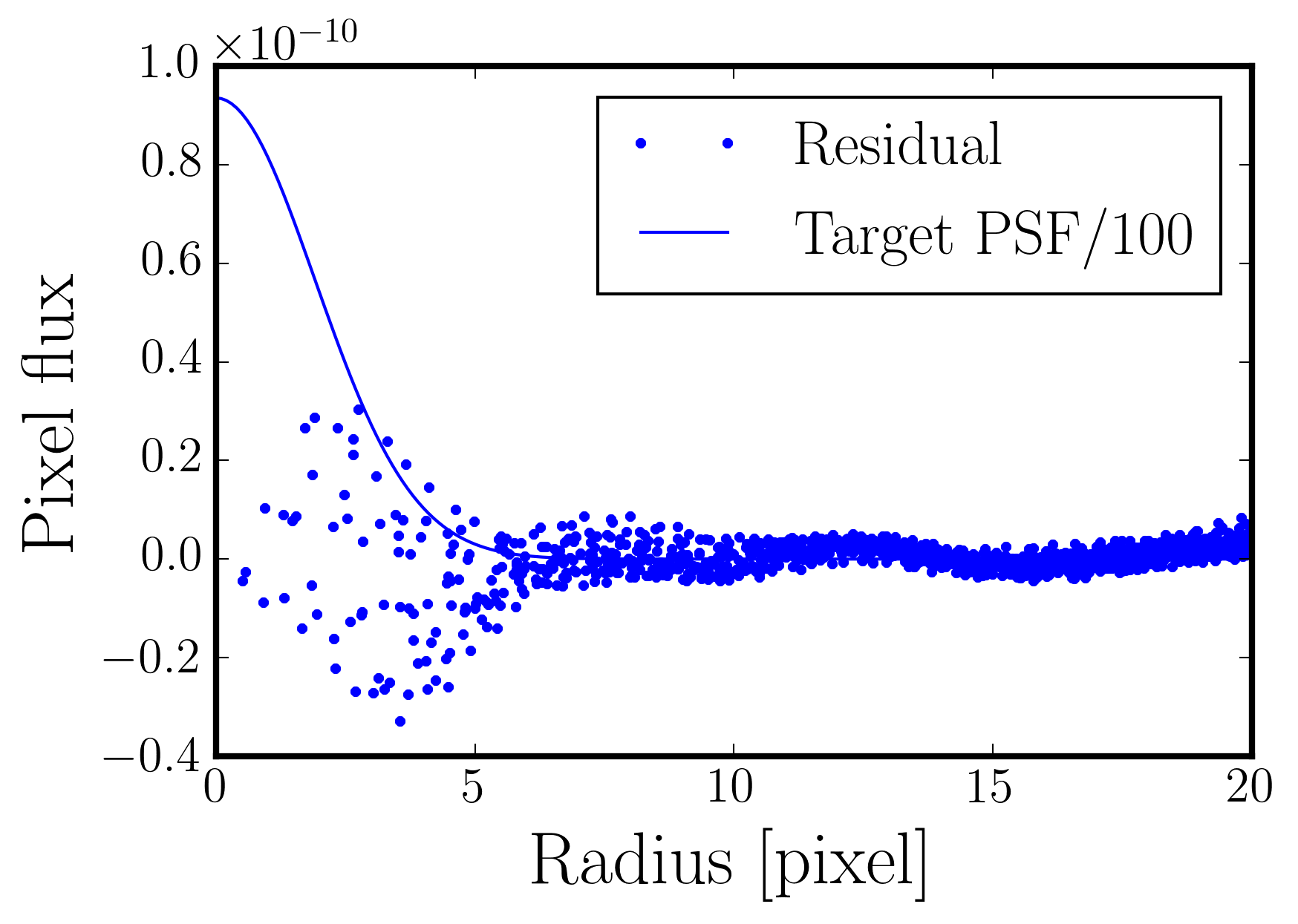}
\caption{
The effect of the residual PSF non-Gaussianity correction step described in \S\ref{sec:tweak}.
The panels in the top row  show the residual PSF flux around a bright star after PSF Gaussianization, both as an image (left) and with pixel values plotted as a function of distance from the centre of the star (right). For reference the target Gaussian PSF profile is plotted as a solid line, scaled down by a factor of 100.
The bottom row shows the same star after the correction step, and demonstrates that the large-radius residual flux is removed effectively by this perturbative procedure.}
\label{fig:gpsftweak}
\end{figure}

Because the shapelet description of the original PSF is not perfect -- in particular, shapelets asymptotically drop off faster than real PSFs -- some residual star flux remains at large radii after the convolution with the Gaussianization kernel. To correct for this, the Gaussianized co-added frames are improved further by means of a `tweaking' step. To this end, for each star on the Gaussianized image we measure the difference $\Delta P(\x)=P(\x)-G(\x)$ of its flux-normalized image $P$ to the target Gaussian $G$. We then fit this residual PSF using a spatially variable shapelet expansion as above, but with a scale radius that is a factor of two larger than before in order to capture more of the large-scale flux into the PSF model. The Gaussianized image is then convolved with a kernel $(1-\Delta P\otimes^{-1}G)$ that corrects, to first order in $\Delta P$, this residual non-Gaussianity of the PSF.
Here $\otimes^{-1}G$ indicates deconvolution by the target Gaussian PSF $G$ of dispersion $\beta_{\rm p}$, which can be done analytically since $\Delta P$ is expressed as a shapelet series with scale radius $\beta>\beta_{\rm p}$.
Fig.~\ref{fig:gpsftweak} shows that this process works well.

\subsection[GAaP photometry]{GA{\sevensize A}P photometry}
\label{app:gaapphot}
Once the PSF is standardized to a Gaussian, it is possible to perform aperture photometry in such a way that the answer is independent of the seeing. The central result of this `Gaussian Aperture and PSF' photometry is the relation between the pre- and post-seeing Gaussian-aperture weighted flux when the PSF is a Gaussian of dispersion $p$:
\begin{eqnarray}
\int \d\x
\left[\frac{1}{2\pi p^2} \int \d\x' I(\x') 
{\rmn e}^{-\frac12\vert\x-\x'\vert^2/p^2}\right]
{\rmn e}^{-\frac12\x^\mat{T} \W^{-1}\x}\nonumber\\
=
\frac{\det(\W)^{\frac12}}{\det(\W+p^2\mat{1})^{\frac12}}
\int \d\x\ I(\x) {\rmn e}^{-\frac12\x^\mat{T} (\W+p^2\mat{1})^{-1} \x} .
\label{eq:gaap}
\end{eqnarray}
Here the expression in square brackets is the PSF-convolved image of a source with intrinsic, pre-seeing surface brightness distribution $I(\x)$, and $\W$ is an `aperture matrix' that defines the shape, orientation and size of the Gaussian aperture. 
For a Gaussian aperture of major and minor axis dispersion $a$ and $b$, with the major axis oriented an angle $\alpha$ from the $x$-axis, we have
\be
\W=
\left(\!\!
\begin{array}{cc}
a^2\cos^2\alpha+b^2\sin^2\alpha & (a^2-b^2)\sin\alpha\cos\alpha\\
(a^2-b^2)\sin\alpha\cos\alpha&a^2\sin^2\alpha+b^2\cos^2\alpha
\end{array} 
\!\!\right).
\ee

Eq.~\ref{eq:gaap} shows that, when the PSF is Gaussian, a Gaussian-weighted aperture flux on the PSF-convolved image with aperture matrix $\W$ can be related directly to the intrinsic Gaussian-weighted aperture flux of the pre-seeing image with aperture matrix $\W+p^2\mat{1}$. Conversely, photometry for the intrinsic source with aperture matrix $\W_{\rm int}$ is simply done by applying the aperture matrix $\W_{\rm int}-p^2{\mat{1}}$ to a Gaussianized-PSF image, independent of the PSF size $p$ it was convolved to, and normalizing appropriately. We therefore {\em define} the \textsc{GAaP} flux $F_\W$ with reference to an aperture matrix $\W$ on the pre-seeing image $I(\x)$, but {\em measure} it from a PSF-Gaussianized image $G(\x)$:
\begin{eqnarray} 
F_\W 
&\equiv& 
\int\d\x\ I(\x){\rmn e}^{-\frac12\x^\mat{T} \W^{-1}\x} \nonumber\\
&=& 
\frac{\det(\W)^\frac12}{\det(\W-p^2\mat{1})^\frac12}
\int\d\x\ G(\x){\rmn e}^{-\frac12\x^\mat{T} (\W-p^2\mat{1})^{-1}\x}
\label{eq:Fgaap}
\end{eqnarray}
 For our variable-seeing, multi-band dataset, \textsc{GAaP} photometry therefore provides a way to obtain fluxes which pertain to the same part of the galaxy at all wavelengths, from which true colours can be derived. 

Since aperture photometry is a linear combination of pixel values it is straightforward to take account of the noise covariance $C^G$ (see \S\ref{sec:noisecov}) of the Gaussianized image when determining the uncertainty on the fluxes derived. The variance on a \textsc{GAaP} flux measurement is given by
\be
\hbox{Var}(F_\W)
=
\frac{2^\frac12 \pi\det(\W)}{\det(\W-p^2\mat{1})^\frac12}
\int\d \x\ C^G(\x) {\rmn e}^{-\frac14\x^\mat{T} (\W-p^2\mat{1})^{-1}\x}.
\label{eq:Fgaaperr}
\ee
With some work it is possible to take advantage of the shapelet formulation of $C^G$ in this equation to do the integral analytically. 

We use a dedicated code for the \textsc{GAaP} photometry. It first reads the PSF-Gaussianized image, the noise auto-correlation function map, the weight map corresponding to the original co-added image, as well as a list which contains each source's position together with the axis lengths $(a,b)$ and position angles $\alpha$ of the corresponding pre-seeing apertures (all in world coordinates). It then transforms these to pixel coordinates, performs the aperture photometry according to the second line of Eq.~\ref{eq:Fgaap}, and calculates the error bar following eq.~\ref{eq:Fgaaperr}. The same input list is used for runs on the images obtained for the tile with different filters.
These multi-band fluxes and corresponding errors are then fed into \textsc{bpz} for photometric redshift estimation, and merged into the master \textsc{SExtractor} and \textsc{lens}fit catalogue of the tile.

\begin{figure}
\centerline{
\includegraphics[width=0.8\hsize]{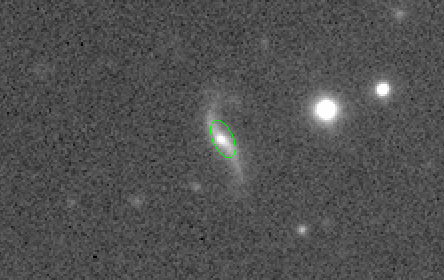}
}
\vspace{0.3mm}
\centerline{
\includegraphics[width=0.8\hsize]{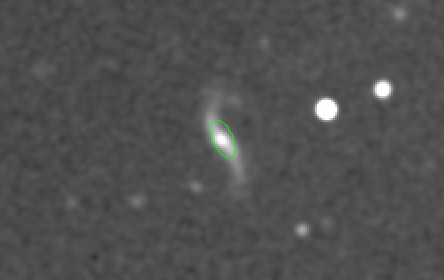}
}
\putfig{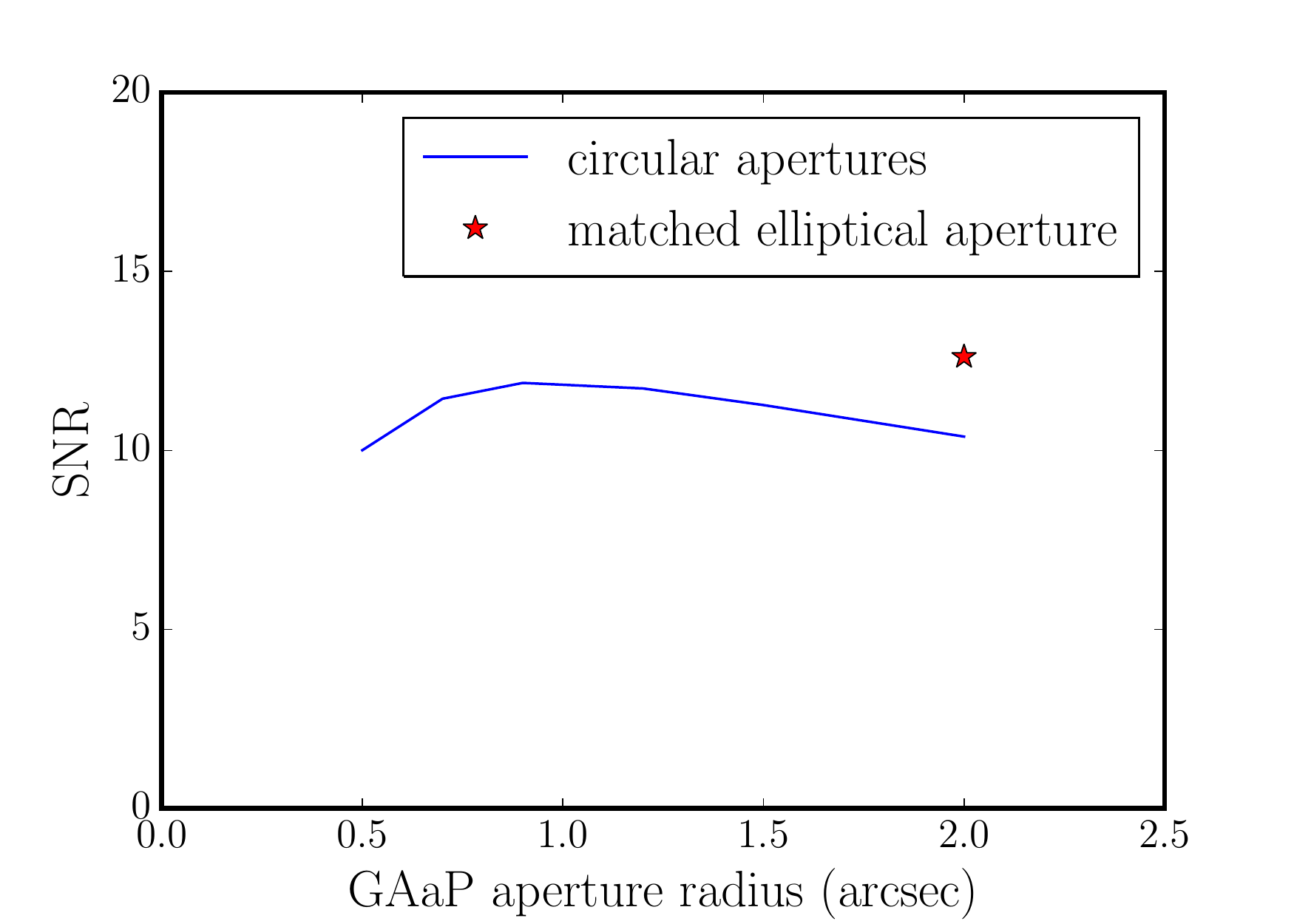}
\caption{\textsc{GAaP} apertures on an $r\simeq20$ galaxy. Top: co-added image, with $2\arcsec\times0.99\arcsec$ (1-$\sigma$) aperture overlaid. Middle: same, but PSF-Gaussianized image is shown. Bottom: Signal-to-noise ratio (SNR) for different circular apertures (solid curve) and for the aperture shown in the other panels (star symbol), for a source as the one illustrated but ten times fainter. The Figure illustrates the broad maximum of the SNR curve as well as the enhancement that follows from approximately matching the aperture size and shape to the source.}
\label{fig:gaaperr}
\end{figure}

While for isolated sources any aperture matrix can be used to obtain unbiased estimates of the \textsc{GAaP} flux, the optimal SNR is obtained when the shape and orientation of the post-seeing aperture defined by $\mat{W}-p^2\mat{1}$ is matched to the source. It turns out that the SNR of the flux has a rather broad maximum as a function of aperture size (Fig.~\ref{fig:gaaperr}), with the details depending on the exact Gaussianization kernel at the location of the source. This means that it is possible to select a compromise aperture which yields nearly optimal SNR in all bands even if these have different seeing. After some experimentation we chose an aperture $\W_{\rm int}$ which is derived from the \textsc{SExtractor} RMS axis lengths of the source on the $r$-band image, where we have the best seeing, by adding 0.7\arcsec\ in quadrature to the axis lengths $a$ and $b$. We keep the position angle $\alpha$ aligned with the major axis. This choice ensures that also poorer-seeing observations in the other bands (chiefly $u$ or $i$) can still be photometered with reasonable SNR. To prevent overlaps with neighbouring sources, we maximize $a$ and $b$ at 2\arcsec. 

It is important to realize that these \textsc{GAaP} aperture photometry values are not total fluxes, but rather the flux inside a well-defined tapered aperture. So while they are eminently suited to colour measurements, they do not replace total magnitudes. The exception is very compact sources and stars: in these cases the intrinsic aperture function $\exp(-\frac12\x^\mat{T}\W_{\rm int}^{-1}\x)$ is equal to 1 over the area where the source contributes flux, and hence the \textsc{GAaP} flux {\em does} equal the total flux.

Note that even though, for practical reasons, we carry out the photometry as a two-step process (first manipulating the pixels in the image and then photometering the result) it can also be written as a single photometry step on the original variable-PSF stack (albeit with a complicated aperture function, given by the convolution of the Gaussianization kernel with the Gaussian aperture function). Effectively, therefore, our photometry is still a linear combination of calibrated pixel fluxes, with a tractable error analysis. 

While here we restrict our use of the Gaussianized images to the aperture photometry, they can also be used for other aspects of the analysis, such as improving star-galaxy separation \citep{piladiez/etal:2014}, measurements of galaxy morphology parameters, PSF-corrected galaxy ellipticity measurements, or generating multi-colour images without colour-dependent PSF effects.

\section{Quality control}
\label{app:sanitycheckplots}

\subsection{Object detection and photometry}

\label{app:phot}

\begin{figure*}
 \includegraphics[width=0.94\hsize]{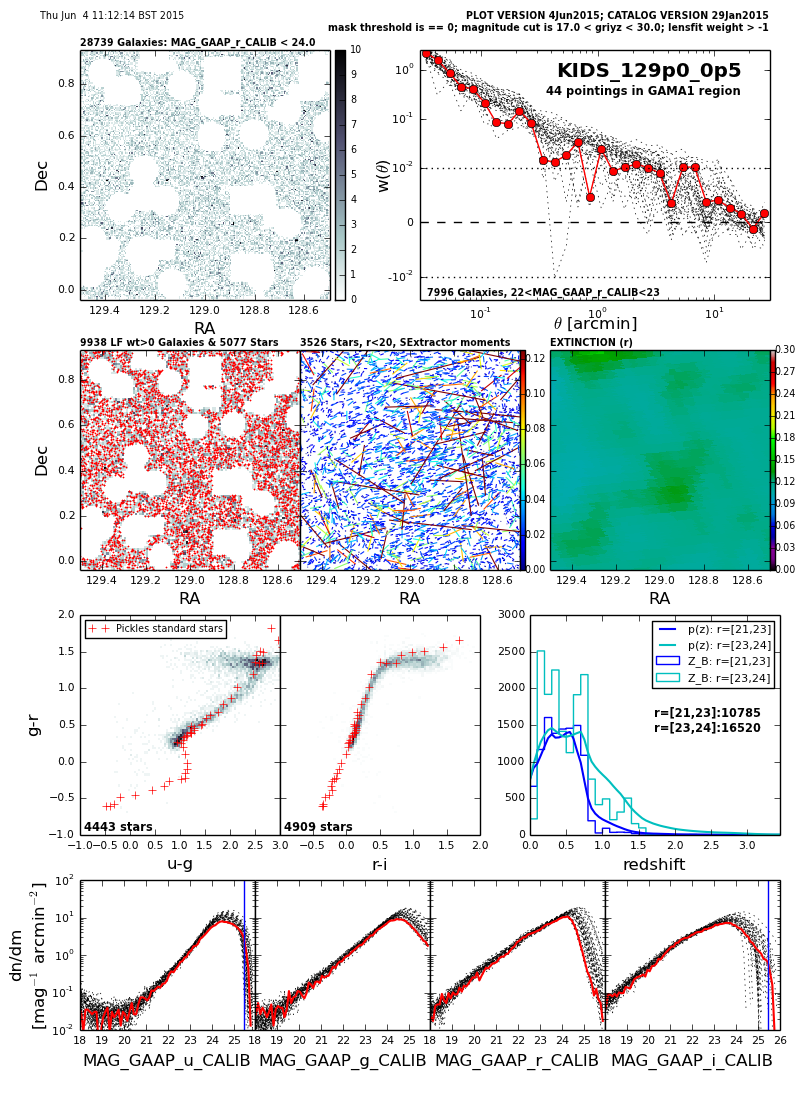}
 \caption{A check plot for photometry.  A single sanity check plot is generated for 
 each pointing for inspection.  See text for details. 
 \label{fig:check_photometry}
}
\end{figure*}

Once an object catalog is available, we inspect and verify the data quality with a
`check plot,' generated for each tile.
Fig.~\ref{fig:check_photometry} shows an example for one of the KiDS fields,
consisting of a collection of twelve plots.
The top row shows the masked distribution of galaxies (left) and the angular
correlation function $w(\theta)$ for a subset of these objects (right).
The angular correlation function plot contains data for the given field (red), overlaid
with those from other tiles within the same region of the survey (dotted black).  
Discrepant correlation functions are a sensitive indicator of spurious object detection and surface density inhomogeneity;
the combination of the two plots in the top row thus flags cases where additional masking
may be necessary.

The three plots in the second row show the stellar distribution (red crosses) and the
distribution of galaxies with valid shape measurements in grey scale (left); the PSF whisker
plot from the co-added image (centre), and the $r$-band extinction map (right).
The plot on the left allows a first check of the star selection: sufficient stars must be available over the entire unmasked area for reliable PSF estimation
(and hence shape measurements).  The PSF whisker plot is that of the co-added image, and will
indicate any issues with astrometry or the star-galaxy separation (the slight star-galaxy
contamination seen in Fig.~\ref{fig:check_photometry} is acceptable).  The extinction
map allows additional diagnosis of the main photometry checks below.

The bottom two rows contain the main photometry checks.  The two left-most plots in 
the third row contain a pair of colour-colour diagrams of the stars (grey scale), with the predicted
stellar loci calculated from Pickles standard star SEDs overlaid (red crosses).  An error in the
zero-point magnitude can be detected by a shift in the stellar loci, while a broadening or lack
of stellar loci indicates a problem with the star-galaxy separation.
The right-most plot in this row shows the photometric redshift distributions for bright
(blue) and faint (cyan) objects, both for the photometric redshift point estimates $z_{\rm B}$ (histograms)
and the stacked $p(z)$ probability distribution (smooth curves).  The photometric redshift distribution is
sensitive to zero-point errors in photometry.  The bottom row shows the galaxy number density per
magnitude $\d n/\d m$ in the four survey bands $ugri$. Similarly to the
$w(\theta)$ plot, these distributions (shown in red) are compared to those of neighbouring tiles, plotted with dotted black lines.
A horizontal shift of $\d n/\d m$ indicates a possible zero-point error, while a vertical offset signifies an unusually low number count, most likely the result of a processing failure.

\subsection{PSF modelling}
\label{app:psfmod}
\begin{figure*}
 \includegraphics[width=0.94\hsize]{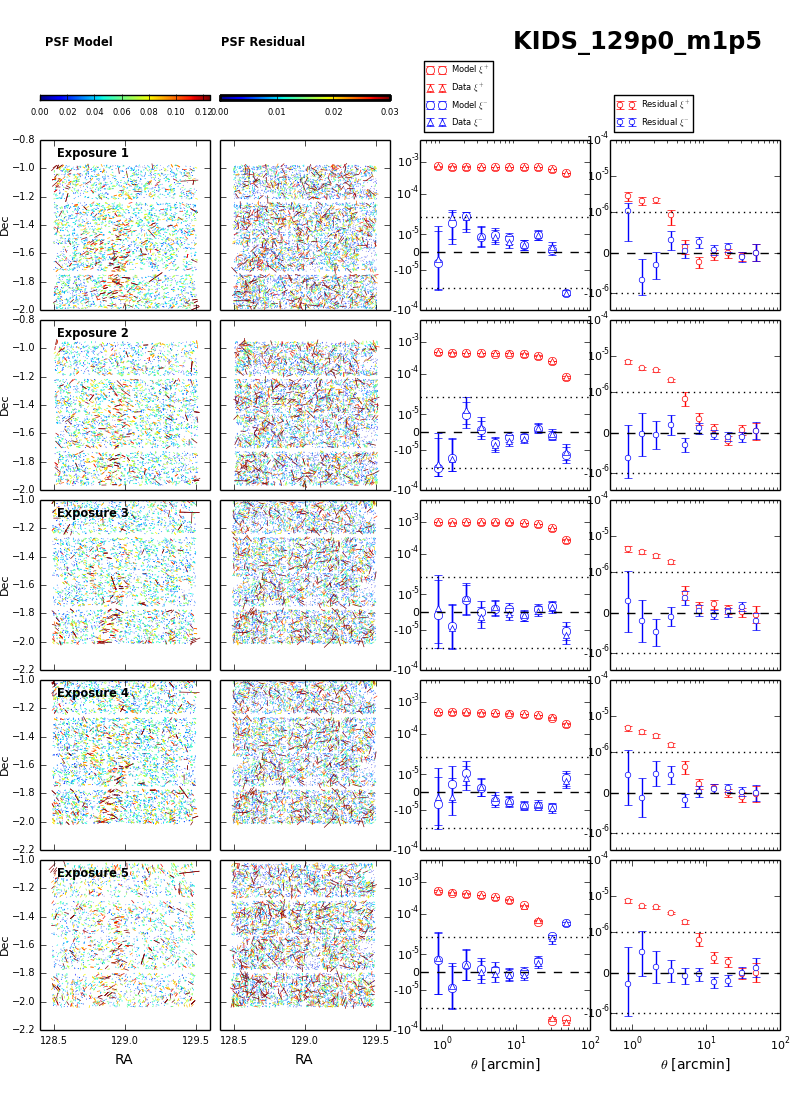}
 \caption{A check plot for PSF modelling.  A single check plot is generated for
 each pointing for inspection.  See text for details.
 \label{fig:check_PSF}
}
\end{figure*}

Fig.~\ref{fig:check_PSF} shows an example check-plot for one of the KiDS fields, designed to inspect the quality of the PSF modelling for the five sub-exposures of each field (each row displays the data for each sub-exposure).  There are four panels from left to right.  The first panel shows the PSF model, where the length and colour of the tick mark shows the direction and strength of the PSF ellipticity (defined in Eq.~\ref{eqn:estar}).  The second panel shows the residual PSF ellipticity $\delta \epsilon_{\rm PSF}$, as defined under equation~\ref{eqn:psfsize}.
Here the range of the ellipticity-scale is reduced to enhance the residuals.  The example shown is a typical KiDS observation revealing a rather noisy PSF residual, which at first sight might not appear useful.  This type of visualisation does however reveal the rare occasions where a poorly constrained polynomial model of the PSF (for example in a heavily masked region of the image) becomes ill-behaved.  The two right-hand panels compress the information in the two left-hand panels using the two-point PSF ellipticity correlation function, where the estimator is given in Eq.~\ref{eqn:xipm_est}. Between the dotted lines, at $\xi=2.10^{-5}$ (left) and $10^{-6}$ (right) the log-log scale becomes log-linear; the dashed line signifies zero correlation.  We see that the model accurately predicts the amplitude and angular dependence of the two-point PSF ellipticity correlation function, and that the PSF distortion for KiDS contributes predominantly to the $\xi_+$ correlation, only signal leaking into the $\xi_-$ correlation on large scales.  The final panel on the right shows the two-point PSF residual ellipticity correlation function which is two orders of magnitude lower than the PSF distortion and within the requirements for a future cosmological analysis of the full KiDS area, as discussed in \S\ref{sec:psfres}.

\section{K\lowercase{i}DS Shear and Photometric Redshift Catalogues:  source list columns}
\label{app:data}
The following table lists the columns that are present in the KiDS-DR2 shear and photometric redshift catalogues that are made publicly available to download from \url{http://kids.strw.leidenuniv.nl}.  We provide three catalogues, one for each GAMA field G09, G12 and G15, in \textsc{ldac-fits}\footnote{\textsc{ldac} tables can be read with most \textsc{fits} reading tools, or with specialized \textsc{Theli} tools from \url{http://marvinweb.astro.uni-bonn.de/data_products/THELIWWW/} .} format. The survey configuration is shown in fig.~1 of \citet{viola/etal:2015}.  Please note that the sources and masks in these catalogues do not correspond exactly to those in the ESO public data release catalogues presented by \citetalias{dejong/etal:2015}, who did not use the \textsc{Theli} reduction that is the basis of the lensing catalogues described here.
We recommend that users apply the following selection criteria on the masks, the \emph{lens}fit weight, SNR, and the \textsc{bpz} photometric redshift in order to measure a robust lensing signal:  $\verb MAN_MASK  = 0$, $\verb weight > 0$, $0.005 < \verb z_B  < 1.2$ and $ \verb SNratio  > 0$.  Objects that appear in overlapping tiles have already been removed from these region compiled catalogues.  Redshift distributions should be calculated from the full photometric redshift probability distribution $p(z)$, labelled \verb PZ_full .  Multiplicative, \verb m_cor_best , and additive calibration corrections, \verb c1_best  and \verb c2_best , should also be applied to the shear estimates.

In the following we provide additional information on certain columns in the catalogue.
\begin{itemize}
\item
\verb KIDS_TILE  is the name of the KiDS survey tile the source falls on. Searching the ESO archive with this \verb OBJECT  name will link to the \textsc{Astro-WISE} images of this tile.
\item
\verb THELI_NAME  gives the \textsc{Theli} name of the KiDS survey tile. For scripting reasons these names replace the `.' and `-' characters with `p' and `m', respectively.
\item
\verb MAG_GAAP_f_CALIB  gives the \textsc{GAaP} aperture flux in filter \verb f  (for filters $u$, $g$, $r$ and $i$), cross-calibrated to the SDSS photometry (S\ref{sec:photcal}). For each source these photometric parameters are given for a single aperture, specified by the major and minor axis lengths given in the \verb Agaper , \verb Bgaper  and \verb PAgaap  keywords. The fluxes are measured from the \textsc{Astro-WISE} co-added $u$-, $g$-, $r$- and $i$-band images, using the sky positions from the \textsc{Theli} $r$-band source catalogue. Note that these aperture magnitudes are mainly intended to be used for colour measurements, since they refer only to the central regions of the source. They are not total magnitudes except in the case of unresolved sources (see \S~\ref{sec:gaap} and~\ref{app:gaapphot}). Approximate total magnitudes may be obtained by taking the \verb MAG_AUTO  magnitude, which is based on the $r$-band images, applying the extinction correction, and adding the \textsc{GAaP} colours.
\item 
\verb Flag_GAAP_f  is set to $100$ when no \textsc{GAaP} flux could be measured in band \verb f , 0 otherwise.
\item
\verb PZ_full:  This is the full photometric redshift probability distribution $p(z)$ from $0.005 \leq z \leq 3.505$.  There are 70 columns sampling $p(z)$ at intervals of $dz = 0.05$.  The first bin is centred at $z = 0.03$ and we recommend a linear or spline interpolation between the mid-points of each bin to recover a smooth redshift distribution from a sample of galaxies.   There is a final bin not included in these catalogues with $z>3.505$, such that in a small number of cases $p(z)$ will not sum to 1.  In a lensing analysis it is highly unlikely that a galaxy at such a high redshift would have a plausible shape measurement, and we recommend applying a hard prior of zero probability past $z>3.505$.
\item 
\verb e1  or \verb e2: \emph{lens}fit shear estimators.  Note that the \verb e2  component is defined relative to the \verb ALPHA_J2000, \verb DELTA_J2000  co-ordinate grid.  Depending on how users define their angles in this reference frame, they may find that they need to change the sign of \verb e2 .
\item
\verb TILE_PSF_SYS_OK : The star-galaxy ellipticity correlation pass/fail flag, as described in \S\ref{sec:passfail}, which applies on a tile-by-tile basis.  Applying this field selection is required to reduce systematic contamination from the PSF for cosmic shear science.  It is not, however, required for galaxy-galaxy lensing or cluster lensing science.
\item 
\verb m_cor  and \verb m_cor_best : The multiplicative calibration correction which should be applied in an ensemble average, rather than on a galaxy-by-galaxy basis.  \verb m_cor  was the correction used by the early science papers that accompany this release which did not include the correct pixel scale (see \S\ref{sec:resamp} for discussion).  \verb m_cor_best  is the calibration correction used in the cosmic shear analysis in this paper, with the correct pixel scale applied.
\item
\verb c1  and \verb c2 : The additive calibration correction used in the early science papers that accompany this release.  These should be subtracted from the \emph{lens}fit shear estimators \verb e1  and \verb e2 . Some regions in the SNR-size-Strehl parameter space are too sparsely populated to calculate this empirical correction, resulting in a few tens of galaxies raising a flag value for \verb c1  and \verb c2  of $-99$.  These galaxies should either be removed from the analysis, or the correction should not be applied.
\item
\verb c1_best  and \verb c2_best : An improved estimate of the additive calibration correction used in the cosmic shear analysis in this paper.  These should be used in conjunction with the pass/fail field selection \verb TILE_PSF_SYS_OK .  See above for flagged values.
\item
\verb PSF_e1  and  \verb PSF_e1_exp[k] : Model PSF ellipticities at the location of the object, in this case the real part of $\epsilon_{\rmn PSF}$. \verb PSF_e1  is the  average PSF ellipticity over all sub-exposures that included the object, whereas \verb PSF_e1_exp[k]  refers to the single sub-exposure number \verb k .   In cases where objects are not imaged in the sub-exposure, \verb PSF_e1_exp[k]  is given a value of -99.
\end{itemize}

\begin{table*}
\caption{\label{tab:cataloguedesc} Columns provided in the KiDS lensing catalogues.}
\begin{tabular}{lll}
\hline\hline 
Label & Description & Units  \\
\hline 
\verb KIDS_TILE & Name of survey tile  \\
\verb THELI_NAME & \textsc{Theli} name for the tile \\
\\
\multicolumn{3}{l}{\textsc{SExtractor}$^a$ parameters derived from the \textsc{Theli} $r$-band co-added image:}\\[0.3em]
\verb SeqNr & Running object number within the catalogue\\
\verb FLUX_AUTO & $r$-band flux & counts\\
\verb FLUXERR_AUTO & Error on \verb FLUX_AUTO  &counts\\
\verb MAG_AUTO & $r$-band magnitude & mag\\
\verb MAGERR_AUTO & Error on \verb MAG_AUTO  &mag\\
\verb KRON_RADIUS &Scaling radius of the ellipse for magnitude measurements\\
\verb BackGr &Background counts at centroid position & counts\\
\verb Level &Detection threshold above background & counts\\
\verb MU_THRESHOLD &Detection threshold above background & mag\ arcsec$^{-2}$\\
\verb MaxVal  &Peak flux above background &counts\\
\verb MU_MAX  &Peak surface brightness above background&mag\ arcsec$^{-2}$\\
\verb ISOAREA_WORLD &Isophotal area above analysis threshold&deg$^2$\\
\verb Xpos  & Centroid $x$ position in the \textsc{Theli} image &pix\\
\verb Ypos  & Centroid $y$ position in the \textsc{Theli} image &pix\\
\verb ALPHA_J2000 & Centroid sky position right ascension (J2000) &deg\\
\verb DELTA_J2000 & Centroid sky position declination (J2000)  &deg\\
\verb A_WORLD  &Profile RMS along major axis&deg\\
\verb B_WORLD  &Profile RMS along minor axis&deg\\
\verb THETA_J2000 $^a$&Position angle (West of North)$^a$&deg\\
\verb ERRA_WORLD  &World RMS position error along major axis&deg\\
\verb ERRB_WORLD  &World RMS position error along minor axis&deg\\
\verb ERRTHETA_J2000  &Error on \verb THETA_J2000 &deg\\
\verb FWHM_IMAGE  &FWHM assuming a gaussian object profile & pix\\
\verb FWHM_WORLD  &FWHM assuming a gaussian object profile &deg\\
\verb Flag   &\textsc{SExtractor} extraction flags\\
\verb FLUX_RADIUS  &Half-light radius&pix\\
\verb NIMAFLAGS_ISO  &Number of flagged pixels\\
\verb CLASS_STAR  &Star-galaxy classifier\\
\\
\multicolumn{3}{l}{Other parameters derived from the \textsc{Theli} $r$-band co-added image:}\\[0.3em]
\verb MAN_MASK  & Final masking flag value, including both manual and automated masks (0 = no mask)\\ 
\verb SG_FLAG   &Star-galaxy separator from 2nd and 4th image moments (0=star, 1=galaxy)\\
\\
\multicolumn{3}{l}{Parameters derived from the \textsc{Astro-WISE} $u$-, $g$- , $r$- and $i$-band co-added images:}\\[0.3em]
\verb Agaper   &Major axis of \textsc{GAaP} aperture&arcsec\\
\verb Bgaper   &Minor axis of \textsc{GAaP} aperture&arcsec\\
\verb PAgaap   &Position angle of major axis of \textsc{GAaP} aperture (North of West)&deg\\
\verb MAG_GAAP_[ugri]_CALIB & \textsc{GAaP} $[ugri]$ magnitude, zero-point and extinction-calibrated &mag\\
\verb MAGERR_GAAP_[ugri]  & Error on \verb MAG_GAAP_[ugri]_CALIB &mag\\
\verb Flag_GAAP_[ugri]  & Flag for  \verb MAG_GAAP_[ugri]_CALIB \\
\verb EXTINCTION_[ugri]  &Galactic extinction in the $ugri$ band&mag\\
\verb MAG_LIM_[ugri]  & 1-$\sigma$ limiting magnitude in the $ugri$ band&mag\\
\verb Z_B   &\textsc{bpz} redshift estimate; peak of posterior probability distribution\\
\verb Z_B_MIN   &Lower bound of the 95\%\ confidence interval of \verb Z_B \\
\verb Z_B_MAX   &Upper bound of the 95\%\ confidence interval of \verb Z_B \\
\verb T_B   &Spectral type corresponding to \verb Z_B \\
\verb ODDS   &Empirical ODDS of \verb Z_B \\
\verb Z_ML   &\textsc{bpz} maximum likelihood redshift\\
\verb T_ML   &Spectral type corresponding to \verb Z_ML \\
\verb CHI_SQUARED_BPZ  &$\chi^2$ value associated with \verb Z_B \\
\verb BPZ_FILT  &Filters with photometry used in \textsc{bpz}; bit-coded mask\\
\verb NBPZ_FILT  &Number of filters with good photometry used in \textsc{bpz}\\
\verb BPZ_NONDETFILT  &Filters with faint photometry (not used in \textsc{bpz}); bit-coded mask\\
\verb NBPZ_NONDETFILT  &Number of filters with faint photometry\\
\verb BPZ_FLAGFILT  &Filters with flagged photometry (not used in \textsc{bpz}); bit-coded mask\\
\verb NBPZ_FLAGFILT  &Number of filters with flagged photometry\\
\verb PZ_full   &Vector containing the posterior photo-$z$ probability\\
\hline
\multicolumn{3}{l}{Continued on next page}
\end{tabular}
\end{table*}

\begin{table*}
\begin{tabular}{llll}
\hline\hline 
Label & Description & Units  \\
\hline 
\multicolumn{3}{l}{Parameters derived with \emph{lens}fit on the \textsc{Theli} $r$-band sub-exposures:}\\[0.3em]
\verb weight   & inverse variance weight\\
\verb fitclass  & fit class (class=0 means a galaxy, no issue)\\
\verb scalelength  & galaxy model scale length&pix\\
\verb bulge_fraction  &galaxy model bulge fraction\\
\verb model_flux  &galaxy model flux&counts\\
\verb SNratio   &SNR for model fit\\
\verb PSF_e1  &mean ellipticity of PSF, component 1\\
\verb PSF_e2  &mean ellipticity of PSF, component 2\\
\verb PSF_Strehl_ratio & Pseudo-Strehl ratio of PSF (flux fraction in central pixel)\\
\verb catmag   &$r$-band magnitude used to calculate the size prior\\
\verb n_exposures_used &Number of sub-exposures used \\
\verb PSF_e1_exp1  &PSF model ellipticity component 1, on sub-exposure 1\\
\verb PSF_e2_exp1  &PSF model ellipticity component 2, on sub-exposure 1\\
... & ... & ...         \\
\verb PSF_e1_exp5  &PSF model ellipticity component 1, on sub-exposure 5\\
\verb PSF_e2_exp5  &PSF model ellipticity component 2, on sub-exposure 5\\
\verb e1                 &Galaxy ellipticity $\epsilon_1$ (no $c$ or $m$ correction)\\
\verb e2                 &Galaxy ellipticity $\epsilon_2$ (no $c$ or $m$ correction)\\
\verb c1   &Additive bias of $\epsilon_1$ based on SNR, scale length and Strehl\\
\verb c2   &Additive bias of $\epsilon_2$ based on SNR, scale length and Strehl\\
\verb m_cor   &Multiplicative bias of $\epsilon_1$ and $\epsilon_2$ based on SNR, scale length and Strehl\\
\verb c1_best   &Updated additive bias of $\epsilon_1$ based on SNR, scale length and Strehl\\
\verb c2_best   &Updated additive bias of $\epsilon_2$ based on SNR, scale length and Strehl\\
\verb m_cor_best   &Updated multiplicative bias of $\epsilon_1$ and $\epsilon_2$ based on SNR, scale length and Strehl\\
\\
\multicolumn{3}{l}{Parameters derived from systematics tests on the catalogue:}\\[0.3em]
\verb TILE_SYS_OK &Tile pass/fail flag based on star-galaxy ellipticity correlation (pass=1, fail=0)\\
\hline
\end{tabular}
\\
$^a$ These catalogues were created with \textsc{SExtractor} version 2.2.2. Note that from version 2.4.6 onwards the definition of \verb THETA_J2000  was changed from West-of-North to East-of-North, i.e., the sign flipped.
\label{lastpage}
\end{table*}

\end{document}